\documentclass[preprint-reprint
 amsmath,amssymb,
 aps,
]{revtex4-1}

\usepackage{graphicx}
\usepackage{dcolumn}
\usepackage{bm}

\begin{document}


\title{Growing complex network of citations of scientific papers- measurements and modeling\\}

\author{Michael Golosovsky}
\email{michael.golosovsky@mail.huji.ac.il}
\author{Sorin Solomon}%
\affiliation{The Racah Institute of Physics, The Hebrew University of Jerusalem, 91904 Jerusalem, Israel\\
}%
\date{\today}

\begin{abstract}
To quantify the mechanism of a complex network growth we focus on the network of citations of scientific papers and  use a combination of the theoretical and experimental tools to uncover microscopic details of this network growth. Namely, we develop a stochastic model of citation dynamics based on copying/redirection/triadic closure mechanism.  In a complementary and coherent way, the model  accounts both for statistics of references of scientific papers and for their citation dynamics. Originating in empirical measurements, the model is cast in such a way that it can be verified quantitatively in every aspect. Such verification is performed by measuring citation dynamics of Physics papers.  The measurements revealed nonlinear citation dynamics, the nonlinearity being intricately related to network topology. The nonlinearity has far-reaching consequences including non-stationary citation distributions, diverging citation trajectory of similar papers, runaways or "immortal papers" with infinite citation lifetime  etc.  Thus, our most important finding is nonlinearity in complex network growth.  In a more specific context, our results can be a basis for quantitative probabilistic prediction of citation dynamics of individual papers and of the journal impact factor.
\begin{description}
\item[PACS numbers]89.75.-k, 02.50.Ey, 01.75.+m
\end{description}
\end{abstract}
\maketitle
\section{\label{sec:intro}Introduction}
Complex networks became  objects of Physics research after advent of the Internet, appearance of large information databases, and mapping of genetic and metabolic networks.  Network topology has been thoroughly studied \cite{Newman2010,Barabasi2015} and the current research shifts more to temporal and evolving networks \cite{Holme2012} and dynamic processes in networks, such as network growth. The paradigm for complex network growth is the cumulative advantage mechanism invented by de Solla Price.\cite{Price1976} The most quantified complex network in his time was citation network. It exhibited an intriguing power-law degree distribution which was considered as an evidence of the scale-free behavior. de Solla Price sought to explain this behavior and postulated that citation network grows by addition of new papers that cite older papers  with  probability
\begin{equation}
\lambda_{i}\propto (K_{i}+K_{0})
\label{Price}
\end{equation}
where $K_{i}$ is the number of citations of the target paper $i$ and $K_{0}$ is an unspecified  constant. de Solla Price showed that the linear growth rule captured by Eq. \ref{Price} generates networks with the power-law degree distribution.  With  appearance of Internet and vigorous advent of network science, a similar rule was invented by Barabasi \cite{Barabasi2002} who  suggested that Eq. \ref{Price} is the generic growth rule of complex networks. The Barabasi-Albert model or preferential attachment is also known colloquially as a "rich get richer" or Matthew effect.\cite{Perc2014}

Equation \ref{Price} was  generalized to include aging and nonlinearity,\cite{Dorogovtsev2000,Krapivsky2001}
\begin{equation}
\lambda_{i}=A(t)(K_{i}+K_{0})^{\delta}
\label{Barabasi}
\end{equation}
Here, $A(t)$ is the aging function, common to all nodes, $K_{0}$ is  the initial attractivity, and $\delta$ is the growth exponent.  The measurements on many complex networks \cite{Perc2014} verified Eq. \ref{Barabasi} and showed ubiquity of  networks with $\delta\sim 1$  and  power-law degree distributions.

Although Eq. \ref{Barabasi} successfully describes the complex network growth,  it presents some conceptual difficulties. Indeed, Eq. \ref{Barabasi} encodes an empirical rule assuming that each node in the network garners new links with the rate proportional  to its current degree, implying that all nodes  differ only in one dimension. This yields similar growth dynamics of the nodes of the same age, while in reality there is  a huge diversity in their growth trajectories.

Bianconi and Barabasi \cite{Bianconi2001} added  a new dimension  to node description - fitness. This notion replaced the egalitarian picture according to which all nodes are born equal  by a  picture where each node is born with some intrinsic propensity of growth.  The corresponding growth rule  \cite{Wang2013} (see also  Refs. \cite{Kong2008,Medo2011}) is
\begin{equation}
\lambda_{i}=\eta_{i}A(t)(K_{i}+K_{0})
\label{fitness-Wang}
\end{equation}
where $\eta_{i}$ is the  node fitness, an empirical parameter introduced on top of the preferential attachment (which is also empirical rule). To be less empiric, several authors \cite{Menczer2004,Papadopoulos2012,Bramoulle2012,Yun2015} added more physical sense to Eq. \ref{fitness-Wang} and replaced the fitness by the  node similarity (homophily). This notion captures the fact that a new node  tends to link to the nodes  with similar content rather than to a randomly chosen node.  Technically, this line of reasoning results in  Eq. \ref{fitness-Wang} where $\eta_{i}$ is replaced  by $\eta_{ij}$, the latter quantifying the similarity between the two connecting nodes.\cite{Servedio2004}

Still,  Eq. \ref{fitness-Wang} contains too many empirical parameters  that prompt for microscopic explanation. The need for such explanation  becomes evident after realizing that Eqs. \ref{Price}-\ref{fitness-Wang} are \emph{global}. In order that a new node  attaches preferentially to most popular nodes it shall be familiar with the whole network.  This global picture is unrealistic and many efforts have been spent to elucidate the \emph{local} microscopic mechanism staying behind Eqs. \ref{Price}-\ref{fitness-Wang}.

The most popular local  mechanism is the copying rule.\cite{Kleinberg1999} The Refs. \cite{Vazquez2003,Evans2005} demonstrated that  Eq. \ref{Price}  can evolve from  this rule which is also  known as recursive search \cite{Vazquez2001}, link copying or redirection  \cite{Krapivsky2001,Krapivsky2005,Ren2012}, random walk/local search \cite{Jackson2007,Goldberg2015},  triple/triangle formation \cite{Wu2009}, transitive triples \cite{Itzhack2010}, or triadic closure \cite{Martin2013}. A similar rule operates in social networks \cite{Pennock2002,Jackson2007}, epidemic-like propagation of ideas \cite{Goffmann1964,Bruckner1990,Bettencourt2006},  diffusion of innovations \cite{Bass2004}, and citation dynamics \cite{Vitanov2012}.
This rule assumes that a new node  performs random and recursive searches:  first, it attaches to a randomly chosen node, secondly, it copies some links of the latter. This results in the following dynamic equation:
\begin{equation}
\lambda_{i}=A(t)[cK_{0}+(1-c)K_{i}]
\label{triple}
\end{equation}
where $A(t)$ is the aging factor, the first and the second  addends in the parentheses correspond to the random and recursive searches, respectively, and the parameter $c$ regulates the relative weights of the two. Similar two-term growth equations were suggested by Refs. \cite{Pennock2002,Menczer2004,Shao2006,Peterson2010}.
Equation \ref{triple} is formally identical to Eq. \ref{Price} with  $K_{0}$  characterizing the probability of random search.   The intuition behind the second addend  is as follows: if some node $i$ has $K_{i}$ links, the probability to  find it through recursive search is increased by  a factor $K_{i}$.  Thus Eq. \ref{triple} seems to provide a natural explanation for  the preferential attachment mechanism.

However, the parameters of Eq. \ref{triple} were never measured systematically. For example, it is not known whether  time dependences of the random and recursive search are the same or  different, whether the probability of recursive search is the same for all nodes of same age or not. Our  goal is to  measure dynamic parameters  of some real network, to establish its microscopic growth rules, and to compare them to existing models. We consider an iconic example of a growing network - citations to scientific papers, having in mind that the models of network growth were originally suggested  in relation to this very network.\cite{SollaPrice1965} Despite some specificity (it is ordered, acyclic, and does not allow rewiring and link deletion), citation network is a well-documented prototypical directed network. Following Ref. \cite{Borner2004} we  adopt a comprehensive approach  where we consider its growth from two perspectives: the perspective of the author and the perspective of the cited paper. The former approach focuses on the composition of the reference list of a paper, the latter approach focuses on the papers that cite a given paper. We establish duality between these two approaches and formulate a stochastic model that accounts both for citation dynamics and for the age composition of the reference lists of the papers.

\section{Citation dynamics from the author's perspective}
The  composition of the reference lists  of scientific papers is the clue to citation analysis. While citation dynamics of a paper is determined by several factors: popularity of the research field, journal impact factor, preferences and tastes of citing authors, etc.; the reference list  derives from one factor: decision of a research team or even single author who chooses the references  basing on their relevance and age. We focus here on the age of the references and do not consider their content, although this can be very important.\cite{Menczer2004} Our goal  is to measure and to model the age composition of the reference lists of papers. 

The author writing a research paper  reads scientific journals or media articles, searches databases, finds relevant papers and selects some of them  as references. These are direct references.\cite{direct}  Then the author studies the reference lists of the preselected papers, picks up  relevant references, reads them, adds  some of them into his reference list, and continues recursively. These are indirect references. The latter can also emerge if the author finds each reference independently. Since old references are usually seminal studies, the  most recent references will probably cite them as well.  Without inquiring too much into the process of reference list arrangement, we classify the references into two classes: direct and indirect. The former are those that are not cited by any other references in the reference list of the paper, the latter are those cited by one or more  preselected  references. The causality principle requires the indirect references to be older than their preselected sources.

\begin{figure}[ht]
\begin{center}
\includegraphics*[width=0.6\textwidth]{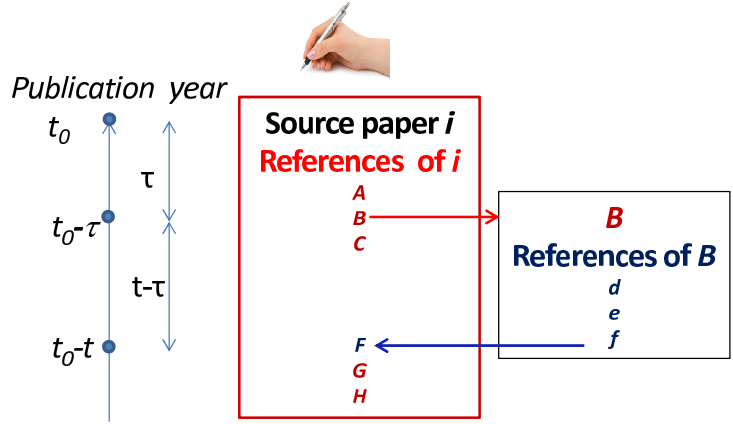}
\caption{Cartoon scenario which is the basis of our model.  Consider a source paper \emph{i} published in  year $t_{0}$  and its list of references \emph{A-H} arranged in descending chronological order. Each preselected paper \emph{B}  can bring several of its references to the reference list of \emph{i}. The probability  that a secondary reference \emph{f}  published in year $t_{0}-t$   appears in the reference list of \emph{i} depends on $\tau$, the time lag between  publication years of \emph{i} and \emph{B}, and on $s$, the number  of  preselected papers citing \emph{f}.
}
\label{fig:cartoon}
\end{center}
\end{figure}

Basing on the causality principle we develop an analytical model that accounts for the  age composition of the average reference list. To this end we consider a specific implementation of the referencing process -  recursive search.\cite{Kleinberg1999,Vazquez2001,Vazquez2003,Evans2005,Krapivsky2001,
Krapivsky2005,Ren2012,Jackson2007,Goldberg2015,Wu2009,Itzhack2010,Martin2013} shown in Fig.  \ref{fig:cartoon}. Consider a parent paper \emph{i} published in year $t_{0}$ and one of its references \emph{B} published in year $t_{0}-\tau$. Once \emph{i} cites   \emph{B},  it can cite any paper \emph{f} from the reference list of the latter. The probability of such indirect citation depends on a variety of factors, the most important being $\tau$- the time lag between  publication years of \emph{i} and \emph{B}. We also account for multiplicity: if some paper \emph{f} appears in the reference lists of $s$ preselected papers, its probability to appear in the reference list of the source paper \emph{i} is obviously increased. We assume that $P_{f}=P(\tau,s,R_{0}(t_{0}-\tau))$ where $R_{0}(t_{0}-\tau)$ is the length of the reference list of \emph{B}.

Following Ref. \cite{Wu2009} we develop this scenario into analytical model accounting for the  age structure of a typical reference list. Indeed, consider  the papers  in one scientific field that were published in the same year $t_{0}$. The average number of references   published in year $t_{0}-t$ that appear in the reference list of a paper published in year $t_{0}$ consist of direct and indirect references,
\begin{equation}
R(t_{0},t_{0}-t)=R_{dir}(t_{0},t_{0}-t)+R_{indir}(t_{0},t_{0}-t)
\label{ref-sum}
\end{equation}
where  $R_{dir}(t_{0},t_{0}-t)$ is an empirical function taken from measurements while $R_{indir}(t_{0},t_{0}-t)$ is determined from the causality principle. Indeed,  the reference list of the paper \emph{B} published in year $t_{0}-\tau$ (Fig. \ref{fig:cartoon}) contains $R(t_{0}-\tau, t_{0}-t)$ references published in year $t_{0}-t$, each of which can be picked up by the paper \emph{i} with probability $P(\tau,s,R_{0}(t_{0}-\tau))$.  The  number of such indirect references  published in  year $t_{0}-t$ is the sum of contributions of all preselected references, namely
\begin{equation}
R_{indir}(t_{0},t_{0}-t)=\sum_{\tau=1}^{t}R(t_{0}-\tau,t_{0}-t)\overline{P(\tau,s,R_{0}(t_{0}-\tau))}R(t_{0},t_{0}-\tau)
\label{ref}
\end{equation}
where  averaging is performed over multiplicity $s$. Equations  \ref{ref-sum},\ref{ref} express $R_{indir}$ through empirical functions $R_{dir}$ and $P$ that shall be taken from measurements.

We also consider the reduced age distribution of references,\cite{Geller1981,Stinson1987,Nakamoto1988,Redner2004,Roth2012,Glanzel2004,Bouabid2013,Sen2014} \begin{equation}
r(t)=\frac{R(t_{0},t_{0}-t)}{R_{0}(t_{0})}=r_{dir}(t)+r_{indir}(t)
\label{ref-reduced}
\end{equation}
where $R_{0}(t_{0})$ is the average length of the reference list of the papers published in year $t_{0}$.
 \begin{figure}[ht]
\includegraphics*[width=0.4\textwidth]{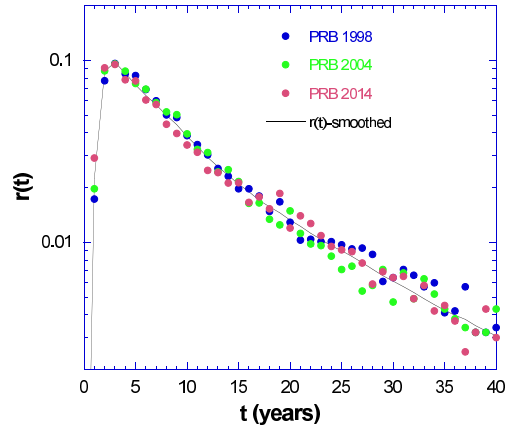}
\caption{Reduced age distribution of references, namely  a fraction of  references published in  year $t_{0}-t$ that appear in the reference list of a paper published in year $t_{0}$.  Red, blue, and green circles stay for three sets of research papers published in July issues of the Physical Review B in 1998, 2004, and 2014, correspondingly.   Similar to previous studies,\cite{Geller1981,Redner2004} we observe that $r(t)$ for all publication years collapse onto a single dependence (besides $t=1$). Continuous line was obtained by averaging and smoothing the data.}
\label{fig:ref}
\end{figure}
Figure \ref{fig:ref} shows that $r(t)$ is almost independent of the publication year $t_{0}$. Having in mind this remarkable observation we transform Eq. \ref{ref} to achieve
\begin{equation}
r_{indir}(t)=\sum_{\tau=1}^{t}r(t-\tau)T(\tau)r(\tau)
\label{ref-reduced-eq}
\end{equation}
Due to the properties of convolution Eq. \ref{ref-reduced-eq} can be  recast as
\begin{equation}
r_{indir}(t)=\sum_{\tau=1}^{t}r(t-\tau)T(t-\tau)r(\tau).
\label{ref-reduced-eq1}
\end{equation}
where $T(\tau)=\overline{P(\tau,s,R_{0}(t_{0}-\tau))}R_{0}(t_{0}-\tau)$. Since $r(t)$ and, obviously, $r_{indir}(t)$  do not depend on $t_{0}$,   Eq. \ref{ref-reduced-eq} yields that $T$ is also independent of $t_{0}$, hence $\overline{P(\tau,s,R_{0}(t_{0}-\tau))}= \frac{\overline{T(\tau,s)}}{R_{0}(t_{0}-\tau)}$.  Our measurements indicate that $T\propto e^{-\gamma\tau}$, $R_{0}(t)\propto e^{\beta t}$, hence, the latter expression can be recast as $\overline{P_{0}}e^{-(\gamma+\beta)\tau}$ where $P_{0}=\frac{T(s)}{R_{0}(t_{0})}$.

To calibrate the model we performed dedicated measurements and chose a small but representative set of papers in one field  which we analyzed  manually.  Namely, we somehow arbitrary chose 21 research papers published in the Physical Review B in  2014 and analyzed their first-generation and second-generation  references using Scopus database. We identified direct references of each parent paper as those appearing only in the first generation, and indirect references as those appearing in both generations of references. We arranged  the unified reference list of these 21 parent papers in chronological order, counted the number of direct and indirect references published in each year, and divided them by the total number of references.  The average reference list includes  $35\%$  direct  and $65\%$ indirect references, the half of the latter appearing in reference lists of several preselected references. This conforms with previous estimates: Refs. \cite{Clough2014,Goldberg2015} report, correspondingly, 67-78$\%$  and  80$\%$  indirect references in the reference lists of high-energy Physics preprints; Ref. \cite{Bramoulle2012} reports  56.4$\%$ indirect references in the American Physical Society publications (only APS to APS references were counted);  Ref. \cite{Chen2007} found 40-50 $\%$ indirect references in the Physical Review publications (only PR to PR references were counted).

\begin{figure}[ht]
\begin{center}
\includegraphics*[width=0.4\textwidth]{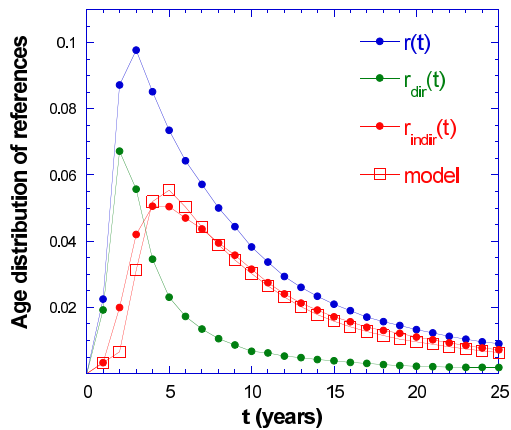}
\caption{Time dependence of  $R_{dir},R_{indir}$, and $R=R_{dir}+R_{indir}$, the number of direct, indirect, and total  references in the reference list of an average Physical Review B paper published in 2014. (SM-B). Open squares show model prediction based on Eq. \ref{ref} with exponential kernel.
}
\label{fig:dir_indir_refs}
\end{center}
\end{figure}

Figure \ref{fig:dir_indir_refs} shows the measured functions $R_{dir}(t),R_{indir}(t)$, and  $R(t)=R_{dir}(t)+R_{indir}(t)$. We observe that $R_{dir}(t)$ achieves its maximum at $t=2$ yr and then slowly decays. We succeeded to fit $R_{indir}(t)$  using Eq. \ref{ref} with $R(t)$ from Fig. \ref{fig:dir_indir_refs} and the exponential kernel $\overline{P(\tau,s,R_{0}(t_{0}-\tau))}=0.34e^{-1.2(\tau-1)}$. Slowly decaying $R_{dir}(t)$ and quickly decaying $P(\tau)$  contrast previous speculation of Ref. \cite{Simkin2007} who assumed that $R_{dir}(t)$ decays fast  while $P(\tau)$ has a long tail.
\section{\label{sec:duality}Reference-citation duality}
\begin{figure}[ht]
\begin{center}
\includegraphics*[width=0.5\textwidth]{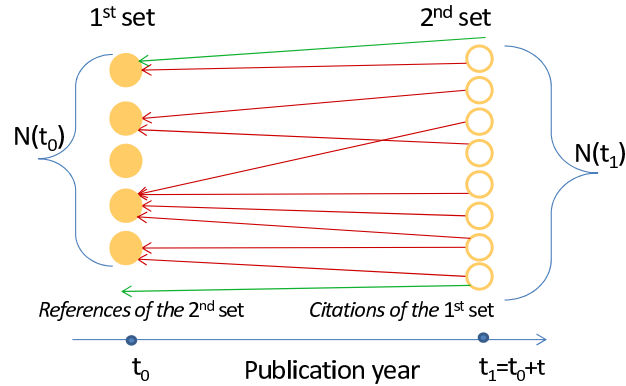}
\caption{Reference-citation duality. The open and filled  circles show all papers in one research field that  were published in  years $t_{0}$ and $t_{1}=t_{0}+t$, correspondingly. Red links between these two sets  are shown by arrows. With respect to the first set, these links are citations, with respect to the second set  they are references. Green lines show interdisciplinary citations and references.
}
\label{fig:duality}
\end{center}
\end{figure}

Our further goal is the extension of this model  to citations. We  consider   all papers in one research field  that were published in one year. These papers represent a directed network. Figure \ref{fig:duality} shows  two sets of papers published in years $t_{0}$ and $t_{0}+t$, correspondingly. The links between the two sets can be considered either as citations or references. Indeed, the reference and citation networks are dual, since one paper's citation is another paper's reference.
To explore mathematical consequences of this duality we introduce $N_{publ}(t_{0})$ and $N_{publ}(t_{0}+t)$- the  number of papers in each set, $M(t_{0},t_{0}+t)$- the mean number of citations garnered in year $t_{0}+t$ by a paper of the first set, and $R(t_{0}+t,t_{0})$-  the average number of references published in year $t_{0}+t$ that appear in the reference list of the papers of the second set.  We assume that all citing papers  belong to the same research field and neglect interdisciplinary papers, books, and other references/citations  which are not research papers. Under this assumption, the  number of  papers that cite the first set and that were  published in  year $t_{0}+t$ is equal to the  number of references  in the reference lists of the papers of the second set published in  year $t_{0}$, namely,
\begin{equation}
N_{publ}(t_{0})M(t_{0},t_{0}+t)=N_{publ}(t_{0}+t)R(t_{0}+t,t_{0})
\label{duality0}
\end{equation}

The annual growth of the number of publications and of the reference list length is nearly exponential,(SM-A)
\begin{equation}
N_{publ}(t_{0}+t)\approx N_{publ}(t_{0})e^{\alpha t}, R_{0}(t_{0}+t)\approx R_{0}(t_{0})e^{\beta t}
\label{exp-growth}
\end{equation}
Equations \ref{ref-reduced}, \ref{exp-growth} yield  $R(t_{0}+t, t_{0})=R(t_{0}, t_{0}-t)e^{\beta t}$. Then Eqs. \ref{duality0}, \ref{exp-growth}  yield
\begin{equation}
M(t_{0},t_{0}+t)=R(t_{0},t_{0}-t)e^{(\alpha+\beta)t}.
\label{duality}
\end{equation}
which is the mathematical expression of the  reference-citation duality relating synchronous (retrospective) and diachronous (prospective) citation distributions.\cite{Geller1981,Stinson1987,Nakamoto1988,Redner2004,Roth2012,Glanzel2004,Bouabid2013,Sen2014}

\begin{figure}[ht]
\begin{center}
\includegraphics*[width=0.35\textwidth]{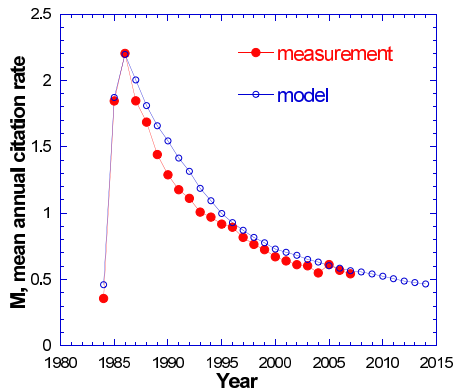}
\caption{$M(t)$, the mean annual number of citations (red circles). The measurements were performed over 40,195 Physics papers published in 1984 (overviews excluded). The blue circles show model prediction based on Eq. \ref{duality} with  $r(t)$ from Fig. \ref{fig:dir_indir_refs}(b), $R_{0}=20.5$, and $\alpha+\beta=0.046$ yr$^{-1}$.
}
\label{fig:mean-ref}
\end{center}
\end{figure}

\begin{figure}[ht]
\begin{center}
\includegraphics*[width=0.4\textwidth]{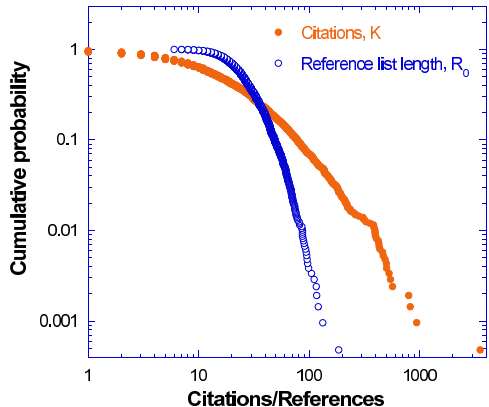}
\caption{Cumulative distribution of the reference list lengths $R_{0}$ (open blue circles) and of the number of citations $K$ (red filled circles) \emph{for the same set of papers} (all 2078 Physical Review B papers published in 1984). Citations were counted in 2014. While both distributions have almost the same mean, $R_{0}$  exhibits a relatively narrow  bell-shaped distribution while citation distribution  is very wide and has a fat tail.
}
\label{fig:cit-ref-dist}
\end{center}
\end{figure}

Our measurements (Fig. \ref{fig:mean-ref}) validate Eq. \ref{duality} and show that $M(t_{0},t_{0}+t)$ and $R(t_{0},t_{0}-t)$ are similar. However, there is a subtle  qualitative difference between these two dependences. Indeed, in the long-time limit $R(t) \propto (t-0.8)^{-1.5}$ in such a way that $\int_{0}^{t}R(\tau)d\tau$ converges as $t\rightarrow\infty$. The function $M(t)$  decays slower than $R(t)$ due to exponential factor $e^{(\alpha+\beta)t}$  (Eq. \ref{duality}). Although the exponent $\alpha+\beta$ is very small, it changes the condition of convergence and for Physics papers the integral $\int_{0}^{t}M(\tau)d\tau$ diverges as $t\rightarrow\infty$.

Basing on Eqs. \ref{ref-reduced-eq1},\ref{duality} we develop equation for $M(t)$.  To this end we substitute $R(t_{0},t_{0}-t)=M(t_{0},t_{0}+t)e^{-(\alpha+\beta)t}$ into Eqs.  \ref{ref-reduced},\ref{ref-reduced-eq1} and after simple algebra we find
\begin{equation}
M(t_{0},t_{0}+t)=M_{dir}(t_{0},t_{0}+t)+\sum_{\tau=1}^{t}M(t_{0}+\tau,t_{0}+t)\overline{P(t-\tau,s,R_{0}(t_{0}+t-\tau))}M(t_{0},t_{0}+\tau)
\label{M}
\end{equation}
where
\begin{equation}
M_{dir}(t_{0},t_{0}+t)=r_{dir}(t)R_{0}(t_{0})e^{(\alpha+\beta)t}
\label{duality2}
\end{equation}
While Eqs. \ref{ref-reduced-eq1},\ref{M} are similar, there is a  profound difference between statistics of citations and references. Figure \ref{fig:cit-ref-dist} shows that the statistical distribution of citations and of the reference list lengths  (in- and out-degrees in network language)  are very different:  citation distribution  is very broad and has a fat tail, while the reference list length distribution is a relatively narrow bell-shaped curve. The WWW exhibits a similar asymmetry between in- and out-degree distributions.\cite{Broder2000} Narrow $R_{0}$ distribution implies that  $R(t)$ represents the age composition of the reference list of an average paper. Broad $M$-distribution indicates that Eq. \ref{M} describes citation dynamics only in the mean-field  approximation;   citation dynamics of  individual papers can be qualitatively different from the mean.
\section{Citation dynamics from the  perspective of cited paper}
\begin{figure}[ht]
\includegraphics*[width=0.5\textwidth]{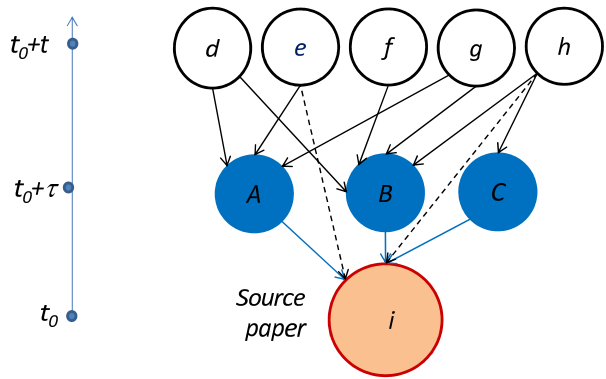}
\caption{A fragment of citation network showing two generations of papers citing the source paper \emph{i}.  The papers \emph{A,B,C} cite \emph{i} and do not cite any other paper citing \emph{i}. These are direct citations and they belong to the first generation of citing papers. The  papers \emph{d-h} cite  first-generation citing papers and they are the second-generation citing papers.  The  papers \emph{e,h}  cite \emph{i} and one of its citing papers. These are indirect references and they belong to both generations of citing papers. The solid and dashed lines link the source paper  with its  direct and indirect  citing papers. Each indirect citation closes a triangle  in which the source paper \emph{i} is a vertex. The paper  \emph{h} cites two first-generation citing papers \emph{B} and \emph{C} while \emph{e} cites only one such paper \emph{A}, therefore  \emph{h} will cite the source paper \emph{i} with higher probability  than \emph{e}.
}
\label{fig:citation-model}
\end{figure}

To  model  citation dynamics of individual papers  we reformulate our scenario (Fig. \ref{fig:cartoon}) in terms of citations.  Figure \ref{fig:citation-model} shows a parent paper \emph{i} published in  year $t_{0}$ and its  citations garnered in subsequent years. The papers \emph{A,B,C} cite the paper \emph{i} after finding it through media, journals, or databases and they represent direct citations. The papers \emph{d-h} cite \emph{A,B,C} and these are second-generation citing papers. Consider  papers  \emph{A} and \emph{e}  whereas \emph{i} is cited by \emph{A} and \emph{A} is cited by \emph{e}. The author of \emph{e} finds \emph{i} in the reference list of \emph{A} and cites it with some  probability $P$. The author of \emph{h} cites the source paper \emph{i} with higher probability since he may pick it up  from two preselected papers \emph{B} and \emph{C} rather than from one.

To quantify this scenario we assume that $k_{i}$, the annual  number of citations garnered by a source paper \emph{i}, is a discrete random variable following a time-inhomogeneous Poisson process \cite{Golosovsky2012} with the probability distribution
\begin{equation}
Poiss(k_{i})=\frac{(\lambda_{i})^{k_{i}}}{k_{i}!}e^{-\lambda_{i}}
\label{Poisson}
\end{equation}
where  Poissonian rate $\lambda_{i}$ is derived from our model.

The model considers all  $k_{i}(\tau)$ first-generation citing papers published in year $t_{0}+\tau$  and the trail of second-generation citing papers published in a later year $t_{0}+t$.   We denote by  $N^{II}_{i}(t_{0}+\tau,t_{0}+t)$ and $M^{II}_{i}(t_{0}+\tau,t_{0}+t)$, correspondingly, the average number of the second-generation citing papers and citations  per one first-generation citing paper published in year $t_{0}+\tau$. While the numbers of the first-generation citations and citing papers  are equal, the number of the second-generation citing papers differs from that of citations  since  one second-generation citing paper can cite several first-generation citing papers. For example, Fig. \ref{fig:citation-model} shows that the paper \emph{i} has  5 second-generation citing papers \emph{d,e,f,g,h} and 8 second-generation citations (black links).  To characterize this multiplicity we introduce a new parameter $s_{i}=\frac{M^{II}_{i}}{N^{II}_{i}}$ which quantifies the average number of paths connecting  a second-generation citing paper  to its parent paper.

Each  second-generation citing paper can cite the source paper \emph{i} (indirectly) with  probability $P(t-\tau,s_{i},R_{0}(t_{0}+t-\tau))$. The corresponding latent  citation rate is
\begin{equation}
\lambda_{i}(t_{0},t_{0}+t)=\lambda_{i}^{dir}(t_{0},t_{0}+t)+\sum_{\tau=1}^{t}N^{II}_{i}(t_{0}+\tau,t_{0}+t)P(t-\tau,s_{i},R_{0}(t_{0}+t-\tau)) k_{i}(t_{0},t_{0}+\tau)
\label{paper-all}
\end{equation}
As expected, there is a close correspondence between Eq. \ref{paper-all} and Eq. \ref{M}. Basing on our measurements of the reference age composition and using Eq. \ref{exp-growth} we find that
\begin{equation}  P(t-\tau,s_{i},R_{0}(t_{0}+t-\tau))=P_{0}e^{-(\gamma-\beta)(t-\tau)}
\label{probability-all}
\end{equation}
where  $P_{0}=\frac{T_{0}(s_{i})}{R_{0}(t_{0})}$.  Our  goal is to quantify Eqs. \ref{paper-all},\ref{probability-all} through dedicated measurements and to find $\lambda_{i}^{dir},N^{II}_{i}$,$P_{0}$.
\section{\label{sec:second-generation}Measurements of citation dynamics of individual papers}
Citation trajectories of individual papers are by no means similar and this is reflected in a very broad citation distribution (Fig. \ref{fig:cit-ref-dist}).  To make meaningful measurements and to minimize scatter we chose to operate with groups of similar papers.  Our measurements were designed basing on the following assumption:  the papers that belong to the same field,  were published in the same year,  \emph{and garnered  the same number of citations in the long-time limit}-  have more or less similar citation dynamics. \subsection{Second-generation citations and citing papers}

\begin{figure}[!ht]
\includegraphics*[width=0.3 \textwidth]{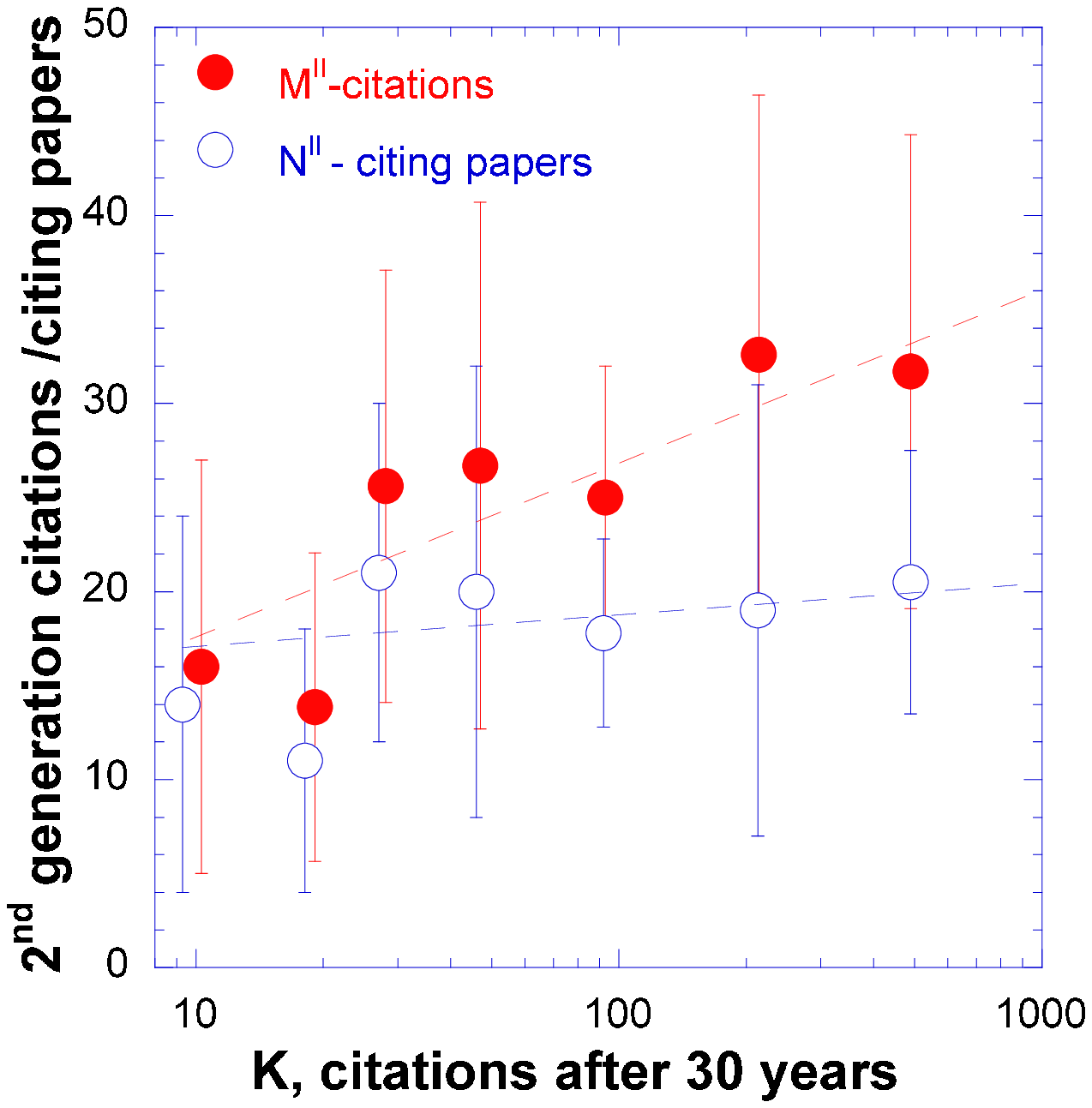}
\includegraphics*[width=0.35\textwidth]{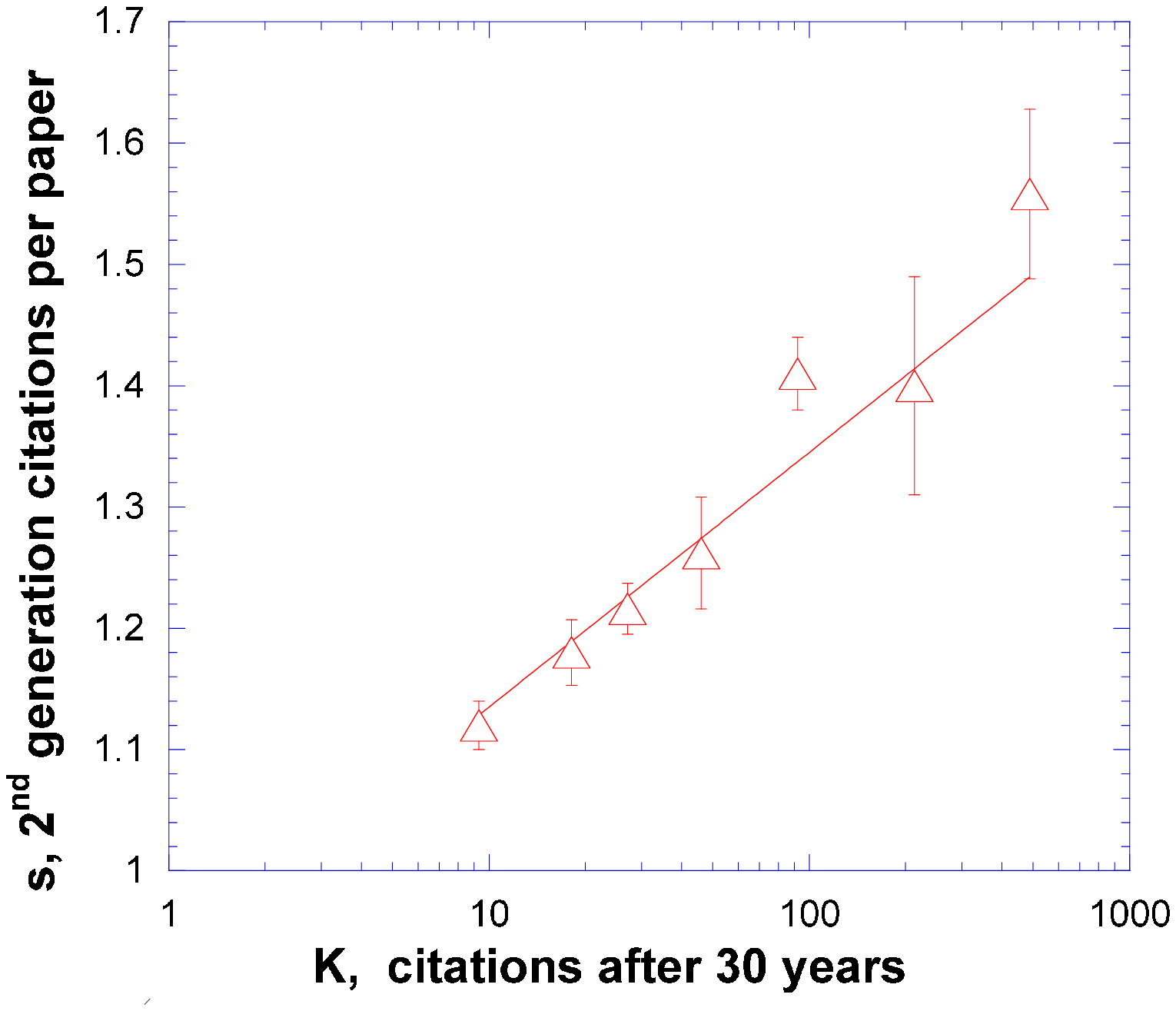}
\begin{center}
\caption{Second-generation citations  and citing papers  for 108 Physical Review B papers published in 1984.  The filled and open circles show average $M^{II}=\overline{M^{II}_{i}}$ and $N^{II}=\overline{N^{II}_{i}}$  for the groups of papers with the same $K_{i}$, the total number of citations  garnered by the parent paper by the end of 2013 i.e., 30 years after publication. $M^{II}=k_{nn}$ increases with $K$, indicating that citation network of Physics papers is assortative while $N^{II}$  is nearly independent of $K$.  The dashed lines are the guides to the eye. The right panel shows $s=<\frac{M^{II}_{i}}{N^{II}_{i}}>$ where the average  is over the group of papers with the same $K$. $s$ quantifies  the average number of paths connecting a second-generation citing paper to its parent paper and it increases logarithmically with $K$. The  line is the guide to the eye.
}
\label{fig:M-N}
\end{center}
\end{figure}

We considered  108  Physics papers published in the Physical Review B in 1984  and arranged them into  several groups, each of which consisting of papers that garnered approximately the same number of  citations $K$  by the end of 2013 i.e., 30 years after publication.  For every parent paper $i$ we counted  second-generation citations and citing papers that were published by the end of 2013,  divided these counts by $K_{i}$, and found $M^{II}_{i}$ and $N^{II}_{i}$. Then we calculated $M^{II}=\overline{M^{II}_{i}}$ and $N^{II}=\overline{N^{II}_{i}}$, the average over each group of papers with the same $K$. Figure \ref{fig:M-N} shows that $N^{II}$ is nearly independent of, while $M^{II}$ slowly increases with $K$.  In the language of network science $M^{II}=k_{nn}$, the average nearest-neighbor connectivity. Increasing $M^{II}(K)$ dependence implies  that highly-cited parent papers have highly-cited descendants, i.e. citation network is assortative, as it was already observed by Ref. \cite{Barabasi2015} for the network of PR to PR citations. (It should be noted that  we excluded  overviews that are hubs of citation network and strongly affect degree assortativity).

It is important to note that $M^{II}_{i}$ and $N^{II}_{i}$  for the same paper  are correlated   and large $N^{II}_{i}$ usually means large $M^{II}_{i}$. Indeed, while there is a large scatter in $M^{II}_{i}$ and $N^{II}_{i}$ numbers within each group, Fig. \ref{fig:M-N}b shows that the scatter of their ratio $s_{i}=\frac{M^{II}_{i}}{N^{II}_{i}}$ is much smaller. We introduce $s(K)=\overline{s_{i}}$, the mean $s$  over the group of papers with the same $K$. Figure \ref{fig:M-N}b shows that $s$ grows logarithmically with $K$ from $s\sim 1$ for low-cited papers to $s=1.5-1.7$ for highly-cited papers.  In other words, the low-cited  source papers are connected to their second-generation descendant mostly by single paths, while the highly-cited source papers are connected to each of their second-generation descendants by multiple paths.  The difference between the neighborhoods of the low-cited and highly-cited papers may arise from the saturation effect: the  descendants of low-cited papers constitute only a small fraction of all papers in their research field,  while the descendants of  highly-cited papers constitute a considerable fraction of it. (SM-E).

\subsection{Probability of indirect citation}
\begin{figure*}[!ht]
\includegraphics*[width=0.8\textwidth]{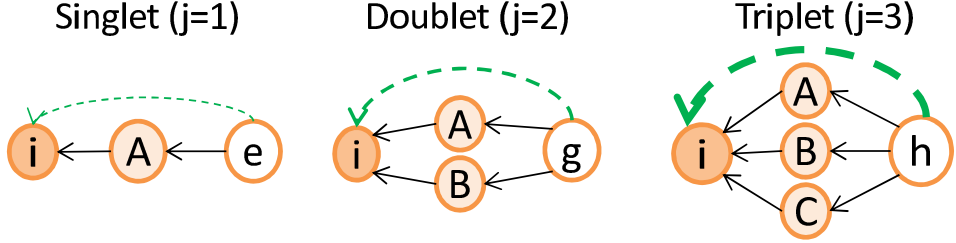}
\caption{Network motifs. The circles show papers, continuous lines show direct citations, dashed lines show indirect citations. $i$- parent paper, $A,B,C$ - first-generation citing papers, $e,g,h$-  second generation citing papers. Table \ref{tab:table1} indicates that the probability of papers $e,g,h$ to cite the parent paper $i$ (indirectly) increases nonlinearly with the multiplicity $j$.}
\label{fig:motifs}
\end{figure*}

Our next goal is to find out how the probability of indirect citation depends on $s$, the number of paths connecting the second-generation citing paper to the source paper.  To this end we chose three representative  Physical Review B  papers that were published in 1984 and  gained  100 citations by the end of 2013. We studied two generations of their citing papers while limiting ourselves only to descendants of the direct citations and disregarding  indirect citations  bringing another indirect citation.  For each  parent paper we pinpointed direct citations  (first generation) and the  papers that cite them (second generation).  We built two-generation citation map, and identified the network motifs consisting of $j$-multiplets such as singlet ($j=1$), doublet ($j=2$), triplet ($j=3$) etc. (Fig. \ref{fig:motifs}).  The number of direct citations of the parent paper is $K_{dir}$, the number of the   second-generation citing papers is $K^{II}=N^{II}K_{dir}$, and the number of the latter associated with $j$-multiplets is  $f_{j}K^{II}$, in such a way that $\sum_{j}f_{j}=1$. Among the second-generation citing papers associated with  $j$-multiplets we counted all those that cite the parent paper. These are indirect citations. The number of the latter is $\pi_{j}f_{j}K^{II}$ where  $\pi_{j}$ is the probability of indirect citation of the parent paper by a second-generation paper which is already connected to it by $j$ paths. (Since $\pi_{j}$ are counted for different multiplets,  $\sum_{j}\pi_{j}\neq 1$).
\begin{table*}
\caption{\label{tab:table1}
Second-generation citing papers: multiplets and their contribution to indirect citations}
\begin{ruledtabular}
\begin{tabular}{lrlr}
\multicolumn{1}{c}{\textrm{Multiplet}}&\textrm{$f_{j}$, fraction among second-}
&\textrm{$\pi_{j}$, probability}&\textrm{$\pi_{j}f_{j}$, contribution}\\
\multicolumn{1}{c}{\textrm{}}&\textrm{generation citations}
&\textrm{of indirect citation}&\textrm{to indirect citations}\\
\colrule
singlet (j=1) & 88$\%$ & 0.054 & 56$\%$ \\
doublet (j=2)  & 9$\%$  & 0.28 & 30$\%$\\
triplet (j=3)  & 2$\%$ & 0.57 & 13$\%$\\
\end{tabular}
\end{ruledtabular}
\label{table:1}
\end{table*}

Table \ref{tab:table1} lists $f_{j}$ and $\pi_{j}$. As expected, $f_{j}$ decreases and $\pi_{j}$ increases with $j$.  The growth of $\pi_{j}$ with $j$ is nonlinear and this is nontrivial. Indeed, if each second-generation citation were having the same probability of inducing indirect citation of the parent paper,  the latter should increase linearly with the number of paths connecting the citing paper to its ancestor, namely,  $\pi_{j}\propto j$. Table \ref{tab:table1}  indicates that  $\pi_{j}$ rather follows quadratic dependence,  $\pi_{j}\propto j^{2}$, suggestive of  multipath interference. This also means that the contribution of higher multiplets (doublets, triplets, etc.) to the total number of indirect citations is disproportionately high. Indeed, Table \ref{table:1} shows that while higher multiplets constitute only 12$\%$ of the second-generation citations, they contribute  44$\%$ of indirect citations.

The number of higher-order multiplets is closely related to the parameter $s$ (Fig. \ref{fig:M-N}). Indeed,  $s=\sum_{j}jf_{j}$ where $j$ is the multiplicity  and $f_{j}$ is the fraction of $j$-multiplets.  If we limit ourselves only to singlets and doublets, then $f_{1}+f_{2}=1$ and $s=1+f_{2}$. We consider now the probability of indirect citation $P_{0}$ (Eq. \ref{probability-all}). Obviously, $P_{0}\propto\sum_{j}\pi_{j}f_{j}=\pi_{1}f_{1}+\pi_{2}f_{2}$. Table \ref{tab:table1} shows that  $\pi_{2}\approx4\pi_{1}$, hence
\begin{equation}
P_{0}\propto\pi_{1}[1+3(s-1)],
\label{3s}
\end{equation}
 while in the absence of multipath interference we would have $P_{0}\propto\pi_{1}s$.

\subsection{\label{sec:direct-indirect}Dynamics of direct and indirect citations}
We focused on one research field- Physics, one research journal- Physical Review B (SM-B), and one publication year - 1984.  We performed our analysis manually and selected small but representative groups of original  research papers  that garnered the same number of citations $K$ by the year 2013 (14 papers with 10 citations, 10 papers with 20 citations, 10 papers with 30 citations, and 3 papers with 100 citations, the mix of theoretical and experimental papers). We measured  citation dynamics of these 37 papers using the Thomson-Reuters  Web of Science database.  For each  parent paper we considered the first- and the second-generation citing papers (overviews excluded, self-citations included) and identified direct citations as those appearing only in the first generation  and indirect citations  as those appearing  in both generations of citations (Fig. \ref{fig:citation-model}).
\begin{figure}[!ht]
\includegraphics*[width=0.3\textwidth]{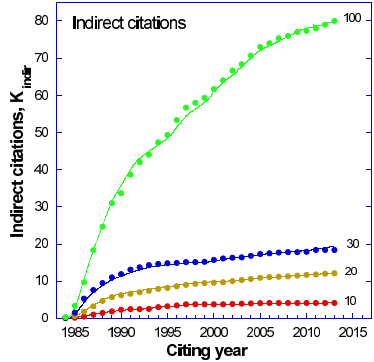}
\caption{Indirect citations  of the Physical Review B papers published in 1984.  Each set of points represents  cumulative  citations  averaged over the groups  of  papers that garnered  the same number of citations  $K$ by the end of 2013 ($K=$10, 20, 30, and 100).  Continuous lines are fits to the second addend in Eq. \ref{paper-all}  with $P_{0}(K)$ as a fitting parameter for each group and $N_{i}^{II}(t-\tau)$   from  Fig. \ref{fig:mean-ref}.
}
\label{fig:indirect}
\end{figure}

\emph{Indirect citations.}
Figure \ref{fig:indirect}  shows  time dependence of the cumulative number of indirect citations for  four groups of papers. These  are well accounted for by Eq. \ref{paper-all} with necessary ramifications. Namely, we replaced there $N^{II}_{i}(t_{0}+\tau,t_{0}+t)$ by $M(t-\tau)/\overline{s}$  where $M(t-\tau)$ dependence was taken from  Fig. \ref{fig:mean-ref} and  $\overline{s}=1.2$ is the average over all Physics papers published in 1984. We assumed that the probability of indirect citation is given by Eq. \ref{probability-all} where only the factor $P_{0}(s_{i})$ depends on paper's individuality. We substituted into Eq. \ref{probability-all}  $\gamma=1.2$ yr$^{-1}$ and $\beta=0.02$yr$^{-1}$ as found in our study of references (Fig. \ref{fig:ref}), and considered $P_{0}$ as the fitting parameter for each group. Figure \ref{fig:indirect} demonstrates that this approach reproduces time dependence of indirect citations fairly well.

\begin{figure}[!ht]
\includegraphics*[width=0.3\textwidth]{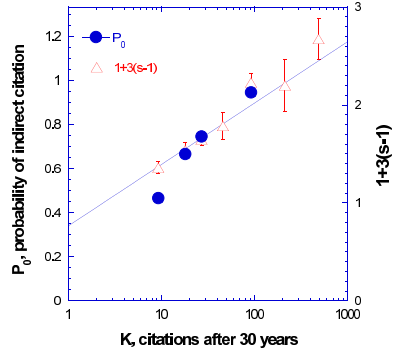}
\caption{$P_{0}$, the probability of indirect citation of the source paper by a second-generation citing paper versus $K$, the number of citations garnered by the source paper after 30 years.  We also plot $1+3(s-1)$ where the values of $s$ were taken from Fig. \ref{fig:M-N}. The straight line shows logarithmic dependence given by Eq. \ref{probability}.
}
\label{fig:ratio-probability}
\end{figure}

Figure \ref{fig:ratio-probability} shows the empirically-found $P_{0}(K)$ dependence.  Since our model  suggests that $P_{0}$ depends only on $s$ and its dependence on $K$  stems from the $s(K)$ dependence, we plot on Fig. \ref{fig:ratio-probability} the expression captured by Eq. \ref{3s} with $s$-values taken from  Fig. \ref{fig:M-N}b. After proper scaling of the vertical axes the both sets of data overlap, as expected from the model. These considerations yield empirical dependence $P_{0}(s)=0.44(1+3(s-1))$. Since $s$ depends on $K$, this is equivalent to
\begin{equation}
P_{0}(K)=0.34(1+0.82\log K)
\label{probability}
\end{equation}
Two factors contribute to this $P_{0}(K)$-dependence: assortativity of the citation network (increasing $s(K)$- dependence) and interference between transitive triples (Table \ref{tab:table1}).

\begin{figure}[!ht]
\includegraphics*[width=0.3\textwidth]{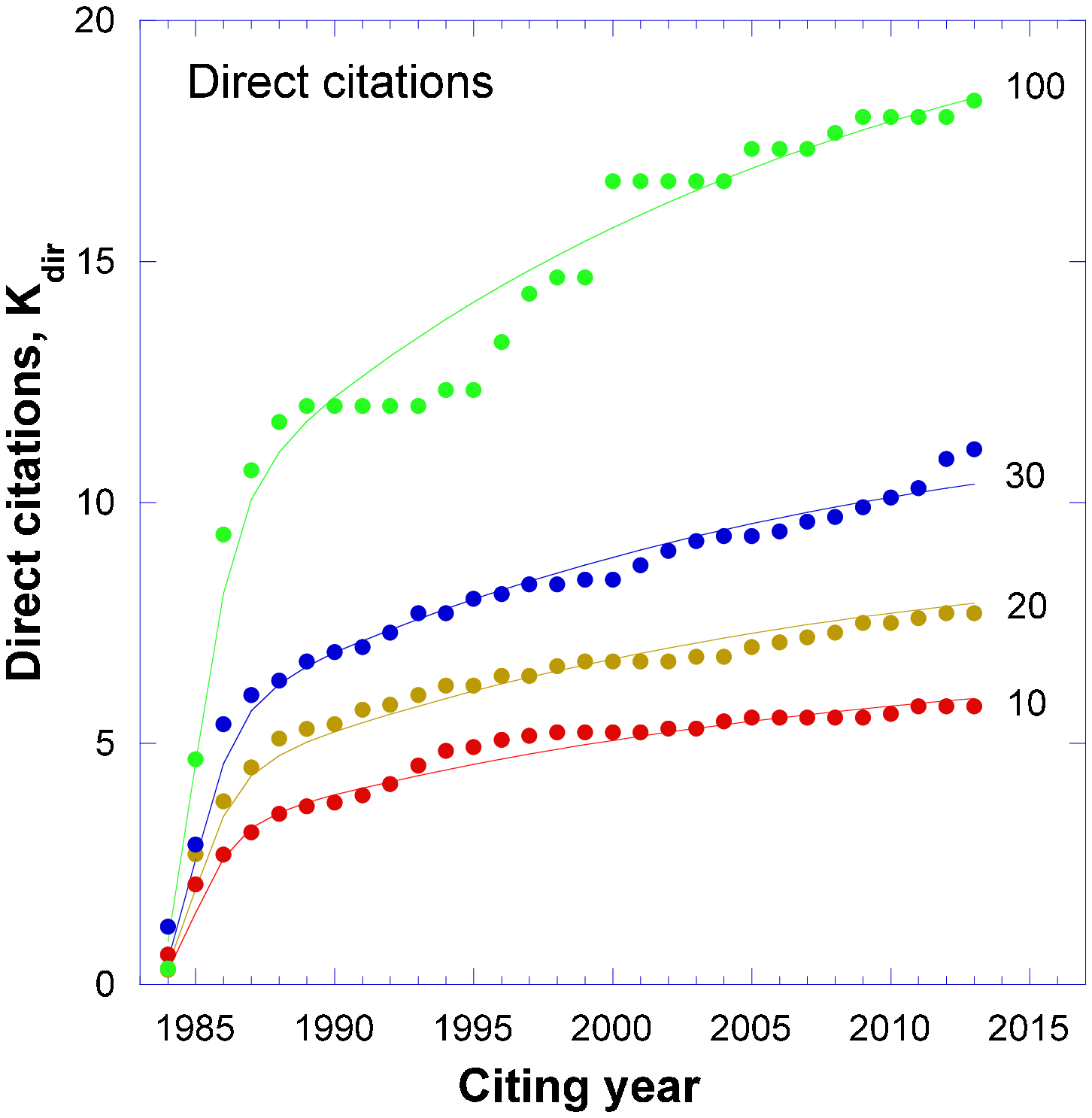}
\includegraphics*[width=0.32\textwidth]{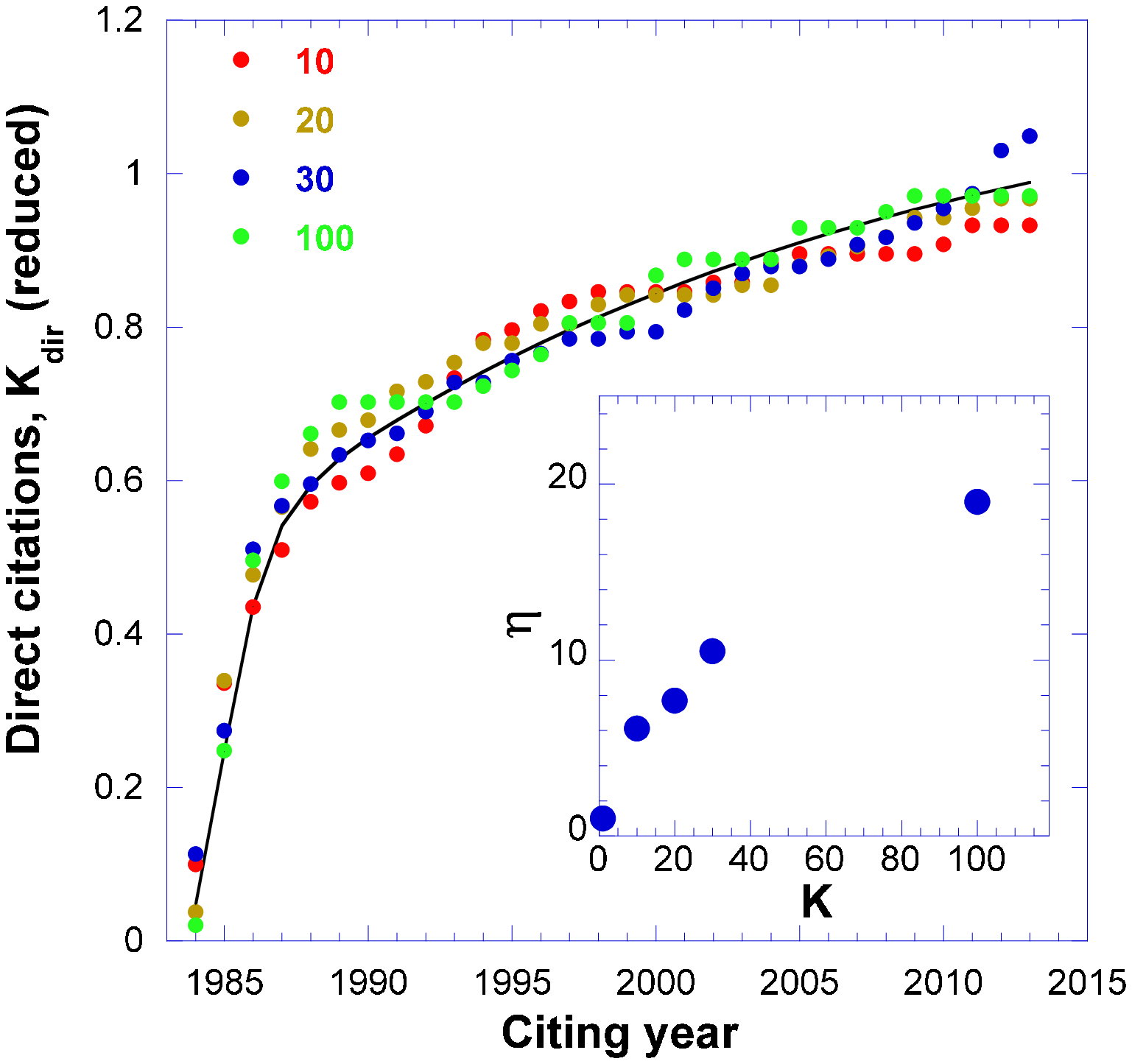}
\includegraphics*[width=0.31\textwidth]{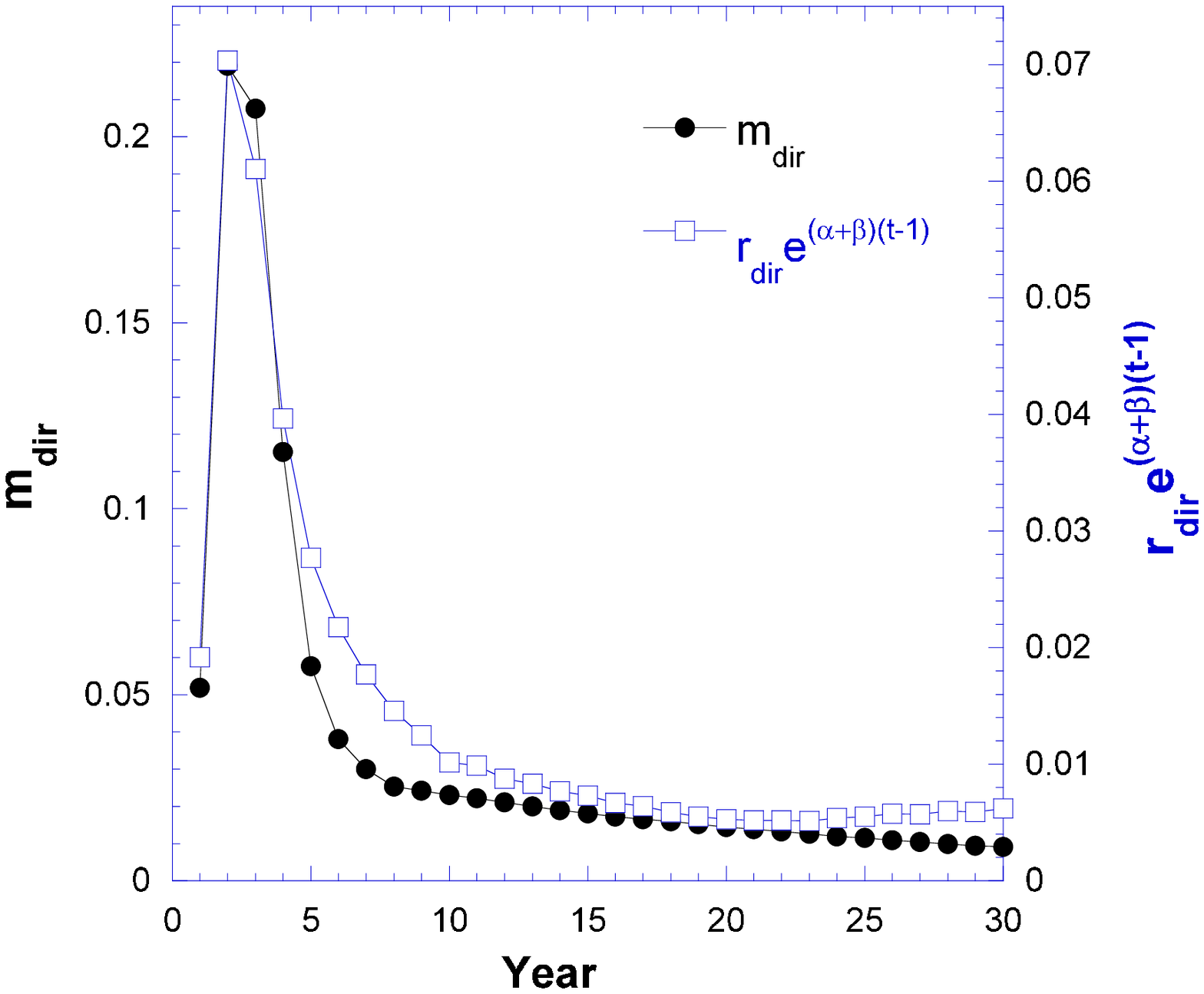}
\caption{(a) Direct citations  of the Physical Review B papers shown in Fig. \ref{fig:indirect}. Continuous lines show $\eta(K)\sum_{\tau=1}^{t}m(\tau)$ dependences  where $m(t)$ is the same function for all groups and $\eta(K)$, the fitness,  is the fitting parameter. (b) Scaled data of the Fig. \ref{fig:direct}a. Continuous black line shows $\sum_{\tau=1}^{t}m(\tau)$ dependence which was obtained by averaging and smoothing the scaled data. The inset shows $\eta(K)$. (c) Filled circles show  $m(t)$ found by differentiation of the continuous black curve in (b). Open squares  show prediction of Eq. \ref{duality2} with $r_{dir}(t)$  from Fig. \ref{fig:dir_indir_refs}(b)  and $\alpha+\beta=0.046$ yr$^{-1}$.
}
\label{fig:direct}
\end{figure}

\emph{Direct citations.}
Figure \ref{fig:direct} shows time dependence  of $K_{dir}(t)$, the  number of the cumulative direct citations for the groups of papers shown in Fig. \ref{fig:indirect}.  These dependences  collapse onto a single curve and can be represented as $K_{dir}(t)=\eta(K)\sum_{\tau=1}^{t}m(\tau)$, in such a way that
\begin{equation}
k_{dir}(t)=\eta(K)m(t)
\label{dir-measured}
\end{equation}
where  $\eta(K)$ is a characteristic number for each group which we name  fitness, and $m(t)$ is the same function for all groups. Since $K_{dir}$ does not come to saturation even after 30 years, in order to uniquely define $m(t)$ we adopted  constraint: $\sum_{\tau=1}^{t=30}m(\tau)=1$. Under this constraint  $\eta$ is the number of direct citations after 30 years.

The averaging of Eq. \ref{paper-all} over all papers shall give Eq. \ref{M}. In view of Eq. \ref{duality2} this yields $\overline{\eta_{i}}m(t)=r_{dir}(t)R_{0}(t_{0})e^{(\alpha+\beta)t}$.  Figure \ref{fig:direct}c shows that $m(t)$ agrees well with this expression.
\section{\label{sec:final}Stochastic model of citation dynamics and comparison to measurements}
We introduce Eqs. \ref{probability}, \ref{dir-measured} into Eq. \ref{paper-all}. The kernel becomes now $P_{0}(K)F(t-\tau)$ where $P_{0}(K)$ absorbs all $K$-dependent factors  and  $F(t-\tau)=N^{II}(t-\tau)e^{-(\gamma-\beta)(t-\tau)}$  absorbs all time-dependent factors (note, that as Fig. \ref{fig:M-N} shows,  $N^{II}$ is almost independent of $K$).  Finally, we make a crucial assumption  that $K$ in Eq. \ref{probability} is the \emph{current} number of citations of the parent paper, namely $K=K(t)$. Thus we obtain our key result- a nonlinear stochastic dynamic equation for the latent citation rate of a paper \emph{i}
\begin{equation}
\lambda_{i}(t)= \eta_{i}m_{dir}(t)+\sum_{\tau=1}^{t}P_{0}(K_{i})F(t-\tau)k_{i}(\tau).
\label{paper-new}
\end{equation}
Here, $m_{dir}(t)$ is the time-dependent direct citation rate,  $P_{0}(K_{i})$ is the probability of indirect citation of a source paper by a second-generation citing paper, $F(t)$ is the average  number of the second-generation citing papers per one first-generation citing paper, $k_{i}$ is the time-dependent annual citation rate, $K_{i}$ is the cumulative number of citations at time $t$, and $\eta_{i}$ is an empirical number characterizing each paper (fitness).  The probability distribution of additional citations is given by the Poisson distribution, $Poiss(k_{i})=\frac{(\lambda_{i})^{k_{i}}}{(k_{i})!}e^{-\lambda_{i}}$. Equation \ref{paper-new} relates  $\lambda_{i}(t)$, the latent citation rate of the paper $i$ at time $t$,   to  fitness $\eta_{i}$, recent citation rate $k_{i}$, and the number of cumulative citations $K_{i}$. It depends on $K_{i}(t)$ and $ k_{i}(t)$ at all previous times, hence it describes a non-Markovian process with memory.\cite{Rosvall2014}

The  functions $m_{dir}(t)$, $P_{0}(K)$, and $F(t)$ are the same for all papers published in one year in one field and they are taken from measurements.  In particular,  $m_{dir}(t)$  is taken from Fig. \ref{fig:direct}c,  $P_{0}(K)$ is taken from Eq. \ref{probability}. $F(t)$, as found in our measurements  for Physics papers published in 1984,  is $F(t=1)=0.089, F(2)=0.138, F(3)=0.046, F(4)=0.012, F(5)=0.0035..$. Initially, $F$ grows with time as the paper receives more recognition (there is approximately one year delay between the publication of the paper and its first citation) and then  decays exponentially.  This obsolescence is strong, hence the memory of the citation process is restricted to a few years.

Equation \ref{paper-new} describes a self-exciting Hawkes process. Similar equations appear   in the renewal theory, in the context of Bellman-Harris branching (cascade) processes \cite{Harris2002}, in population dynamics (the age-dependent birth-death process with immigration  \cite{Ebeling1986}), dynamics of viewing behaviour of YouTube users,\cite{Crane2008}  social networking sites (resharing),\cite{Cheng2014} and viral information spreading.\cite{Iribarren2011} The novel feature   is  a nonlinear kernel  $P_{0}(K)$.

\begin{figure}[!ht]
\begin{center}
\includegraphics*[width=0.4\textwidth]{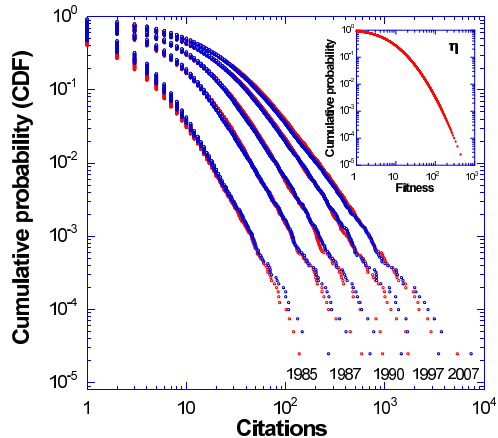}
\caption{ Annual cumulative citation distributions for  40,195 Physics papers published in 1984.  Red symbols show  measured data, blue symbols show results of stochastic simulation based on the Poisson process with the rate given by Eq. \ref{paper-new} and initial conditions set by $\eta_{i}$. The inset shows lognormal $\eta_{i}$- distribution with $\mu=1.62$ and $\sigma=1.1$.
}
\label{fig:simulation}
\end{center}
\end{figure}

While early models of  complex networks growth were validated  by comparing measured and simulated aggregate characteristics, such as  degree distribution, Eq. \ref{paper-new} is  the next-generation model which is much more detailed and the comparison to real data is more demanding.  To the best of our knowledge, the methodology of comparing stochastic model/simulation to stochastic data is not well-established. Following Ref. \cite{Medo2014} we believe that the proper validation of a stochastic model shall include multidimensional analysis. In what follows we verify our model in several dimensions:
\begin{itemize}
        \item Cumulative citation distribution
        \item Citation trajectories of individual papers
        \item Stochastic component of the citation dynamics
        \item Autocorrelation of citation trajectories
        \item The number of uncited papers
 \end{itemize}
Namely, we measured the above aspects of citation dynamics of a large ensemble of papers and compared them to model prediction. The results are shown in the supplementary material (SM-C) and we demonstrate here  only  citation distributions. Figure \ref{fig:simulation} shows cumulative citation distributions for  40,195 papers published in 83 leading Physics journals in 1984 (overviews excluded, self-citations included) measured using Thomson-Reuters  Web of Science database.
Our goal is to simulate these distributions using  Eq. \ref{paper-new}. To this end we need to know  $\eta_{i}$- the number of direct citations in the long-time limit. We sidestepped the difficulty of measuring $\eta_{i}$ for each paper and assumed  lognormal distribution, $\frac{1}{\sqrt{2\pi}\sigma \eta_{i}}\exp{-\left(\frac{(\ln{\eta_{i}}-\mu)^{2}}{2\sigma^{2}}\right )}$, where $\mu$ and $\sigma$ are the fitting parameters.  We run dynamic simulation for  40,195 papers using Eq. \ref{paper-new} with  $\gamma-\beta=1.2$ yr$^{-1}$, as found in our measurements of indirect references and citations; $m_{dir}(t)$ from the Fig. \ref{fig:direct}c; $F(t)$ from our measurements of the mean citation dynamics (Fig. \ref{fig:mean-ref}) and Sec. \ref{sec:final}, and $P_{0}(K)$  from Eq. \ref{probability}.  Figure \ref{fig:simulation} shows excellent  agreement between the  measured  and simulated citation distributions. In fact, we were able to achieve almost the same agreement by using a simplified numerical scheme in which  Eq. \ref{paper-new} is considered as an autoregressive model.  We looked for the minimal model that can faithfully represent our measurements and found that the first-order  model is inadequate while the second-order autoregressive model
\begin{equation}
\lambda_{i}(t)=\eta_{i}m_{dir}(t)+[1+0.82\log K_{i}(t)][0.09k_{i}(t)+0.19k_{i}(t-1)]
\label{paper-new1}
\end{equation}
is quite satisfactory. Here $t=1$ corresponds to the publication year.

The best fitting parameters for the fitness distribution are  $\mu=1.62$ and $\sigma=1.1$. The test for their validity  comes from inspection of  citations garnered during  1-3 years after  publication. These citations are mostly direct, hence a fair correspondence between the measured and simulated earliest citation distributions validates our  $\eta_{i}$- distribution.

\begin{figure}[!ht]
\includegraphics*[width=0.33\textwidth]{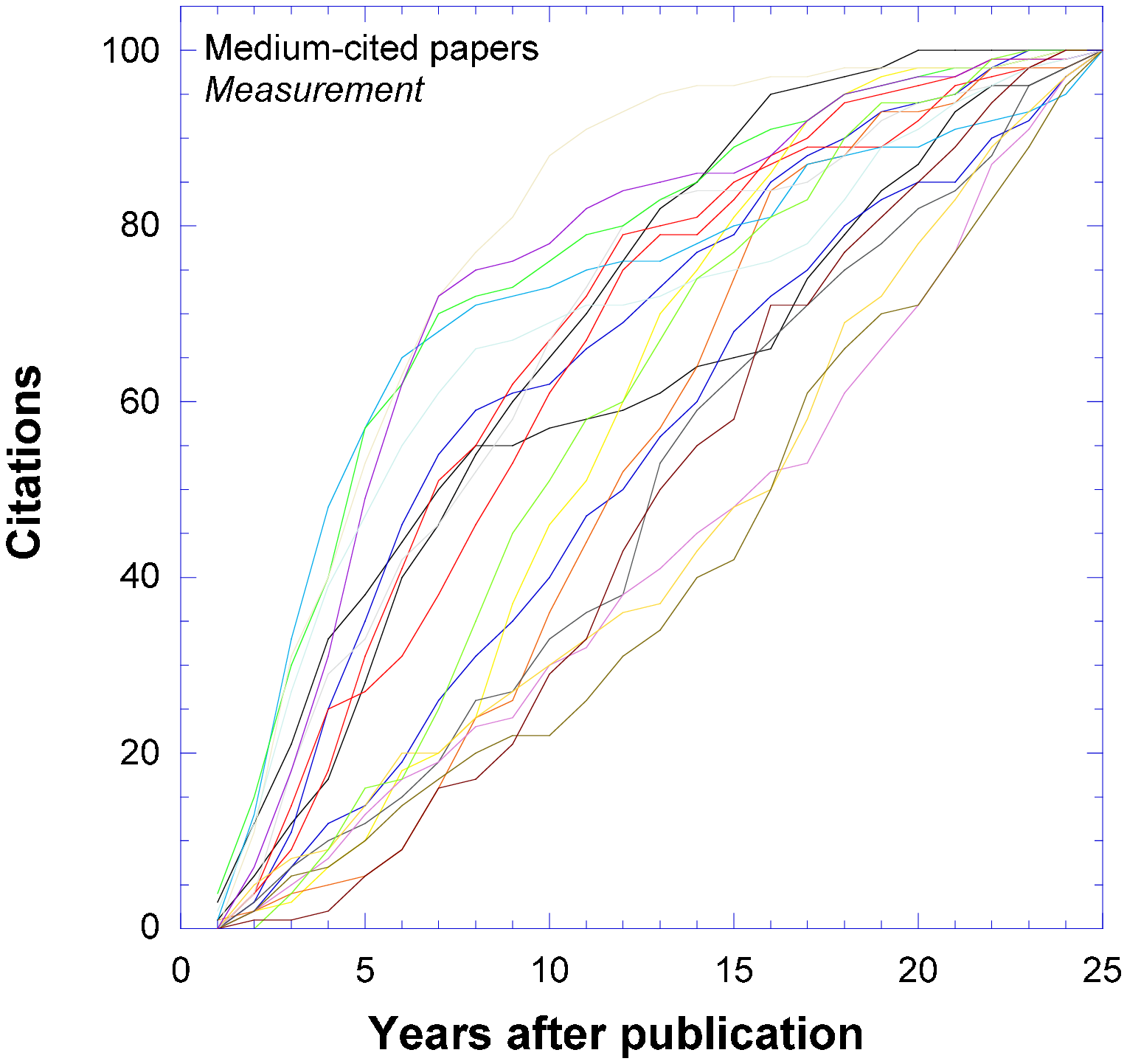}
\includegraphics*[width=0.35\textwidth]{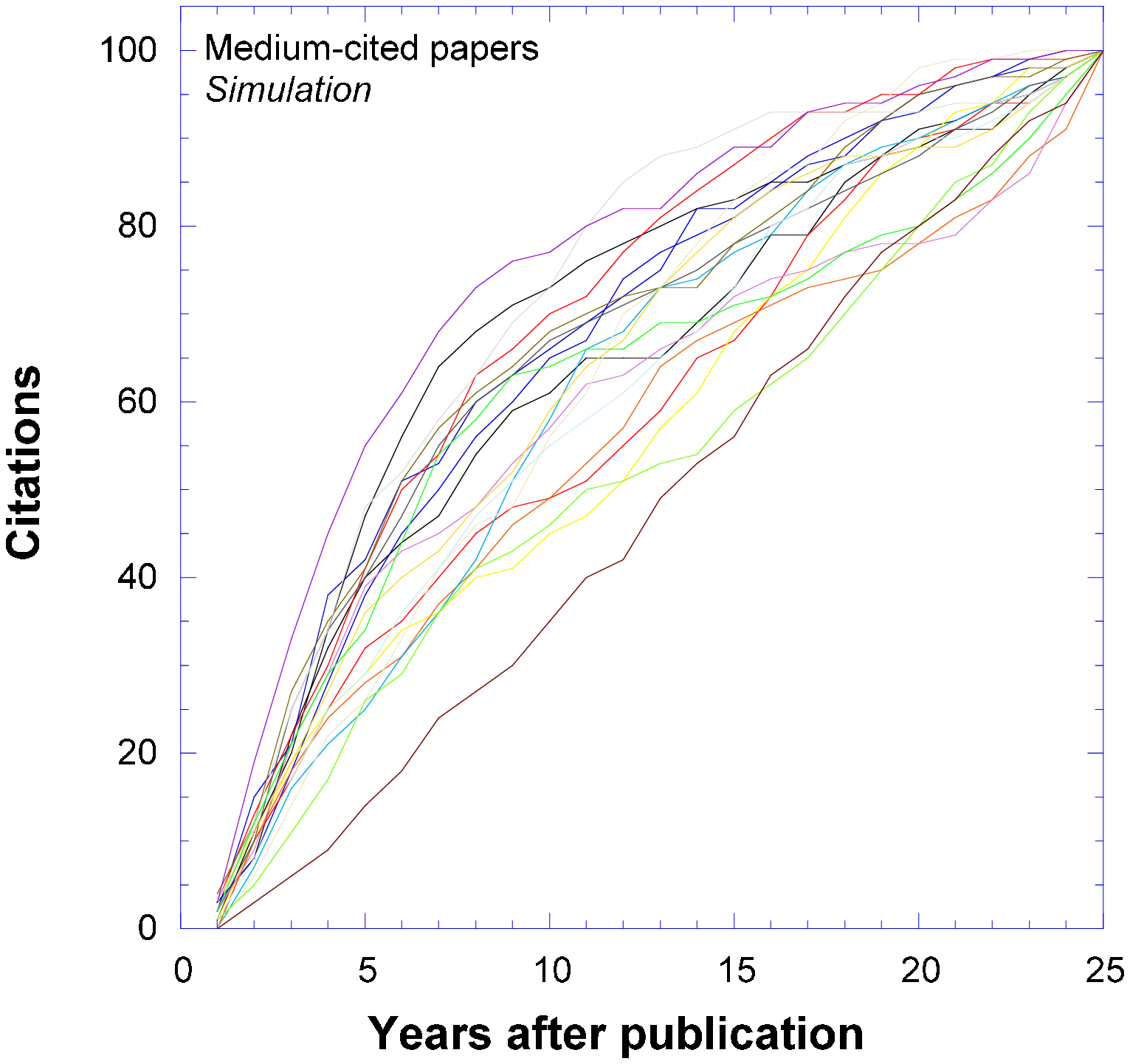}
\caption{Citation dynamics of the Physics papers that were published in 1984  and accrued 99 citations in subsequent 25 years.  Stochastic numerical simulation based on our model  correctly predicts the shape of citation trajectories.}
\label{fig:trajectories1}
\end{figure}

\begin{figure}[!ht]
\includegraphics*[width=0.33\textwidth]{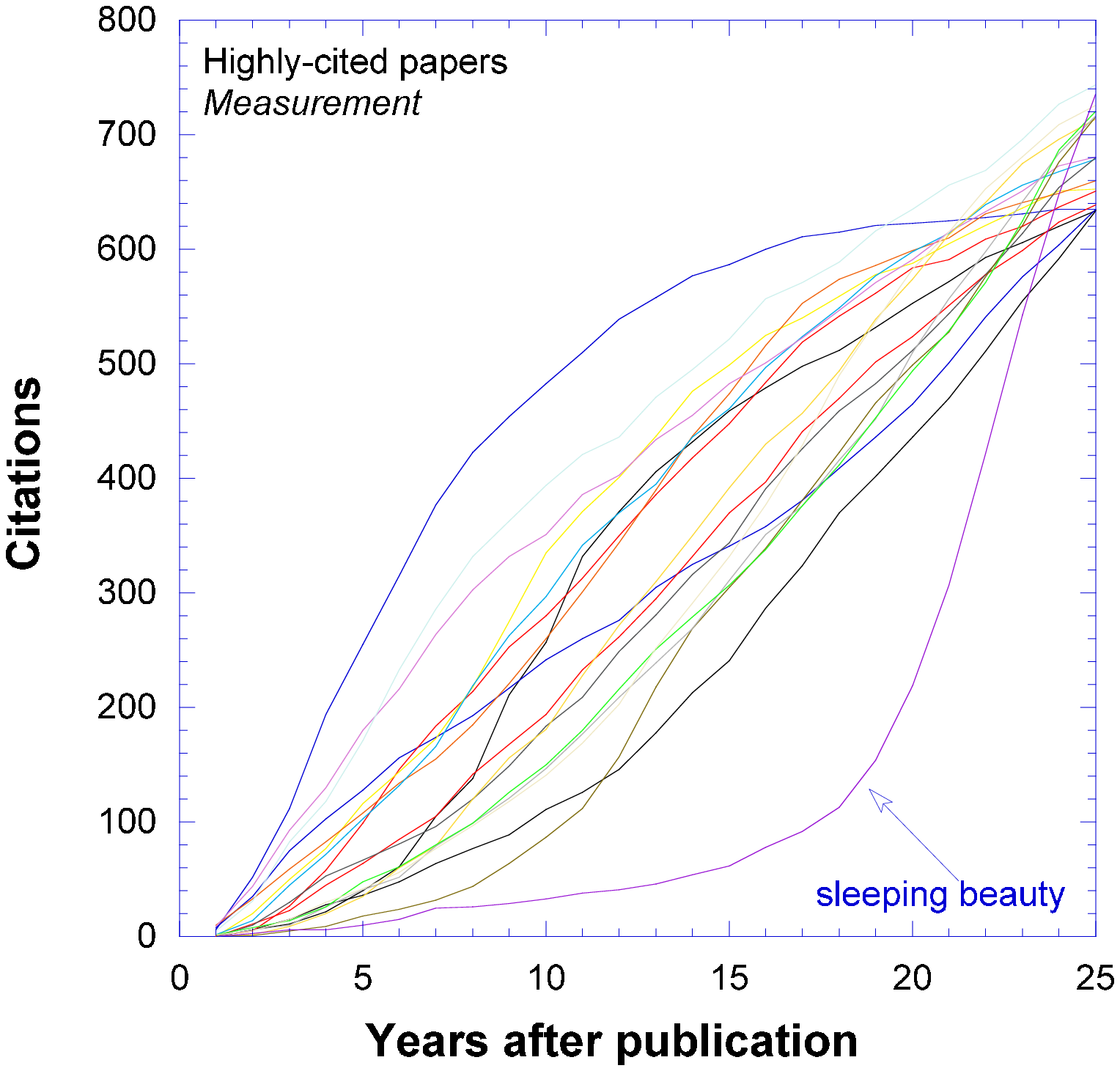}
\includegraphics*[width=0.35\textwidth]{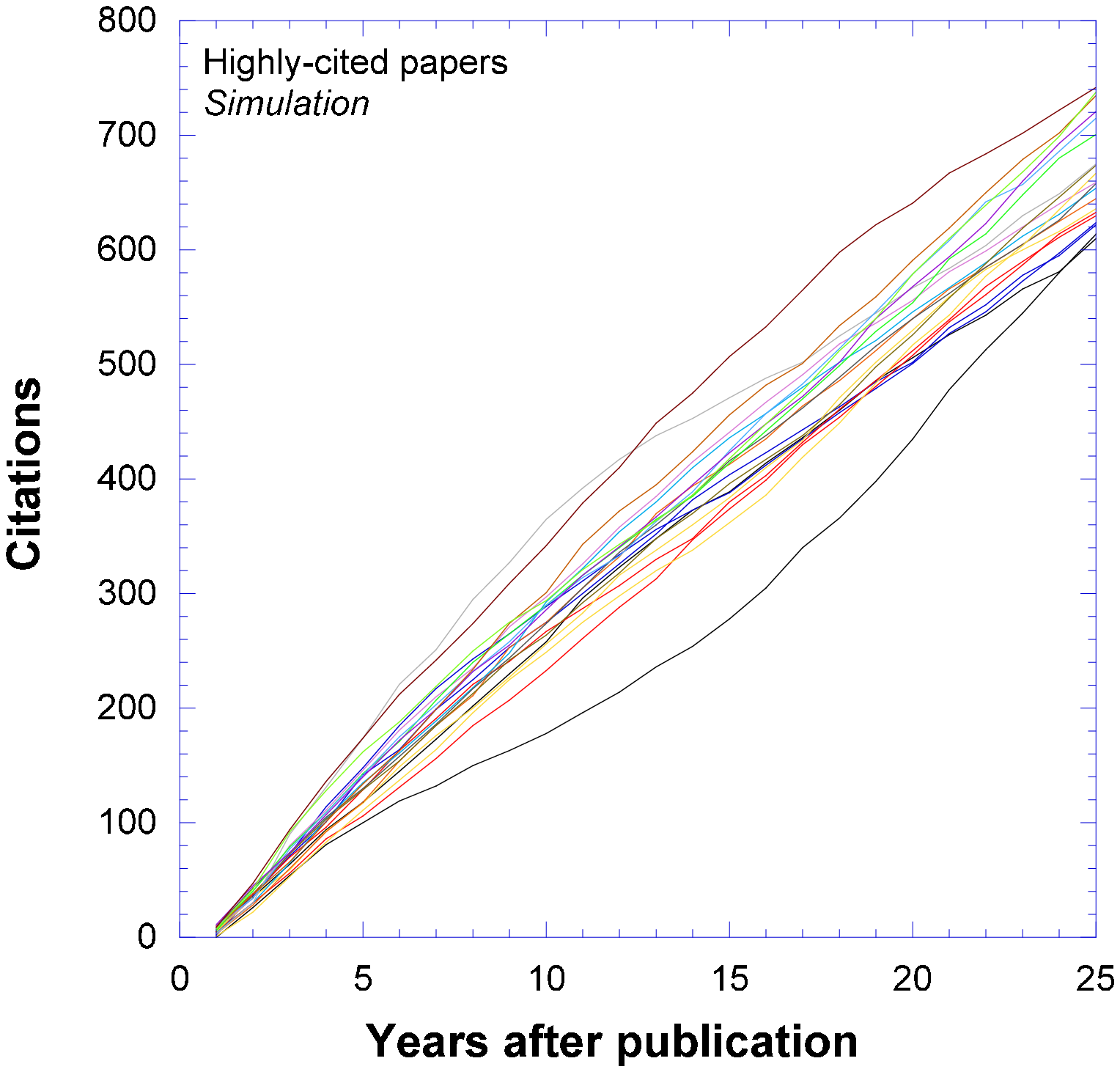}
\caption{Citation dynamics of the Physics papers that were published in 1984  and accrued 600-750 citations in subsequent 25 years.  The spread of measured trajectories exceeds the spread of simulated trajectories. The blue arrow indicates a sleeping beauty- the paper with delayed recognition. Our model does not predict sleeping beauties.}
\label{fig:trajectories2}
\end{figure}

At the next step we compared the measured and simulated citation trajectories of the Physics papers that were published in 1984.  For moderately-cited papers (Fig. \ref{fig:trajectories1}a) the measured and simulated  trajectories look  similar- they are jerky and the fluctuations are of the same size, the spread in trajectories is also the same. For highly-cited papers (Fig. \ref{fig:trajectories2}b) both sets of trajectories are smooth, but the spread of the measured trajectories exceeds that of the simulated ones.


In summary, we found that Eq. \ref{paper-new} with lognormal fitness distribution reproduces citation dynamics of a large ensemble of Physics papers fairly well. This includes  aggregate characteristics (citation distributions) and microscopic dynamics (the number of uncited papers, the mean and the fluctuating parts of citation trajectories of individual papers, citation lifetime, etc.- see SM-C). While our model correctly reproduces citation trajectories of the low- and moderately-cited papers, it underestimates the variability of citation trajectories of the highly-cited papers.
\section{\label{sec:analysis}Continuous approximation of the model}
To  better understand Eq. \ref{paper-new} we analyze its continuous approximation, namely, we  disregard stochasticity and replace the latent citation rate $\lambda_{i}$  by the actual citation rate, $k_{i}$, which we consider  as a continuous variable. The time is also continuous, hence we replace the sum by the integral. For a moment we forget that $P_{0}(K)$ is time-dependent (through $K(t)$) and replace the kernel $P_{0}(K)F(t-\tau)$ by the exponent $qe^{-\gamma(t-\tau)}$ where all time dependences are absorbed in  $\gamma$ and  all prefactors are absorbed in $q$. (We use this approximation here with purely pedagogical purposes, it can not be used for quantitative estimates since it does not account for the time delay between the publication of the parent paper and its citations.) The continuous approximation of  Eq. \ref{paper-new} is thus
\begin{equation}
k_{i}(t)=\eta_{i} m_{dir}(t)+q_{i}\int_{0}^{t}k_{i}e^{-\gamma(t-\tau)} d\tau.
\label{SM-dynamics1}
\end{equation}
where   $q=1.09P_{0}(K)$, $P_{0}(K)$ is given by Eq. \ref{probability}, and $K_{i}(t)=\int_{0}^{t}k_{i}(\tau)d\tau$. Dynamic behavior described by Eq. \ref{SM-dynamics1} results from the interplay between the positive feedback   captured by the factor $q(K_{i})$  and  the  obsolescence characterized by  $\gamma$.

For small $\gamma$  Eq. \ref{SM-dynamics1}  reduces to the models of Refs.  \cite{Pennock2002,Menczer2004,Shao2006,Peterson2010}
\begin{equation}
k_{i}(t)\approx p_{i}(t)+qK_{i}(t)
\label{dynamics-Bass}
\end{equation}
where $p(t)=\eta_{i} m(t)$. Equation \ref{dynamics-Bass} is nothing else but the Bass equation for diffusion of innovations  in infinite market.\cite{Bass2004}   Citations correspond to adopters, direct citations correspond to innovators, and indirect citations correspond to imitators.  The connection to the Bass model is not occasional since each paper is a new product whose penetration to the market of ideas is gauged by the number of citations. The novelty here is the nonlinear $q(K)$ dependence which is not unexpected:  the nonlinear coefficient of imitation $q(K)$  indicates increased probability of adoption of a new product if several neighbors in the  network already adopted it, some kind of social reinforcement.\cite{Centola2010}

For  large $\gamma$, the main contribution to the integral in Eq. \ref{SM-dynamics1} comes from  recent citations garnered between $t$ and  $t-1/\gamma$. We approximate $k(\tau)$ by $k(t)-(t-\tau) \frac{dk}{d\tau}|_{t}$, perform integration in this short time window, and after some algebra arrive at
\begin{equation}
k_{i}(t)\approx \eta_{i} m(t)+\frac{q_{i}}{\gamma}k_{i}(t-1/\gamma)
\label{dynamics-short}
\end{equation}
Equation \ref{dynamics-short} is the first-order autoregressive model of citation dynamics  where  time delay is $1/\gamma$ and $q/\gamma$ is the first-order autoregressive parameter.\cite{Golosovsky2012}  In distinction to Eq. \ref{dynamics-Bass} that attributes equal weight to all past citations, Eq. \ref{dynamics-short}  puts more weight to recent citations, and it describes citation dynamics of scientific publications  more realistically than Eq. \ref{dynamics-Bass}.  Similar models  were suggested by Refs. \cite{Simkin2007,Wang2008}. Ref. \cite{Gleeson2014} showed that dynamics of Facebook installation decisions is  also biased toward recent rather than cumulative popularity.

We come back to Eq. \ref{SM-dynamics1}. Although its analytical solution is unknown, we can gain some intuition if we consider  $q=const$ (this is justified since $q$ increases very slowly with $K$). We introduce a new variable $y_{i}=\int_{0}^{t}k_{i}e^{\gamma\tau}d\tau$, substitute it into Eq. \ref{SM-dynamics1}, and obtain
\begin{equation}
\frac{dy}{dt}=\eta m_{dir}(t)e^{\gamma t} +qy.
\label{y}
\end{equation}
where  index $i$ has been dropped for the ease of readability. Equation \ref{y} is easily integrated and its solution is $y=\eta e^{qt}\int_{0}^{t} m_{dir}(\tau)e^{(\gamma-q)\tau}d\tau$. We recall that $k=\frac{dy}{dt}e^{-\gamma t}$  and obtain
\begin{equation}
k(t)=\eta\left[m_{dir}(t)+q\int_{0}^{t}m_{dir}(\tau)e^{-(\gamma-q)(t-\tau})d\tau\right].
\label{SM-dynamics2}
\end{equation}
The total number of citations is 
\begin{widetext}
\begin{equation}
K(t)=\eta\left[\int_{0}^{t}m_{dir}(\tau)d\tau
+q\int_{0}^{t}d\tau\int_{0}^{\tau}m_{dir} (\tau')e^{-(\gamma-q)(\tau-\tau'})d\tau'\right].
\label{SM-dynamics3}
\end{equation}
\end{widetext}
Equation \ref{SM-dynamics3} indicates that each direct citation, captured by the term $\eta m_{dir}(t)$, induces a cascade of indirect citations that decays if  $\gamma>q$ and propagates if $\gamma<q$. The former case corresponds to ordinary papers, the latter case corresponds to runaways.  At the beginning of the paper's citation career, $K$ is small and $q/\gamma\sim 0.3<1$.  Since $\gamma=1.2$ yr$^{-1}$ and $m_{dir}(\tau)$ decays with $\tau$ slower than exponentially, we perform inner integration in Eq. \ref{SM-dynamics3} assuming $m_{dir}(\tau')=m_{dir}(\tau)$ and arrive at
\begin{equation}
K(t)\approx\frac{\eta\gamma}{\gamma-q}\int_{0}^{t}m_{dir}(\tau)d\tau
\label{SM-dynamics4}
\end{equation}
Since $q\propto P_{0}(K)$ and slowly grows with $K$ (Eq. \ref{probability}), we rearrange Eq. \ref{SM-dynamics4} to gather all $K$-dependent terms together,
\begin{equation}
K\left(1-\frac{q(K)}{\gamma}\right)\approx\eta \int_{0}^{t}m_{dir}(\tau)d\tau
\label{SM-dynamics5}
\end{equation}
The left-hand side  of Eq. \ref{SM-dynamics5}  depends nonmonotonously on $K$ and achieves a maximum at certain $K_{crit}$. We introduce $\eta_{crit}=K_{crit}\left(1-\frac{q(K_{crit})}{\gamma}\right)$ and consider Eq. \ref{SM-dynamics5} in several limiting cases.
It should be noted  that since we limit ourselves by the 30-year span after paper publication, then $\int_{0}^{t>30}m_{dir}(\tau)d\tau \approx 1$ by definition (Figure \ref{fig:direct}b shows that although this integral that does not come to saturation in 30 years its increase with time after t=30 yr is very slow). In this case Eq. \ref{SM-dynamics5} has stationary solution only for the papers with $\eta<\eta_{crit}$. Then $K(t)\rightarrow K_{\infty}$, indicating that  citation career of such papers eventually saturates.  On the contrary, the number of citations of the papers with $\eta>\eta_{crit}$ diverges, $K(t)\rightarrow \infty$. Citation career  of these runaways prolongs indefinitely.
\begin{figure}[!ht]
\begin{center}
\includegraphics*[width=0.4\textwidth]{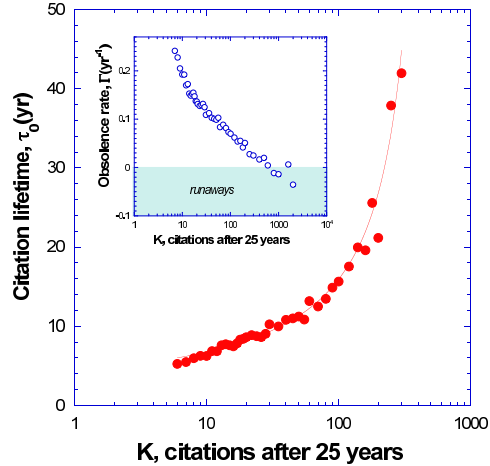}
\caption{\small{Paper longevity (citation lifetime) $\tau_{0}$ versus $K$, the number of citations after 25 years. The measurements  for 40,195 Physics papers published in 1984.  $\tau_{0}$ grows with increasing $K$ and diverges at $K\sim 700$ in such a way that highly-cited papers become runaways. The solid line shows crude approximation suggested by Eq. \ref{lifetime} where $\gamma=1.2$ yr$^{-1}, m_{dir}(t=0)=0.23$ yr$^{-1}$, and $q=0.38(1+0.82\log K)$. The inset shows the obsolescence rate, $\Gamma=\tau_{0}^{-1}$. To reduce fluctuations we binned the data.}
}
\label{fig:lifetime}
\end{center}
\end{figure}
In what follows we estimate the time scale $\tau_{0}$ associated with paper's longevity (citation lifetime).  If citation dynamics of a paper were exponential, then $K(t)=K_{\infty}(1-e^{-\frac{t}{\tau_{0}}})$  where $K_{\infty}$ is the number of paper's citations in the long-time limit. We rearrange this expression to exclude $K_{\infty}$ and find  $K(t)=\tau_{0}(1-e^{-\frac{t}{\tau_{0}}})\frac{dK}{dt}|_{t=0}$. In the long-time limit  when $e^{-\frac{t}{\tau_{0}}}<<1$  we can write $\tau_{0}\approx \frac{K(t)}{\frac{dK}{dt}|_{t=0}}$. We substitute this expression into Eq. \ref{SM-dynamics5} and find
\begin{equation}
\tau_{0}\approx \frac{\gamma-q(K=0)}{\gamma-q(K)}\frac{1}{m_{dir}|_{t=0}}
\label{lifetime}
\end{equation}
Equation \ref{lifetime} implies that $\tau_{0}$  increases with  $K$ and  diverges for  seminal papers for which $q(K)\sim\gamma$ indicating that citation lifetime is longer for highly-cited papers. This complements the famous parable "rich get richer" with the statement "rich live longer". Figure \ref{fig:lifetime} shows that citation lifetime $\tau_{0}$ for 40,195 Physics papers published in 1984  agree fairly well with Eq. \ref{lifetime}. The papers with diverging $\tau_{0}$ are runaways \cite{Golosovsky2012a} or supercritical papers.\cite{Bianconi2001} Similar runaways were  detected in the distribution of the Web page popularity.\cite{Kong2008}

\section{Discussion}
\subsection{Comparison to previous studies}
Our results shall be compared to the Simkin-Roychowdhury mathematical theory of citing \cite{Simkin2007} which is based on the copying algorithm of Krapivsky-Redner.\cite{Krapivsky2005} Ref. \cite{Simkin2007} considered a following scenario: when a scientist writes a manuscript, he picks several recent  papers published in the preceding year, cites them, and also copies some of their references with equal probability. This is the first-generation qualitative model - it provides clever insight, the basic scenario, but it can't be used for qualitative estimates. Our model is based on a much more detailed scenario: when a scientist writes a manuscript, he picks several papers with the probability depending on their publication year, cites them, and  copies some of their references with the probability depending on the publication year of the parent paper and on the local structure of citation network associated with the copied paper. All probabilities were measured.  Thus, our well-calibrated  model builds on Ref. \cite{Simkin2007} but belongs to  next generation of models, those that can be used for quantitative estimates.

\subsection{Nonlinearity and "power-law" degree distributions}
Citations of scientific papers were one of the first examples of the power-law (fat-tailed) distribution.\cite{Price1976} The prevailing notion is that all papers are created equal and the power-law distribution of their citations is created dynamically. Our results  tell different story:  the fat-tail citation distribution is mostly inherited.  Indeed, Fig. \ref{fig:simulation} shows that citation distribution at small $t$ mimics fitness distribution which is already  a fat-tailed distribution.  As time goes on, citation distribution shifts towards higher $K$. Since the kernel in Eq. \ref{paper-new} is nonlinear and increases with $K$, the tail of the distribution  shifts faster than its body.  If initial citation  distribution was concave in the log-log coordinates it straightens with time and becomes similar to the power-law distribution. When the time goes on further, the slope of the distribution gradually decreases as it is clearly seen in Fig. \ref{fig:simulation}.

This observation beats the intuition assuming that the power-law degree distribution is an evidence of the scale-free network.  We show here  that at least for citations, the power-law distribution is not the ultimate but a transient distribution. There is a hidden scale that is defined by  Eq. \ref{probability} and by the relation $P_{0}(K)\sim\gamma$. This scale marks the onset of the runaway behaviour which is  demonstrated in Figs. \ref{fig:trajectories2},\ref{fig:lifetime}.
\subsection{Preferential attachment}
When we started this research we believed that citation network grows following Eqs. \ref{Barabasi}-\ref{triple}. Hence, we  based our model on recursive search which is a specific implementation of the preferential attachment rule captured by Eqs. \ref{Barabasi}-\ref{triple}. Our measurements yielded  Eq. \ref{paper-new}  which is very different and follows more the line of thought of Refs. \cite{Caldarelli2002,Simkin2007} who  focussed entirely on fitness. Does this invalidate the common understanding of the preferential attachment rule as an algorithm according to which a new node performs global search in the whole network to find the most connected nodes? Not necessarily. We demonstrate that preferential attachment exists but it is more subtle than commonly believed.

Our measurements and modeling suggest the following mechanism of network growth. A new node in the network attaches to several  older target nodes that become its nearest neighbors. Then the new node explores its  next-nearest neighbors and preferentially connects to those that are connected  not to one nearest-neighbor but to several nearest-neighbors. The latter are obviously the most connected nodes in the vicinity of a target node (cf. acquaintance immunization \cite{Cohen2003}).  Taken together with the assortativity of citation network, this results in nonlinear attachment probability. Although this algorithm is based on local search, it is one step towards the global search, since it analyzes not only the nearest, but the next-nearest neighbors as well.  Hence, the preferential attachment mechanism pops out explicitly in our model but in different guise- it is captured by the kernel $P_{0}(K)$ in Eq. \ref{paper-new}.
\subsection{\label{sec:prediction}Prediction of citation career}
The models of citation dynamics find application in predicting future citation trajectories of papers \cite{Stringer2008,Wang2013,Uzzi2013} and citation career of the authors.\cite{Acuna2012,Mazloumian2012,Penner2013} Our calibrated model can be used for probabilistic prediction of the number of citations the regular paper can garner in the future. Our formalism can  be also used to pinpoint sleeping beauties/shooting stars at the earliest stage of their citation career.  This task is usually solved by applying some model that extrapolates  citation dynamics of papers basing on their citation history and then focusses on those papers that deviate from model prediction.\cite{Newman2014,Ponomarev2014,Medo2014}  Our model is well-suited for this purpose since it predicts not only the mean citation dynamics but the probability of its deviation from the mean.  We leave for further studies application of our model for forecasting citation behaviour of scientific papers.

What are the limitations of our model?  One  particularly strong assumption is  the constancy of  fitness  along the whole citation career of the paper. Ref. \cite{Kong2008} in their description of the Web-pages popularity also used this assumption and  justified it by measurements.  While the assumption of constant fitness is  reasonable for majority of scientific papers, and is validated by our measurements, there are sleeping beauties (Fig. \ref{fig:trajectories2}a) that can be dormant for a long-time and then achieve a burst of  popularity. Although these papers are rare, they are often associated with scientific breakthroughs, and their importance is incomparable to their abundance. Refs. \cite{Roth2012,Ke2015}  analyzed such papers and found that their peculiar citation trajectory has content-based explanation.\cite{Uzzi2013} Is it possible that such citation bursts can appear  by chance? Although our model describes a Hawkes process  where small deviations from the mean-field behavior can be amplified during prolonged time period  producing bursts,\cite{Onaga2014} we do not believe that our model can generate strong bursts of citations. The reason is  exponentially strong obsolescence (Eq. \ref{probability-all}) that prevents continuous amplification of small fluctuations. We believe that  our model describes a regular science in the sense of T. Kuhn \cite{Kuhn1970} and does not capture exceptional papers associated with serendipitous discoveries, the bursts of scientific activity,  emergence of new and disappearance of old fields - everything that makes the science fun.

\subsection{\label{sec:extension}Extension  to other fields}
How general is our model? While it was calibrated using Physics papers published in 1984,  we  performed similar measurements for Mathematics and Economics papers also published in 1984 and found very similar citation dynamics, including nonlinear kernel and runaways. Hence we have a good reason to believe that our model describes these fields  as well albeit with different parameters.
\begin{itemize}
  \item  We found  lognormal fitness distributions for Physics, Mathematics, and Economics papers  with the same  $\sigma=1.1$ and different $\mu$.
  \item Indirect citations. Nonlinear kernel $P_{0}(K)$  with logarithmic dependence on the number of citations has been found for all three fields.
  \item  Although we didn't measure $m_{dir}(t)$ for Mathematics and Economics we do not expect that it is the same for all research fields. While $r(t)$ and $r_{dir}(t)$ seem to be very similar for different fields,  the growth rates of the number of publications and the reference list length strongly differ.
\end{itemize}

We consider now a more general question- whether our network growth model based on  recursive search with a nonlinear kernel can describe other phenomena besides citation dynamics of scientific papers. Indeed, the  recursive search mechanism  was invoked to account for spreading of ideas, rumors, diseases, and viral marketing  \cite{Goffmann1964,Ishii2012,Bettencourt2006,Leskovec2007,Pastor2015}. Generally, these processes are modeled using  linear  dynamic equations assuming pairwise interactions between the neighbors in the network. The studies of Centola \cite{Centola2007,Centola2010} of the spreading in social networks revealed complex propagation with social reinforcement where simultaneous exposure to several active neighbors in the network is important. Such synergistic effects in propagation on networks were also considered theoretically \cite{Perez2011,Broder2015} and found experimentally  in epidemiology,\cite{Ludlam2011} where susceptibility of a person to infection may depend on the number of infected neighbors. Our studies suggest that such  multiple-node interactions result in nonlinear dynamic equation of complex propagation in networks.  Indeed,  Ref. \cite{Iribarren2011} found  nonlinearity in the dynamics of viral marketing, in particular, it observed the linear relation between transmittivity and fan-out coefficient which is very similar to our observation (Fig. \ref{fig:ratio-probability}) of the linear relation between the number of second-generation citations and the probability of indirect citation. Refs. \cite{Csardi2007,Valverde2007,Sheridan2012} found nonlinearity in  citation dynamics of US patents, Ref. \cite{Zhou2004} found  nonlinearity in their studies of the Internet connectivity and growth. Our finding that even weak dynamic  nonlinearity can lead to runaways can be important for other social networks where spreading processes with social reinforcement occur.

\section{\label{sec:summary}Summary}
We  report  a nonlinear stochastic model of citation dynamics of scientific papers. The model is fully calibrated by measurements of citations dynamics and statistics of references.  The model assumes that the author of a new scientific paper finds relevant papers from the media or journals and cites  them. Then he studies the reference lists of these preselected papers, picks up  relevant papers, cites them as well, and continues this process recursively. If some paper is cited by several preselected papers,  the author chooses it with higher probability than that cited by only one preselected paper. This local rule enables the author to sample the global connectivity of the network.

This recursive search rule results in dynamic nonlinearity.  The nonlinearity is the reason why the ideas advocated in highly-cited papers undergo viral propagation in scientific community, while the low-cited papers affect only a small part of it. Such dynamic  nonlinearity can play important role in viral propagation in social media.

\begin{acknowledgments}
We are grateful to Sidney Redner, Lev Muchnik, Danny Shapiro, and Yoram Louzoun for fruitful discussions, we appreciate instructive correspondence with Mikhail Simkin. We are grateful to Andrea Scharnhorst for constant encouragement and we acknowledge financial support of the EU COST Action TD1210.
\end{acknowledgments}

\bibliography{references_list_sorted}

\begin{thebibliography}{91}%
\makeatletter
\providecommand \@ifxundefined [1]{%
 \@ifx{#1\undefined}
}%
\providecommand \@ifnum [1]{%
 \ifnum #1\expandafter \@firstoftwo
 \else \expandafter \@secondoftwo
 \fi
}%
\providecommand \@ifx [1]{%
 \ifx #1\expandafter \@firstoftwo
 \else \expandafter \@secondoftwo
 \fi
}%
\providecommand \natexlab [1]{#1}%
\providecommand \enquote  [1]{``#1''}%
\providecommand \bibnamefont  [1]{#1}%
\providecommand \bibfnamefont [1]{#1}%
\providecommand \citenamefont [1]{#1}%
\providecommand \href@noop [0]{\@secondoftwo}%
\providecommand \href [0]{\begingroup \@sanitize@url \@href}%
\providecommand \@href[1]{\@@startlink{#1}\@@href}%
\providecommand \@@href[1]{\endgroup#1\@@endlink}%
\providecommand \@sanitize@url [0]{\catcode `\\12\catcode `\$12\catcode
  `\&12\catcode `\#12\catcode `\^12\catcode `\_12\catcode `\%12\relax}%
\providecommand \@@startlink[1]{}%
\providecommand \@@endlink[0]{}%
\providecommand \url  [0]{\begingroup\@sanitize@url \@url }%
\providecommand \@url [1]{\endgroup\@href {#1}{\urlprefix }}%
\providecommand \urlprefix  [0]{URL }%
\providecommand \Eprint [0]{\href }%
\providecommand \doibase [0]{http://dx.doi.org/}%
\providecommand \selectlanguage [0]{\@gobble}%
\providecommand \bibinfo  [0]{\@secondoftwo}%
\providecommand \bibfield  [0]{\@secondoftwo}%
\providecommand \translation [1]{[#1]}%
\providecommand \BibitemOpen [0]{}%
\providecommand \bibitemStop [0]{}%
\providecommand \bibitemNoStop [0]{.\EOS\space}%
\providecommand \EOS [0]{\spacefactor3000\relax}%
\providecommand \BibitemShut  [1]{\csname bibitem#1\endcsname}%
\let\auto@bib@innerbib\@empty
\bibitem [{\citenamefont {Newman}(2010)}]{Newman2010}%
  \BibitemOpen
  \bibfield  {author} {\bibinfo {author} {\bibfnamefont {M.}~\bibnamefont
  {Newman}},\ }\href
  {http://www.ebook.de/de/product/9314383/mark_newman_networks.html} {\emph
  {\bibinfo {title} {Networks}}}\ (\bibinfo  {publisher} {Oxford University
  Press},\ \bibinfo {year} {2010})\BibitemShut {NoStop}%
\bibitem [{\citenamefont {Barabasi}(2015)}]{Barabasi2015}%
  \BibitemOpen
  \bibfield  {author} {\bibinfo {author} {\bibfnamefont {A.-L.}\ \bibnamefont
  {Barabasi}},\ }\href
  {http://www.ebook.de/de/product/24312547/albert_laszlo_barabasi_network_science.html}
  {\emph {\bibinfo {title} {Network Science}}}\ (\bibinfo  {publisher}
  {Cambridge University Press},\ \bibinfo {year} {2015})\BibitemShut {NoStop}%
\bibitem [{\citenamefont {Holme}\ and\ \citenamefont
  {Saramaki}(2012)}]{Holme2012}%
  \BibitemOpen
  \bibfield  {author} {\bibinfo {author} {\bibfnamefont {P.}~\bibnamefont
  {Holme}}\ and\ \bibinfo {author} {\bibfnamefont {J.}~\bibnamefont
  {Saramaki}},\ }\href {\doibase 10.1016/j.physrep.2012.03.001} {\bibfield
  {journal} {\bibinfo  {journal} {Physics Reports}\ }\textbf {\bibinfo {volume}
  {519}},\ \bibinfo {pages} {97} (\bibinfo {year} {2012})}\BibitemShut
  {NoStop}%
\bibitem [{\citenamefont {Price}(1976)}]{Price1976}%
  \BibitemOpen
  \bibfield  {author} {\bibinfo {author} {\bibfnamefont {D.~D.~S.}\
  \bibnamefont {Price}},\ }\href {\doibase 10.1002/asi.4630270505} {\bibfield
  {journal} {\bibinfo  {journal} {Journal of the American Society for
  Information Science}\ }\textbf {\bibinfo {volume} {27}},\ \bibinfo {pages}
  {292} (\bibinfo {year} {1976})}\BibitemShut {NoStop}%
\bibitem [{\citenamefont {Albert}\ and\ \citenamefont
  {Barabasi}(2002)}]{Barabasi2002}%
  \BibitemOpen
  \bibfield  {author} {\bibinfo {author} {\bibfnamefont {R.}~\bibnamefont
  {Albert}}\ and\ \bibinfo {author} {\bibfnamefont {A.-L.}\ \bibnamefont
  {Barabasi}},\ }\href {\doibase 10.1103/RevModPhys.74.47} {\bibfield
  {journal} {\bibinfo  {journal} {Rev. Mod. Phys.}\ }\textbf {\bibinfo {volume}
  {74}},\ \bibinfo {pages} {47} (\bibinfo {year} {2002})}\BibitemShut {NoStop}%
\bibitem [{\citenamefont {Perc}(2014)}]{Perc2014}%
  \BibitemOpen
  \bibfield  {author} {\bibinfo {author} {\bibfnamefont {M.}~\bibnamefont
  {Perc}},\ }\href {\doibase 10.1098/rsif.2014.0378} {\bibfield  {journal}
  {\bibinfo  {journal} {Journal of The Royal Society Interface}\ }\textbf
  {\bibinfo {volume} {11}} (\bibinfo {year} {2014}),\
  10.1098/rsif.2014.0378}\BibitemShut {NoStop}%
\bibitem [{\citenamefont {Dorogovtsev}\ and\ \citenamefont
  {Mendes}(2000)}]{Dorogovtsev2000}%
  \BibitemOpen
  \bibfield  {author} {\bibinfo {author} {\bibfnamefont {S.~N.}\ \bibnamefont
  {Dorogovtsev}}\ and\ \bibinfo {author} {\bibfnamefont {J.~F.~F.}\
  \bibnamefont {Mendes}},\ }\href {\doibase 10.1103/physreve.62.1842}
  {\bibfield  {journal} {\bibinfo  {journal} {Physical Review E}\ }\textbf
  {\bibinfo {volume} {62}},\ \bibinfo {pages} {1842} (\bibinfo {year}
  {2000})}\BibitemShut {NoStop}%
\bibitem [{\citenamefont {Krapivsky}\ and\ \citenamefont
  {Redner}(2001)}]{Krapivsky2001}%
  \BibitemOpen
  \bibfield  {author} {\bibinfo {author} {\bibfnamefont {P.~L.}\ \bibnamefont
  {Krapivsky}}\ and\ \bibinfo {author} {\bibfnamefont {S.}~\bibnamefont
  {Redner}},\ }\href {\doibase 10.1103/physreve.63.066123} {\bibfield
  {journal} {\bibinfo  {journal} {Physical Review E}\ }\textbf {\bibinfo
  {volume} {63}},\ \bibinfo {pages} {066123} (\bibinfo {year}
  {2001})}\BibitemShut {NoStop}%
\bibitem [{\citenamefont {Bianconi}\ and\ \citenamefont
  {Barabasi}(2001)}]{Bianconi2001}%
  \BibitemOpen
  \bibfield  {author} {\bibinfo {author} {\bibfnamefont {G.}~\bibnamefont
  {Bianconi}}\ and\ \bibinfo {author} {\bibfnamefont {A.-L. A.-L.}\
  \bibnamefont {Barabasi}},\ }\href {\doibase 10.1103/PhysRevLett.86.5632}
  {\bibfield  {journal} {\bibinfo  {journal} {Phys. Rev. Lett.}\ }\textbf
  {\bibinfo {volume} {86}},\ \bibinfo {pages} {5632} (\bibinfo {year}
  {2001})}\BibitemShut {NoStop}%
\bibitem [{\citenamefont {Wang}\ \emph {et~al.}(2013)\citenamefont {Wang},
  \citenamefont {Song},\ and\ \citenamefont {Barabasi}}]{Wang2013}%
  \BibitemOpen
  \bibfield  {author} {\bibinfo {author} {\bibfnamefont {D.}~\bibnamefont
  {Wang}}, \bibinfo {author} {\bibfnamefont {C.}~\bibnamefont {Song}}, \ and\
  \bibinfo {author} {\bibfnamefont {A.-L.}\ \bibnamefont {Barabasi}},\ }\href
  {http://www.sciencemag.org/content/342/6154/127.abstract N2 - The lack of
  predictability of citation-based measures frequently used to gauge impact,
  from impact factors to short-term citations, raises a fundamental question:
  Is there long-term predictability in citation patterns? Here, we derive a
  mechanistic model for the citation dynamics of individual papers, allowing us
  to collapse the citation histories of papers from different journals and
  disciplines into a single curve, indicating that all papers tend to follow
  the same universal temporal pattern. The observed patterns not only help us
  uncover basic mechanisms that govern scientific impact but also offer
  reliable measures of influence that may have potential policy implications.}
  {\bibfield  {journal} {\bibinfo  {journal} {Science}\ }\textbf {\bibinfo
  {volume} {342}},\ \bibinfo {pages} {127} (\bibinfo {year}
  {2013})}\BibitemShut {NoStop}%
\bibitem [{\citenamefont {Kong}\ \emph {et~al.}(2008)\citenamefont {Kong},
  \citenamefont {Sarshar},\ and\ \citenamefont {Roychowdhury}}]{Kong2008}%
  \BibitemOpen
  \bibfield  {author} {\bibinfo {author} {\bibfnamefont {J.~S.}\ \bibnamefont
  {Kong}}, \bibinfo {author} {\bibfnamefont {N.}~\bibnamefont {Sarshar}}, \
  and\ \bibinfo {author} {\bibfnamefont {V.~P.}\ \bibnamefont {Roychowdhury}},\
  }\href {http://www.pnas.org/content/105/37/13724.abstract N2 - We use
  sequential large-scale crawl data to empirically investigate and validate the
  dynamics that underlie the evolution of the structure of the web. We find
  that the overall structure of the web is defined by an intricate interplay
  between experience or entitlement of the pages (as measured by the number of
  inbound hyperlinks a page already has), inherent talent or fitness of the
  pages (as measured by the likelihood that someone visiting the page would
  give a hyperlink to it), and the continual high rates of birth and death of
  pages on the web. We find that the web is conservative in judging talent and
  the overall fitness distribution is exponential, showing low variability. The
  small variance in talent, however, is enough to lead to experience
  distributions with high variance: The preferential attachment mechanism
  amplifies these small biases and leads to heavy-tailed power-law (PL) inbound
  degree distributions over all pages, as well as over pages that are of the
  same age. The balancing act between experience and talent on the web allows
  newly introduced pages with novel and interesting content to grow quickly and
  surpass older pages. In this regard, it is much like what we observe in
  high-mobility and meritocratic societies: People with entitlement continue to
  have access to the best resources, but there is just enough screening for
  fitness that allows for talented winners to emerge and join the ranks of the
  leaders. Finally, we show that the fitness estimates have potential practical
  applications in ranking query results.} {\bibfield  {journal} {\bibinfo
  {journal} {Proceedings of the National Academy of Sciences}\ }\textbf
  {\bibinfo {volume} {105}},\ \bibinfo {pages} {13724} (\bibinfo {year}
  {2008})}\BibitemShut {NoStop}%
\bibitem [{\citenamefont {Medo}\ \emph {et~al.}(2011)\citenamefont {Medo},
  \citenamefont {Cimini},\ and\ \citenamefont {Gualdi}}]{Medo2011}%
  \BibitemOpen
  \bibfield  {author} {\bibinfo {author} {\bibfnamefont {M.}~\bibnamefont
  {Medo}}, \bibinfo {author} {\bibfnamefont {G.}~\bibnamefont {Cimini}}, \ and\
  \bibinfo {author} {\bibfnamefont {S.}~\bibnamefont {Gualdi}},\ }\href
  {\doibase 10.1103/PhysRevLett.107.238701} {\bibfield  {journal} {\bibinfo
  {journal} {Phys. Rev. Lett.}\ }\textbf {\bibinfo {volume} {107}},\ \bibinfo
  {pages} {238701} (\bibinfo {year} {2011})}\BibitemShut {NoStop}%
\bibitem [{\citenamefont {Menczer}(2004)}]{Menczer2004}%
  \BibitemOpen
  \bibfield  {author} {\bibinfo {author} {\bibfnamefont {F.}~\bibnamefont
  {Menczer}},\ }\href {\doibase 10.1073/pnas.0307554100} {\bibfield  {journal}
  {\bibinfo  {journal} {Proceedings of the National Academy of Sciences}\
  }\textbf {\bibinfo {volume} {101}},\ \bibinfo {pages} {5261} (\bibinfo {year}
  {2004})}\BibitemShut {NoStop}%
\bibitem [{\citenamefont {Papadopoulos}\ \emph {et~al.}(2012)\citenamefont
  {Papadopoulos}, \citenamefont {Kitsak}, \citenamefont {Serrano},
  \citenamefont {Bogu{\~{n}}{\'{a}}},\ and\ \citenamefont
  {Krioukov}}]{Papadopoulos2012}%
  \BibitemOpen
  \bibfield  {author} {\bibinfo {author} {\bibfnamefont {F.}~\bibnamefont
  {Papadopoulos}}, \bibinfo {author} {\bibfnamefont {M.}~\bibnamefont
  {Kitsak}}, \bibinfo {author} {\bibfnamefont {M.~{\'{A}}.}\ \bibnamefont
  {Serrano}}, \bibinfo {author} {\bibfnamefont {M.}~\bibnamefont
  {Bogu{\~{n}}{\'{a}}}}, \ and\ \bibinfo {author} {\bibfnamefont
  {D.}~\bibnamefont {Krioukov}},\ }\href {\doibase 10.1038/nature11459}
  {\bibfield  {journal} {\bibinfo  {journal} {Nature}\ }\textbf {\bibinfo
  {volume} {489}},\ \bibinfo {pages} {537} (\bibinfo {year}
  {2012})}\BibitemShut {NoStop}%
\bibitem [{\citenamefont {Bramoull{\'{e}}}\ \emph {et~al.}(2012)\citenamefont
  {Bramoull{\'{e}}}, \citenamefont {Currarini}, \citenamefont {Jackson},
  \citenamefont {Pin},\ and\ \citenamefont {Rogers}}]{Bramoulle2012}%
  \BibitemOpen
  \bibfield  {author} {\bibinfo {author} {\bibfnamefont {Y.}~\bibnamefont
  {Bramoull{\'{e}}}}, \bibinfo {author} {\bibfnamefont {S.}~\bibnamefont
  {Currarini}}, \bibinfo {author} {\bibfnamefont {M.~O.}\ \bibnamefont
  {Jackson}}, \bibinfo {author} {\bibfnamefont {P.}~\bibnamefont {Pin}}, \ and\
  \bibinfo {author} {\bibfnamefont {B.~W.}\ \bibnamefont {Rogers}},\ }\href
  {\doibase 10.1016/j.jet.2012.05.007} {\bibfield  {journal} {\bibinfo
  {journal} {Journal of Economic Theory}\ }\textbf {\bibinfo {volume} {147}},\
  \bibinfo {pages} {1754} (\bibinfo {year} {2012})}\BibitemShut {NoStop}%
\bibitem [{\citenamefont {Yun}\ \emph {et~al.}(2015)\citenamefont {Yun},
  \citenamefont {Kim},\ and\ \citenamefont {Jeong}}]{Yun2015}%
  \BibitemOpen
  \bibfield  {author} {\bibinfo {author} {\bibfnamefont {J.}~\bibnamefont
  {Yun}}, \bibinfo {author} {\bibfnamefont {P.-J.}\ \bibnamefont {Kim}}, \ and\
  \bibinfo {author} {\bibfnamefont {H.}~\bibnamefont {Jeong}},\ }\href
  {\doibase 10.1371/journal.pone.0117388} {\bibfield  {journal} {\bibinfo
  {journal} {{PLOS} {ONE}}\ }\textbf {\bibinfo {volume} {10}},\ \bibinfo
  {pages} {e0117388} (\bibinfo {year} {2015})}\BibitemShut {NoStop}%
\bibitem [{\citenamefont {Servedio}\ \emph {et~al.}(2004)\citenamefont
  {Servedio}, \citenamefont {Caldarelli},\ and\ \citenamefont
  {Butt{\`{a}}}}]{Servedio2004}%
  \BibitemOpen
  \bibfield  {author} {\bibinfo {author} {\bibfnamefont {V.~D.~P.}\
  \bibnamefont {Servedio}}, \bibinfo {author} {\bibfnamefont {G.}~\bibnamefont
  {Caldarelli}}, \ and\ \bibinfo {author} {\bibfnamefont {P.}~\bibnamefont
  {Butt{\`{a}}}},\ }\href {\doibase 10.1103/physreve.70.056126} {\bibfield
  {journal} {\bibinfo  {journal} {Physical Review E}\ }\textbf {\bibinfo
  {volume} {70}},\ \bibinfo {pages} {056126} (\bibinfo {year}
  {2004})}\BibitemShut {NoStop}%
\bibitem [{\citenamefont {Kleinberg}\ \emph {et~al.}(1999)\citenamefont
  {Kleinberg}, \citenamefont {Kumar}, \citenamefont {Raghavan}, \citenamefont
  {Rajagopalan},\ and\ \citenamefont {Tomkins}}]{Kleinberg1999}%
  \BibitemOpen
  \bibfield  {author} {\bibinfo {author} {\bibfnamefont {J.~M.}\ \bibnamefont
  {Kleinberg}}, \bibinfo {author} {\bibfnamefont {R.}~\bibnamefont {Kumar}},
  \bibinfo {author} {\bibfnamefont {P.}~\bibnamefont {Raghavan}}, \bibinfo
  {author} {\bibfnamefont {S.}~\bibnamefont {Rajagopalan}}, \ and\ \bibinfo
  {author} {\bibfnamefont {A.~S.}\ \bibnamefont {Tomkins}},\ }in\ \href
  {http://dl.acm.org/citation.cfm?id=1765751.1765753} {\emph {\bibinfo
  {booktitle} {Proceedings of the 5th Annual International Conference on
  Computing and Combinatorics}}},\ \bibinfo {series and number} {COCOON'99}\
  (\bibinfo  {publisher} {Springer-Verlag},\ \bibinfo {address} {Berlin,
  Heidelberg},\ \bibinfo {year} {1999})\ pp.\ \bibinfo {pages}
  {1--17}\BibitemShut {NoStop}%
\bibitem [{\citenamefont {Vazquez}(2003)}]{Vazquez2003}%
  \BibitemOpen
  \bibfield  {author} {\bibinfo {author} {\bibfnamefont {A.}~\bibnamefont
  {Vazquez}},\ }\href {\doibase 10.1103/PhysRevE.67.056104} {\bibfield
  {journal} {\bibinfo  {journal} {Phys. Rev. E}\ }\textbf {\bibinfo {volume}
  {67}},\ \bibinfo {pages} {056104} (\bibinfo {year} {2003})}\BibitemShut
  {NoStop}%
\bibitem [{\citenamefont {Evans}\ and\ \citenamefont
  {Saramaki}(2005)}]{Evans2005}%
  \BibitemOpen
  \bibfield  {author} {\bibinfo {author} {\bibfnamefont {T.~S.}\ \bibnamefont
  {Evans}}\ and\ \bibinfo {author} {\bibfnamefont {J.~P.}\ \bibnamefont
  {Saramaki}},\ }\href {\doibase 10.1103/PhysRevE.72.026138} {\bibfield
  {journal} {\bibinfo  {journal} {Phys. Rev. E}\ }\textbf {\bibinfo {volume}
  {72}},\ \bibinfo {pages} {026138} (\bibinfo {year} {2005})}\BibitemShut
  {NoStop}%
\bibitem [{\citenamefont {Vazquez}(2001)}]{Vazquez2001}%
  \BibitemOpen
  \bibfield  {author} {\bibinfo {author} {\bibfnamefont {A.}~\bibnamefont
  {Vazquez}},\ }\href {http://stacks.iop.org/0295-5075/54/i=4/a=430} {\bibfield
   {journal} {\bibinfo  {journal} {EPL (Europhysics Letters)}\ }\textbf
  {\bibinfo {volume} {54}},\ \bibinfo {pages} {430} (\bibinfo {year}
  {2001})}\BibitemShut {NoStop}%
\bibitem [{\citenamefont {Krapivsky}\ and\ \citenamefont
  {Redner}(2005)}]{Krapivsky2005}%
  \BibitemOpen
  \bibfield  {author} {\bibinfo {author} {\bibfnamefont {P.~L.}\ \bibnamefont
  {Krapivsky}}\ and\ \bibinfo {author} {\bibfnamefont {S.}~\bibnamefont
  {Redner}},\ }\href {\doibase 10.1103/PhysRevE.71.036118} {\bibfield
  {journal} {\bibinfo  {journal} {Phys. Rev. E}\ }\textbf {\bibinfo {volume}
  {71}},\ \bibinfo {pages} {036118} (\bibinfo {year} {2005})}\BibitemShut
  {NoStop}%
\bibitem [{\citenamefont {Ren}\ \emph {et~al.}(2012)\citenamefont {Ren},
  \citenamefont {Shen},\ and\ \citenamefont {Cheng}}]{Ren2012}%
  \BibitemOpen
  \bibfield  {author} {\bibinfo {author} {\bibfnamefont {F.-X.}\ \bibnamefont
  {Ren}}, \bibinfo {author} {\bibfnamefont {H.-W.}\ \bibnamefont {Shen}}, \
  and\ \bibinfo {author} {\bibfnamefont {X.-Q.}\ \bibnamefont {Cheng}},\ }\href
  {\doibase 10.1016/j.physa.2012.02.001} {\bibfield  {journal} {\bibinfo
  {journal} {Physica A: Statistical Mechanics and its Applications}\ }\textbf
  {\bibinfo {volume} {391}},\ \bibinfo {pages} {3533} (\bibinfo {year}
  {2012})}\BibitemShut {NoStop}%
\bibitem [{\citenamefont {Jackson}\ and\ \citenamefont
  {Rogers}(2007)}]{Jackson2007}%
  \BibitemOpen
  \bibfield  {author} {\bibinfo {author} {\bibfnamefont {M.~O.}\ \bibnamefont
  {Jackson}}\ and\ \bibinfo {author} {\bibfnamefont {B.~W.}\ \bibnamefont
  {Rogers}},\ }\href {\doibase 10.1257/aer.97.3.890} {\bibfield  {journal}
  {\bibinfo  {journal} {American Economic Review}\ }\textbf {\bibinfo {volume}
  {97}},\ \bibinfo {pages} {890} (\bibinfo {year} {2007})}\BibitemShut
  {NoStop}%
\bibitem [{\citenamefont {Goldberg}\ \emph {et~al.}(2015)\citenamefont
  {Goldberg}, \citenamefont {Anthony},\ and\ \citenamefont
  {Evans}}]{Goldberg2015}%
  \BibitemOpen
  \bibfield  {author} {\bibinfo {author} {\bibfnamefont {S.~R.}\ \bibnamefont
  {Goldberg}}, \bibinfo {author} {\bibfnamefont {H.}~\bibnamefont {Anthony}}, \
  and\ \bibinfo {author} {\bibfnamefont {T.~S.}\ \bibnamefont {Evans}},\ }\href
  {\doibase 10.1007/s11192-015-1737-9} {\bibfield  {journal} {\bibinfo
  {journal} {Scientometrics}\ }\textbf {\bibinfo {volume} {105}},\ \bibinfo
  {pages} {1577} (\bibinfo {year} {2015})}\BibitemShut {NoStop}%
\bibitem [{\citenamefont {Wu}\ and\ \citenamefont {Holme}(2009)}]{Wu2009}%
  \BibitemOpen
  \bibfield  {author} {\bibinfo {author} {\bibfnamefont {Z.-X.}\ \bibnamefont
  {Wu}}\ and\ \bibinfo {author} {\bibfnamefont {P.}~\bibnamefont {Holme}},\
  }\href {\doibase 10.1103/PhysRevE.80.037101} {\bibfield  {journal} {\bibinfo
  {journal} {Phys. Rev. E}\ }\textbf {\bibinfo {volume} {80}},\ \bibinfo
  {pages} {037101} (\bibinfo {year} {2009})}\BibitemShut {NoStop}%
\bibitem [{\citenamefont {Itzhack}\ \emph {et~al.}(2010)\citenamefont
  {Itzhack}, \citenamefont {Muchnik}, \citenamefont {Erez}, \citenamefont
  {Tsaban}, \citenamefont {Goldenberg}, \citenamefont {Solomon},\ and\
  \citenamefont {Louzoun}}]{Itzhack2010}%
  \BibitemOpen
  \bibfield  {author} {\bibinfo {author} {\bibfnamefont {R.}~\bibnamefont
  {Itzhack}}, \bibinfo {author} {\bibfnamefont {L.}~\bibnamefont {Muchnik}},
  \bibinfo {author} {\bibfnamefont {T.}~\bibnamefont {Erez}}, \bibinfo {author}
  {\bibfnamefont {L.}~\bibnamefont {Tsaban}}, \bibinfo {author} {\bibfnamefont
  {J.}~\bibnamefont {Goldenberg}}, \bibinfo {author} {\bibfnamefont
  {S.}~\bibnamefont {Solomon}}, \ and\ \bibinfo {author} {\bibfnamefont
  {Y.}~\bibnamefont {Louzoun}},\ }\href {\doibase 10.1016/j.physa.2010.07.011}
  {\bibfield  {journal} {\bibinfo  {journal} {Physica A: Statistical Mechanics
  and its Applications}\ }\textbf {\bibinfo {volume} {389}},\ \bibinfo {pages}
  {5308} (\bibinfo {year} {2010})}\BibitemShut {NoStop}%
\bibitem [{\citenamefont {Martin}\ \emph {et~al.}(2013)\citenamefont {Martin},
  \citenamefont {Ball}, \citenamefont {Karrer},\ and\ \citenamefont
  {Newman}}]{Martin2013}%
  \BibitemOpen
  \bibfield  {author} {\bibinfo {author} {\bibfnamefont {T.}~\bibnamefont
  {Martin}}, \bibinfo {author} {\bibfnamefont {B.}~\bibnamefont {Ball}},
  \bibinfo {author} {\bibfnamefont {B.}~\bibnamefont {Karrer}}, \ and\ \bibinfo
  {author} {\bibfnamefont {M.~E.~J.}\ \bibnamefont {Newman}},\ }\href {\doibase
  10.1103/physreve.88.012814} {\bibfield  {journal} {\bibinfo  {journal} {Phys.
  Rev. E}\ }\textbf {\bibinfo {volume} {88}},\ \bibinfo {pages} {012814}
  (\bibinfo {year} {2013})}\BibitemShut {NoStop}%
\bibitem [{\citenamefont {Pennock}\ \emph {et~al.}(2002)\citenamefont
  {Pennock}, \citenamefont {Flake}, \citenamefont {Lawrence}, \citenamefont
  {Glover},\ and\ \citenamefont {Giles}}]{Pennock2002}%
  \BibitemOpen
  \bibfield  {author} {\bibinfo {author} {\bibfnamefont {D.~M.}\ \bibnamefont
  {Pennock}}, \bibinfo {author} {\bibfnamefont {G.~W.}\ \bibnamefont {Flake}},
  \bibinfo {author} {\bibfnamefont {S.}~\bibnamefont {Lawrence}}, \bibinfo
  {author} {\bibfnamefont {E.~J.}\ \bibnamefont {Glover}}, \ and\ \bibinfo
  {author} {\bibfnamefont {C.~L.}\ \bibnamefont {Giles}},\ }\href {\doibase
  10.1073/pnas.032085699} {\bibfield  {journal} {\bibinfo  {journal}
  {Proceedings of the National Academy of Sciences}\ }\textbf {\bibinfo
  {volume} {99}},\ \bibinfo {pages} {5207} (\bibinfo {year}
  {2002})}\BibitemShut {NoStop}%
\bibitem [{\citenamefont {Goffmann}\ and\ \citenamefont
  {Newill}(1964)}]{Goffmann1964}%
  \BibitemOpen
  \bibfield  {author} {\bibinfo {author} {\bibfnamefont {W.}~\bibnamefont
  {Goffmann}}\ and\ \bibinfo {author} {\bibfnamefont {V.~A.}\ \bibnamefont
  {Newill}},\ }\href {http://dx.doi.org/10.1038/204225a0} {\bibfield  {journal}
  {\bibinfo  {journal} {Nature}\ }\textbf {\bibinfo {volume} {204}},\ \bibinfo
  {pages} {225} (\bibinfo {year} {1964})}\BibitemShut {NoStop}%
\bibitem [{\citenamefont {Bruckner}\ \emph {et~al.}(1990)\citenamefont
  {Bruckner}, \citenamefont {Ebeling},\ and\ \citenamefont
  {Scharnhorst}}]{Bruckner1990}%
  \BibitemOpen
  \bibfield  {author} {\bibinfo {author} {\bibfnamefont {E.}~\bibnamefont
  {Bruckner}}, \bibinfo {author} {\bibfnamefont {W.}~\bibnamefont {Ebeling}}, \
  and\ \bibinfo {author} {\bibfnamefont {A.}~\bibnamefont {Scharnhorst}},\
  }\href {\doibase 10.1007/BF02019160} {\bibfield  {journal} {\bibinfo
  {journal} {Scientometrics}\ }\textbf {\bibinfo {volume} {18}},\ \bibinfo
  {pages} {21} (\bibinfo {year} {1990})}\BibitemShut {NoStop}%
\bibitem [{\citenamefont {Bettencourt}\ \emph {et~al.}(2006)\citenamefont
  {Bettencourt}, \citenamefont {Cintran-Arias}, \citenamefont {Kaiser},\ and\
  \citenamefont {Castillo-Chavez}}]{Bettencourt2006}%
  \BibitemOpen
  \bibfield  {author} {\bibinfo {author} {\bibfnamefont {L.~M.}\ \bibnamefont
  {Bettencourt}}, \bibinfo {author} {\bibfnamefont {A.}~\bibnamefont
  {Cintran-Arias}}, \bibinfo {author} {\bibfnamefont {D.~I.}\ \bibnamefont
  {Kaiser}}, \ and\ \bibinfo {author} {\bibfnamefont {C.}~\bibnamefont
  {Castillo-Chavez}},\ }\href {\doibase 10.1016/j.physa.2005.08.083} {\bibfield
   {journal} {\bibinfo  {journal} {Physica A: Statistical Mechanics and its
  Applications}\ }\textbf {\bibinfo {volume} {364}},\ \bibinfo {pages} {513}
  (\bibinfo {year} {2006})}\BibitemShut {NoStop}%
\bibitem [{\citenamefont {Bass}(2004)}]{Bass2004}%
  \BibitemOpen
  \bibfield  {author} {\bibinfo {author} {\bibfnamefont {F.~M.}\ \bibnamefont
  {Bass}},\ }\href {\doibase 10.1287/mnsc.1040.0264} {\bibfield  {journal}
  {\bibinfo  {journal} {Management Science}\ }\textbf {\bibinfo {volume}
  {50}},\ \bibinfo {pages} {1825} (\bibinfo {year} {2004})}\BibitemShut
  {NoStop}%
\bibitem [{\citenamefont {N.K.Vitanov}\ and\ \citenamefont
  {M.R.Ausloos}(2012)}]{Vitanov2012}%
  \BibitemOpen
  \bibfield  {author} {\bibinfo {author} {\bibnamefont {N.K.Vitanov}}\ and\
  \bibinfo {author} {\bibnamefont {M.R.Ausloos}},\ }in\ \href {\doibase
  10.1007/978-3-642-23068-4_3} {\emph {\bibinfo {booktitle} {Models of Science
  Dynamics}}},\ \bibinfo {editor} {edited by\ \bibinfo {editor} {\bibfnamefont
  {P.}~\bibnamefont {A.Scharnhorst}, \bibfnamefont {K.Borner}}}\ (\bibinfo
  {publisher} {Springer Berlin Heidelberg},\ \bibinfo {year} {2012})\ pp.\
  \bibinfo {pages} {69--125}\BibitemShut {NoStop}%
\bibitem [{\citenamefont {Shao}\ \emph {et~al.}(2006)\citenamefont {Shao},
  \citenamefont {Zou}, \citenamefont {Tan},\ and\ \citenamefont
  {Jin}}]{Shao2006}%
  \BibitemOpen
  \bibfield  {author} {\bibinfo {author} {\bibfnamefont {Z.-G.}\ \bibnamefont
  {Shao}}, \bibinfo {author} {\bibfnamefont {X.-W.}\ \bibnamefont {Zou}},
  \bibinfo {author} {\bibfnamefont {Z.-J.}\ \bibnamefont {Tan}}, \ and\
  \bibinfo {author} {\bibfnamefont {Z.-Z.}\ \bibnamefont {Jin}},\ }\href
  {\doibase 10.1088/0305-4470/39/9/004} {\bibfield  {journal} {\bibinfo
  {journal} {J. Phys. A: Math. Gen.}\ }\textbf {\bibinfo {volume} {39}},\
  \bibinfo {pages} {2035} (\bibinfo {year} {2006})}\BibitemShut {NoStop}%
\bibitem [{\citenamefont {Peterson}\ \emph {et~al.}(2010)\citenamefont
  {Peterson}, \citenamefont {Presse},\ and\ \citenamefont
  {Dill}}]{Peterson2010}%
  \BibitemOpen
  \bibfield  {author} {\bibinfo {author} {\bibfnamefont {G.~J.}\ \bibnamefont
  {Peterson}}, \bibinfo {author} {\bibfnamefont {S.}~\bibnamefont {Presse}}, \
  and\ \bibinfo {author} {\bibfnamefont {K.~A.}\ \bibnamefont {Dill}},\ }\href
  {\doibase 10.1073/pnas.1010757107} {\bibfield  {journal} {\bibinfo  {journal}
  {Proceedings of the National Academy of Sciences}\ }\textbf {\bibinfo
  {volume} {107}},\ \bibinfo {pages} {16023} (\bibinfo {year}
  {2010})}\BibitemShut {NoStop}%
\bibitem [{\citenamefont {de~Solla~Price}(1965)}]{SollaPrice1965}%
  \BibitemOpen
  \bibfield  {author} {\bibinfo {author} {\bibfnamefont {D.~J.}\ \bibnamefont
  {de~Solla~Price}},\ }\href {\doibase 10.1126/science.149.3683.510} {\bibfield
   {journal} {\bibinfo  {journal} {Science}\ }\textbf {\bibinfo {volume}
  {149}},\ \bibinfo {pages} {510} (\bibinfo {year} {1965})}\BibitemShut
  {NoStop}%
\bibitem [{\citenamefont {Borner}\ \emph {et~al.}(2004)\citenamefont {Borner},
  \citenamefont {Maru},\ and\ \citenamefont {Goldstone}}]{Borner2004}%
  \BibitemOpen
  \bibfield  {author} {\bibinfo {author} {\bibfnamefont {K.}~\bibnamefont
  {Borner}}, \bibinfo {author} {\bibfnamefont {J.~T.}\ \bibnamefont {Maru}}, \
  and\ \bibinfo {author} {\bibfnamefont {R.~L.}\ \bibnamefont {Goldstone}},\
  }\href {\doibase 10.1073/pnas.0307625100} {\bibfield  {journal} {\bibinfo
  {journal} {Proceedings of the National Academy of Sciences}\ }\textbf
  {\bibinfo {volume} {101}},\ \bibinfo {pages} {5266} (\bibinfo {year}
  {2004})}\BibitemShut {NoStop}%
\bibitem [{dir()}]{direct}%
  \BibitemOpen
  \href@noop {} {}\bibinfo {note} {This definition of direct/indirect
  references differs from that of Ref. \cite{Peterson2010} and is more close to
  fresh/old papers of \cite{Simkin2007}, random/ancestor search of
  \cite{Krapivsky2005}, broadcasting/word-of-mouth search of
  \cite{Vitanov2012}, adding/walking of
  \cite{Vazquez2001,Vazquez2003}.}\BibitemShut {Stop}%
\bibitem [{\citenamefont {Geller}\ \emph {et~al.}(1981)\citenamefont {Geller},
  \citenamefont {de~Cani},\ and\ \citenamefont {Davies}}]{Geller1981}%
  \BibitemOpen
  \bibfield  {author} {\bibinfo {author} {\bibfnamefont {N.~L.}\ \bibnamefont
  {Geller}}, \bibinfo {author} {\bibfnamefont {J.~S.}\ \bibnamefont {de~Cani}},
  \ and\ \bibinfo {author} {\bibfnamefont {R.~E.}\ \bibnamefont {Davies}},\
  }\href {\doibase 10.1002/asi.4630320102} {\bibfield  {journal} {\bibinfo
  {journal} {J. Am. Soc. Inf. Sci.}\ }\textbf {\bibinfo {volume} {32}},\
  \bibinfo {pages} {1–15} (\bibinfo {year} {1981})}\BibitemShut {NoStop}%
\bibitem [{\citenamefont {Stinson}\ and\ \citenamefont
  {Lancaster}(1987)}]{Stinson1987}%
  \BibitemOpen
  \bibfield  {author} {\bibinfo {author} {\bibfnamefont {E.~R.}\ \bibnamefont
  {Stinson}}\ and\ \bibinfo {author} {\bibfnamefont {F.}~\bibnamefont
  {Lancaster}},\ }\href {http://jis.sagepub.com/content/13/2/65.abstract N2 -
  Using the literature of human and medical genetics, the results of a
  synchronous citation study of obsolescence over a 19-year period were
  compared with the results of a di achronous citation study. If the first two
  years of synchronous data are excluded, the rate of obsolescence measured
  synchro nously is statistically equivalent to the rate of obsolescence
  measured diachronously.The assumption that synchronous studies need to be cor
  rected to account for the growth of the literature was tested. The data
  collected support the hypothesis put forward by Brookes that the growth of
  the literature and the growth of the number of contributors to that
  literature have opposite effects in the measurement of obsolescence. The
  results of a synchro nous study corrected for the growth of the literature
  and also for the growth in number of contributors were statistically
  equivalent to synchronous data with no corrections whatever.} {\bibfield
  {journal} {\bibinfo  {journal} {Journal of Information Science}\ }\textbf
  {\bibinfo {volume} {13}},\ \bibinfo {pages} {65} (\bibinfo {year}
  {1987})}\BibitemShut {NoStop}%
\bibitem [{\citenamefont {Nakamoto}(1988)}]{Nakamoto1988}%
  \BibitemOpen
  \bibfield  {author} {\bibinfo {author} {\bibfnamefont {H.}~\bibnamefont
  {Nakamoto}},\ }\href {http://hdl.handle.net/1942/837} {\bibfield  {journal}
  {\bibinfo  {journal} {Informetrics}\ }\textbf {\bibinfo {volume} {87/88}},\
  \bibinfo {pages} {157} (\bibinfo {year} {1988})}\BibitemShut {NoStop}%
\bibitem [{\citenamefont {Redner}(2004)}]{Redner2004}%
  \BibitemOpen
  \bibfield  {author} {\bibinfo {author} {\bibfnamefont {S.}~\bibnamefont
  {Redner}},\ }\href@noop {} {\enquote {\bibinfo {title} {Citation statistics
  from more than a century of physical review},}\ }\bibinfo {howpublished}
  {arXiv:physics/0407137} (\bibinfo {year} {2004})\BibitemShut {NoStop}%
\bibitem [{\citenamefont {Roth}\ \emph {et~al.}(2012)\citenamefont {Roth},
  \citenamefont {Wu},\ and\ \citenamefont {Lozano}}]{Roth2012}%
  \BibitemOpen
  \bibfield  {author} {\bibinfo {author} {\bibfnamefont {C.}~\bibnamefont
  {Roth}}, \bibinfo {author} {\bibfnamefont {J.}~\bibnamefont {Wu}}, \ and\
  \bibinfo {author} {\bibfnamefont {S.}~\bibnamefont {Lozano}},\ }\href
  {\doibase http://dx.doi.org/10.1016/j.joi.2011.08.005} {\bibfield  {journal}
  {\bibinfo  {journal} {Journal of Informetrics}\ }\textbf {\bibinfo {volume}
  {6}},\ \bibinfo {pages} {111 } (\bibinfo {year} {2012})}\BibitemShut
  {NoStop}%
\bibitem [{\citenamefont {Glanzel}(2004)}]{Glanzel2004}%
  \BibitemOpen
  \bibfield  {author} {\bibinfo {author} {\bibfnamefont {W.}~\bibnamefont
  {Glanzel}},\ }\href {\doibase 10.1023/B:SCIE.0000034391.06240.2a} {\bibfield
  {journal} {\bibinfo  {journal} {Scientometrics}\ }\textbf {\bibinfo {volume}
  {60}},\ \bibinfo {pages} {511} (\bibinfo {year} {2004})}\BibitemShut
  {NoStop}%
\bibitem [{\citenamefont {Bouabid}\ and\ \citenamefont
  {Larivi\`{e}re}(2013)}]{Bouabid2013}%
  \BibitemOpen
  \bibfield  {author} {\bibinfo {author} {\bibfnamefont {H.}~\bibnamefont
  {Bouabid}}\ and\ \bibinfo {author} {\bibfnamefont {V.}~\bibnamefont
  {Larivi\`{e}re}},\ }\href {\doibase 10.1007/s11192-013-0995-7} {\bibfield
  {journal} {\bibinfo  {journal} {Scientometrics}\ }\textbf {\bibinfo {volume}
  {97}},\ \bibinfo {pages} {695} (\bibinfo {year} {2013})}\BibitemShut
  {NoStop}%
\bibitem [{\citenamefont {Parongama~Sen}(2014)}]{Sen2014}%
  \BibitemOpen
  \bibfield  {author} {\bibinfo {author} {\bibfnamefont {B.~K.~C.}\
  \bibnamefont {Parongama~Sen}},\ }\href
  {http://ukcatalogue.oup.com/product/9780199662456.do} {\emph {\bibinfo
  {title} {Sociophysics: An Introduction}}}\ (\bibinfo  {publisher} {Oxford
  University Press},\ \bibinfo {year} {2014})\BibitemShut {NoStop}%
\bibitem [{\citenamefont {Clough}\ \emph {et~al.}(2014)\citenamefont {Clough},
  \citenamefont {Gollings}, \citenamefont {Loach},\ and\ \citenamefont
  {Evans}}]{Clough2014}%
  \BibitemOpen
  \bibfield  {author} {\bibinfo {author} {\bibfnamefont {J.~R.}\ \bibnamefont
  {Clough}}, \bibinfo {author} {\bibfnamefont {J.}~\bibnamefont {Gollings}},
  \bibinfo {author} {\bibfnamefont {T.~V.}\ \bibnamefont {Loach}}, \ and\
  \bibinfo {author} {\bibfnamefont {T.~S.}\ \bibnamefont {Evans}},\ }\href
  {\doibase 10.1093/comnet/cnu039} {\bibfield  {journal} {\bibinfo  {journal}
  {Journal of Complex Networks}\ }\textbf {\bibinfo {volume} {3}},\ \bibinfo
  {pages} {189} (\bibinfo {year} {2014})}\BibitemShut {NoStop}%
\bibitem [{\citenamefont {Chen}\ \emph {et~al.}(2007)\citenamefont {Chen},
  \citenamefont {Xie}, \citenamefont {Maslov},\ and\ \citenamefont
  {Redner}}]{Chen2007}%
  \BibitemOpen
  \bibfield  {author} {\bibinfo {author} {\bibfnamefont {P.}~\bibnamefont
  {Chen}}, \bibinfo {author} {\bibfnamefont {H.}~\bibnamefont {Xie}}, \bibinfo
  {author} {\bibfnamefont {S.}~\bibnamefont {Maslov}}, \ and\ \bibinfo {author}
  {\bibfnamefont {S.}~\bibnamefont {Redner}},\ }\href {\doibase
  10.1016/j.joi.2006.06.001} {\bibfield  {journal} {\bibinfo  {journal}
  {Journal of Informetrics}\ }\textbf {\bibinfo {volume} {1}},\ \bibinfo
  {pages} {8} (\bibinfo {year} {2007})}\BibitemShut {NoStop}%
\bibitem [{\citenamefont {Simkin}\ and\ \citenamefont
  {Roychowdhury}(2007)}]{Simkin2007}%
  \BibitemOpen
  \bibfield  {author} {\bibinfo {author} {\bibfnamefont {M.~V.}\ \bibnamefont
  {Simkin}}\ and\ \bibinfo {author} {\bibfnamefont {V.~P.}\ \bibnamefont
  {Roychowdhury}},\ }\href {\doibase 10.1002/asi.20653} {\bibfield  {journal}
  {\bibinfo  {journal} {Journal of the American Society for Information Science
  and Technology}\ }\textbf {\bibinfo {volume} {58}},\ \bibinfo {pages} {1661}
  (\bibinfo {year} {2007})}\BibitemShut {NoStop}%
\bibitem [{\citenamefont {Broder}\ \emph {et~al.}(2000)\citenamefont {Broder},
  \citenamefont {Kumar}, \citenamefont {Maghoul}, \citenamefont {Raghavan},
  \citenamefont {Rajagopalan}, \citenamefont {Stata}, \citenamefont {Tomkins},\
  and\ \citenamefont {Wiener}}]{Broder2000}%
  \BibitemOpen
  \bibfield  {author} {\bibinfo {author} {\bibfnamefont {A.}~\bibnamefont
  {Broder}}, \bibinfo {author} {\bibfnamefont {R.}~\bibnamefont {Kumar}},
  \bibinfo {author} {\bibfnamefont {F.}~\bibnamefont {Maghoul}}, \bibinfo
  {author} {\bibfnamefont {P.}~\bibnamefont {Raghavan}}, \bibinfo {author}
  {\bibfnamefont {S.}~\bibnamefont {Rajagopalan}}, \bibinfo {author}
  {\bibfnamefont {R.}~\bibnamefont {Stata}}, \bibinfo {author} {\bibfnamefont
  {A.}~\bibnamefont {Tomkins}}, \ and\ \bibinfo {author} {\bibfnamefont
  {J.}~\bibnamefont {Wiener}},\ }\href {\doibase 10.1016/s1389-1286(00)00083-9}
  {\bibfield  {journal} {\bibinfo  {journal} {Computer Networks}\ }\textbf
  {\bibinfo {volume} {33}},\ \bibinfo {pages} {309} (\bibinfo {year}
  {2000})}\BibitemShut {NoStop}%
\bibitem [{\citenamefont {Golosovsky}\ and\ \citenamefont
  {Solomon}(2012{\natexlab{a}})}]{Golosovsky2012}%
  \BibitemOpen
  \bibfield  {author} {\bibinfo {author} {\bibfnamefont {M.}~\bibnamefont
  {Golosovsky}}\ and\ \bibinfo {author} {\bibfnamefont {S.}~\bibnamefont
  {Solomon}},\ }\href {\doibase 10.1103/physrevlett.109.098701} {\bibfield
  {journal} {\bibinfo  {journal} {Physical Review Letters}\ }\textbf {\bibinfo
  {volume} {109}},\ \bibinfo {pages} {098701} (\bibinfo {year}
  {2012}{\natexlab{a}})}\BibitemShut {NoStop}%
\bibitem [{\citenamefont {Rosvall}\ \emph {et~al.}(2014)\citenamefont
  {Rosvall}, \citenamefont {Esquivel}, \citenamefont {Lancichinetti},
  \citenamefont {West},\ and\ \citenamefont {Lambiotte}}]{Rosvall2014}%
  \BibitemOpen
  \bibfield  {author} {\bibinfo {author} {\bibfnamefont {M.}~\bibnamefont
  {Rosvall}}, \bibinfo {author} {\bibfnamefont {A.~V.}\ \bibnamefont
  {Esquivel}}, \bibinfo {author} {\bibfnamefont {A.}~\bibnamefont
  {Lancichinetti}}, \bibinfo {author} {\bibfnamefont {J.~D.}\ \bibnamefont
  {West}}, \ and\ \bibinfo {author} {\bibfnamefont {R.}~\bibnamefont
  {Lambiotte}},\ }\href {http://dx.doi.org/10.1038/ncomms5630} {\bibfield
  {journal} {\bibinfo  {journal} {Nat Commun}\ }\textbf {\bibinfo {volume}
  {5}},\  (\bibinfo {year} {2014})}\BibitemShut {NoStop}%
\bibitem [{\citenamefont {Harris}(2002)}]{Harris2002}%
  \BibitemOpen
  \bibfield  {author} {\bibinfo {author} {\bibfnamefont {T.~E.}\ \bibnamefont
  {Harris}},\ }\href@noop {} {\emph {\bibinfo {title} {The theory of branching
  processes}}}\ (\bibinfo  {publisher} {Courier Corporation},\ \bibinfo {year}
  {2002})\BibitemShut {NoStop}%
\bibitem [{\citenamefont {Ebeling}\ \emph {et~al.}(1986)\citenamefont
  {Ebeling}, \citenamefont {Engel},\ and\ \citenamefont
  {Mazenko}}]{Ebeling1986}%
  \BibitemOpen
  \bibfield  {author} {\bibinfo {author} {\bibfnamefont {W.}~\bibnamefont
  {Ebeling}}, \bibinfo {author} {\bibfnamefont {A.}~\bibnamefont {Engel}}, \
  and\ \bibinfo {author} {\bibfnamefont {V.~G.}\ \bibnamefont {Mazenko}},\
  }\href {http://www.sciencedirect.com/science/article/pii/0303264786900407}
  {\bibfield  {journal} {\bibinfo  {journal} {Biosystems}\ }\textbf {\bibinfo
  {volume} {19}},\ \bibinfo {pages} {213} (\bibinfo {year} {1986})}\BibitemShut
  {NoStop}%
\bibitem [{\citenamefont {Crane}\ and\ \citenamefont
  {Sornette}(2008)}]{Crane2008}%
  \BibitemOpen
  \bibfield  {author} {\bibinfo {author} {\bibfnamefont {R.}~\bibnamefont
  {Crane}}\ and\ \bibinfo {author} {\bibfnamefont {D.}~\bibnamefont
  {Sornette}},\ }\href {\doibase 10.1073/pnas.0803685105} {\bibfield  {journal}
  {\bibinfo  {journal} {Proceedings of the National Academy of Sciences}\
  }\textbf {\bibinfo {volume} {105}},\ \bibinfo {pages} {15649} (\bibinfo
  {year} {2008})}\BibitemShut {NoStop}%
\bibitem [{\citenamefont {Cheng}\ \emph {et~al.}(2014)\citenamefont {Cheng},
  \citenamefont {Adamic}, \citenamefont {Dow}, \citenamefont {Kleinberg},\ and\
  \citenamefont {Leskovec}}]{Cheng2014}%
  \BibitemOpen
  \bibfield  {author} {\bibinfo {author} {\bibfnamefont {J.}~\bibnamefont
  {Cheng}}, \bibinfo {author} {\bibfnamefont {L.}~\bibnamefont {Adamic}},
  \bibinfo {author} {\bibfnamefont {P.~A.}\ \bibnamefont {Dow}}, \bibinfo
  {author} {\bibfnamefont {J.~M.}\ \bibnamefont {Kleinberg}}, \ and\ \bibinfo
  {author} {\bibfnamefont {J.}~\bibnamefont {Leskovec}},\ }in\ \href {\doibase
  10.1145/2566486.2567997} {\emph {\bibinfo {booktitle} {Proceedings of the
  23rd international conference on World wide web}}}\ (\bibinfo  {publisher}
  {Association for Computing Machinery ({ACM})},\ \bibinfo {year}
  {2014})\BibitemShut {NoStop}%
\bibitem [{\citenamefont {Iribarren}\ and\ \citenamefont
  {Moro}(2011)}]{Iribarren2011}%
  \BibitemOpen
  \bibfield  {author} {\bibinfo {author} {\bibfnamefont {J.~L.}\ \bibnamefont
  {Iribarren}}\ and\ \bibinfo {author} {\bibfnamefont {E.}~\bibnamefont
  {Moro}},\ }\href {\doibase 10.1103/physreve.84.046116} {\bibfield  {journal}
  {\bibinfo  {journal} {Phys. Rev. E}\ }\textbf {\bibinfo {volume} {84}},\
  \bibinfo {pages} {046116} (\bibinfo {year} {2011})}\BibitemShut {NoStop}%
\bibitem [{\citenamefont {Medo}(2014)}]{Medo2014}%
  \BibitemOpen
  \bibfield  {author} {\bibinfo {author} {\bibfnamefont {M.}~\bibnamefont
  {Medo}},\ }\href {\doibase 10.1103/physreve.89.032801} {\bibfield  {journal}
  {\bibinfo  {journal} {Physical Review E}\ }\textbf {\bibinfo {volume} {89}},\
  \bibinfo {pages} {032801} (\bibinfo {year} {2014})}\BibitemShut {NoStop}%
\bibitem [{\citenamefont {Centola}(2010)}]{Centola2010}%
  \BibitemOpen
  \bibfield  {author} {\bibinfo {author} {\bibfnamefont {D.}~\bibnamefont
  {Centola}},\ }\href {\doibase 10.1126/science.1185231} {\bibfield  {journal}
  {\bibinfo  {journal} {Science}\ }\textbf {\bibinfo {volume} {329}},\ \bibinfo
  {pages} {1194} (\bibinfo {year} {2010})}\BibitemShut {NoStop}%
\bibitem [{\citenamefont {Wang}\ \emph {et~al.}(2008)\citenamefont {Wang},
  \citenamefont {Yu},\ and\ \citenamefont {Yu}}]{Wang2008}%
  \BibitemOpen
  \bibfield  {author} {\bibinfo {author} {\bibfnamefont {M.}~\bibnamefont
  {Wang}}, \bibinfo {author} {\bibfnamefont {G.}~\bibnamefont {Yu}}, \ and\
  \bibinfo {author} {\bibfnamefont {D.}~\bibnamefont {Yu}},\ }\href {\doibase
  10.1016/j.physa.2008.03.017} {\bibfield  {journal} {\bibinfo  {journal}
  {Physica A: Statistical Mechanics and its Applications}\ }\textbf {\bibinfo
  {volume} {387}},\ \bibinfo {pages} {4692} (\bibinfo {year}
  {2008})}\BibitemShut {NoStop}%
\bibitem [{\citenamefont {Gleeson}\ \emph {et~al.}(2014)\citenamefont
  {Gleeson}, \citenamefont {Cellai}, \citenamefont {Onnela}, \citenamefont
  {Porter},\ and\ \citenamefont {Reed-Tsochas}}]{Gleeson2014}%
  \BibitemOpen
  \bibfield  {author} {\bibinfo {author} {\bibfnamefont {J.~P.}\ \bibnamefont
  {Gleeson}}, \bibinfo {author} {\bibfnamefont {D.}~\bibnamefont {Cellai}},
  \bibinfo {author} {\bibfnamefont {J.-P.}\ \bibnamefont {Onnela}}, \bibinfo
  {author} {\bibfnamefont {M.~A.}\ \bibnamefont {Porter}}, \ and\ \bibinfo
  {author} {\bibfnamefont {F.}~\bibnamefont {Reed-Tsochas}},\ }\href {\doibase
  10.1073/pnas.1313895111} {\bibfield  {journal} {\bibinfo  {journal}
  {Proceedings of the National Academy of Sciences}\ }\textbf {\bibinfo
  {volume} {111}},\ \bibinfo {pages} {10411} (\bibinfo {year}
  {2014})}\BibitemShut {NoStop}%
\bibitem [{\citenamefont {Golosovsky}\ and\ \citenamefont
  {Solomon}(2012{\natexlab{b}})}]{Golosovsky2012a}%
  \BibitemOpen
  \bibfield  {author} {\bibinfo {author} {\bibfnamefont {M.}~\bibnamefont
  {Golosovsky}}\ and\ \bibinfo {author} {\bibfnamefont {S.}~\bibnamefont
  {Solomon}},\ }\href {\doibase 10.1140/epjst/e2012-01576-4} {\bibfield
  {journal} {\bibinfo  {journal} {The European Physical Journal Special
  Topics}\ }\textbf {\bibinfo {volume} {205}},\ \bibinfo {pages} {303}
  (\bibinfo {year} {2012}{\natexlab{b}})}\BibitemShut {NoStop}%
\bibitem [{\citenamefont {Caldarelli}\ \emph {et~al.}(2002)\citenamefont
  {Caldarelli}, \citenamefont {Capocci}, \citenamefont {Rios},\ and\
  \citenamefont {Mu{\~{n}}oz}}]{Caldarelli2002}%
  \BibitemOpen
  \bibfield  {author} {\bibinfo {author} {\bibfnamefont {G.}~\bibnamefont
  {Caldarelli}}, \bibinfo {author} {\bibfnamefont {A.}~\bibnamefont {Capocci}},
  \bibinfo {author} {\bibfnamefont {P.~D.~L.}\ \bibnamefont {Rios}}, \ and\
  \bibinfo {author} {\bibfnamefont {M.~A.}\ \bibnamefont {Mu{\~{n}}oz}},\
  }\href {\doibase 10.1103/physrevlett.89.258702} {\bibfield  {journal}
  {\bibinfo  {journal} {Phys. Rev. Lett.}\ }\textbf {\bibinfo {volume} {89}},\
  \bibinfo {pages} {258702} (\bibinfo {year} {2002})}\BibitemShut {NoStop}%
\bibitem [{\citenamefont {Cohen}\ \emph {et~al.}(2003)\citenamefont {Cohen},
  \citenamefont {Havlin},\ and\ \citenamefont {ben Avraham}}]{Cohen2003}%
  \BibitemOpen
  \bibfield  {author} {\bibinfo {author} {\bibfnamefont {R.}~\bibnamefont
  {Cohen}}, \bibinfo {author} {\bibfnamefont {S.}~\bibnamefont {Havlin}}, \
  and\ \bibinfo {author} {\bibfnamefont {D.}~\bibnamefont {ben Avraham}},\
  }\href {\doibase 10.1103/physrevlett.91.247901} {\bibfield  {journal}
  {\bibinfo  {journal} {Phys. Rev. Lett.}\ }\textbf {\bibinfo {volume} {91}},\
  \bibinfo {pages} {247901} (\bibinfo {year} {2003})}\BibitemShut {NoStop}%
\bibitem [{\citenamefont {Stringer}\ \emph {et~al.}(2008)\citenamefont
  {Stringer}, \citenamefont {Sales-Pardo},\ and\ \citenamefont
  {Amaral}}]{Stringer2008}%
  \BibitemOpen
  \bibfield  {author} {\bibinfo {author} {\bibfnamefont {M.~J.}\ \bibnamefont
  {Stringer}}, \bibinfo {author} {\bibfnamefont {M.}~\bibnamefont
  {Sales-Pardo}}, \ and\ \bibinfo {author} {\bibfnamefont {L.~A.~N.}\
  \bibnamefont {Amaral}},\ }\href {\doibase 10.1371/journal.pone.0001683}
  {\bibfield  {journal} {\bibinfo  {journal} {{PLoS} {ONE}}\ }\textbf {\bibinfo
  {volume} {3}},\ \bibinfo {pages} {e1683} (\bibinfo {year}
  {2008})}\BibitemShut {NoStop}%
\bibitem [{\citenamefont {Uzzi}\ \emph {et~al.}(2013)\citenamefont {Uzzi},
  \citenamefont {Mukherjee}, \citenamefont {Stringer},\ and\ \citenamefont
  {Jones}}]{Uzzi2013}%
  \BibitemOpen
  \bibfield  {author} {\bibinfo {author} {\bibfnamefont {B.}~\bibnamefont
  {Uzzi}}, \bibinfo {author} {\bibfnamefont {S.}~\bibnamefont {Mukherjee}},
  \bibinfo {author} {\bibfnamefont {M.}~\bibnamefont {Stringer}}, \ and\
  \bibinfo {author} {\bibfnamefont {B.}~\bibnamefont {Jones}},\ }\href
  {http://www.sciencemag.org/content/342/6157/468.abstract N2 - Novelty is an
  essential feature of creative ideas, yet the building blocks of new ideas are
  often embodied in existing knowledge. From this perspective, balancing
  atypical knowledge with conventional knowledge may be critical to the link
  between innovativeness and impact. Our analysis of 17.9 million} {\bibfield
  {journal} {\bibinfo  {journal} {Science}\ }\textbf {\bibinfo {volume}
  {342}},\ \bibinfo {pages} {468} (\bibinfo {year} {2013})}\BibitemShut
  {NoStop}%
\bibitem [{\citenamefont {Acuna}\ \emph {et~al.}(2012)\citenamefont {Acuna},
  \citenamefont {Allesina},\ and\ \citenamefont {Kording}}]{Acuna2012}%
  \BibitemOpen
  \bibfield  {author} {\bibinfo {author} {\bibfnamefont {D.~E.}\ \bibnamefont
  {Acuna}}, \bibinfo {author} {\bibfnamefont {S.}~\bibnamefont {Allesina}}, \
  and\ \bibinfo {author} {\bibfnamefont {K.~P.}\ \bibnamefont {Kording}},\
  }\href {http://dx.doi.org/10.1038/489201a} {\bibfield  {journal} {\bibinfo
  {journal} {Nature}\ }\textbf {\bibinfo {volume} {489}},\ \bibinfo {pages}
  {201} (\bibinfo {year} {2012})}\BibitemShut {NoStop}%
\bibitem [{\citenamefont {Mazloumian}(2012)}]{Mazloumian2012}%
  \BibitemOpen
  \bibfield  {author} {\bibinfo {author} {\bibfnamefont {A.}~\bibnamefont
  {Mazloumian}},\ }\href {\doibase 10.1371/journal.pone.0049246} {\bibfield
  {journal} {\bibinfo  {journal} {PLoS ONE}\ }\textbf {\bibinfo {volume} {7}},\
  \bibinfo {pages} {e49246} (\bibinfo {year} {2012})}\BibitemShut {NoStop}%
\bibitem [{\citenamefont {Penner}\ \emph {et~al.}(2013)\citenamefont {Penner},
  \citenamefont {Pan}, \citenamefont {Petersen}, \citenamefont {Kaski},\ and\
  \citenamefont {Fortunato}}]{Penner2013}%
  \BibitemOpen
  \bibfield  {author} {\bibinfo {author} {\bibfnamefont {O.}~\bibnamefont
  {Penner}}, \bibinfo {author} {\bibfnamefont {R.~K.}\ \bibnamefont {Pan}},
  \bibinfo {author} {\bibfnamefont {A.~M.}\ \bibnamefont {Petersen}}, \bibinfo
  {author} {\bibfnamefont {K.}~\bibnamefont {Kaski}}, \ and\ \bibinfo {author}
  {\bibfnamefont {S.}~\bibnamefont {Fortunato}},\ }\href
  {http://dx.doi.org/10.1038/srep03052} {\bibfield  {journal} {\bibinfo
  {journal} {Scientific Reports}\ }\textbf {\bibinfo {volume} {3}},\ \bibinfo
  {pages} {3052} (\bibinfo {year} {2013})}\BibitemShut {NoStop}%
\bibitem [{\citenamefont {Newman}(2014)}]{Newman2014}%
  \BibitemOpen
  \bibfield  {author} {\bibinfo {author} {\bibfnamefont {M.~E.~J.}\
  \bibnamefont {Newman}},\ }\href {\doibase 10.1209/0295-5075/105/28002}
  {\bibfield  {journal} {\bibinfo  {journal} {{EPL} (Europhysics Letters)}\
  }\textbf {\bibinfo {volume} {105}},\ \bibinfo {pages} {28002} (\bibinfo
  {year} {2014})}\BibitemShut {NoStop}%
\bibitem [{\citenamefont {Ponomarev}\ \emph {et~al.}(2014)\citenamefont
  {Ponomarev}, \citenamefont {Williams}, \citenamefont {Hackett}, \citenamefont
  {Schnell},\ and\ \citenamefont {Haak}}]{Ponomarev2014}%
  \BibitemOpen
  \bibfield  {author} {\bibinfo {author} {\bibfnamefont {I.~V.}\ \bibnamefont
  {Ponomarev}}, \bibinfo {author} {\bibfnamefont {D.~E.}\ \bibnamefont
  {Williams}}, \bibinfo {author} {\bibfnamefont {C.~J.}\ \bibnamefont
  {Hackett}}, \bibinfo {author} {\bibfnamefont {J.~D.}\ \bibnamefont
  {Schnell}}, \ and\ \bibinfo {author} {\bibfnamefont {L.~L.}\ \bibnamefont
  {Haak}},\ }\href {\doibase 10.1016/j.techfore.2012.09.017} {\bibfield
  {journal} {\bibinfo  {journal} {Technological Forecasting and Social Change}\
  }\textbf {\bibinfo {volume} {81}},\ \bibinfo {pages} {49} (\bibinfo {year}
  {2014})}\BibitemShut {NoStop}%
\bibitem [{\citenamefont {Ke}\ \emph {et~al.}(2015)\citenamefont {Ke},
  \citenamefont {Ferrara}, \citenamefont {Radicchi},\ and\ \citenamefont
  {Flammini}}]{Ke2015}%
  \BibitemOpen
  \bibfield  {author} {\bibinfo {author} {\bibfnamefont {Q.}~\bibnamefont
  {Ke}}, \bibinfo {author} {\bibfnamefont {E.}~\bibnamefont {Ferrara}},
  \bibinfo {author} {\bibfnamefont {F.}~\bibnamefont {Radicchi}}, \ and\
  \bibinfo {author} {\bibfnamefont {A.}~\bibnamefont {Flammini}},\ }\href
  {\doibase 10.1073/pnas.1424329112} {\bibfield  {journal} {\bibinfo  {journal}
  {Proceedings of the National Academy of Sciences}\ }\textbf {\bibinfo
  {volume} {112}},\ \bibinfo {pages} {7426} (\bibinfo {year}
  {2015})}\BibitemShut {NoStop}%
\bibitem [{\citenamefont {Onaga}\ and\ \citenamefont
  {Shinomoto}(2014)}]{Onaga2014}%
  \BibitemOpen
  \bibfield  {author} {\bibinfo {author} {\bibfnamefont {T.}~\bibnamefont
  {Onaga}}\ and\ \bibinfo {author} {\bibfnamefont {S.}~\bibnamefont
  {Shinomoto}},\ }\href {\doibase 10.1103/physreve.89.042817} {\bibfield
  {journal} {\bibinfo  {journal} {Physical Review E}\ }\textbf {\bibinfo
  {volume} {89}},\ \bibinfo {pages} {042817} (\bibinfo {year}
  {2014})}\BibitemShut {NoStop}%
\bibitem [{\citenamefont {Kuhn}(1970)}]{Kuhn1970}%
  \BibitemOpen
  \bibfield  {author} {\bibinfo {author} {\bibfnamefont {T.~S.}\ \bibnamefont
  {Kuhn}},\ }\href@noop {} {\emph {\bibinfo {title} {The structure of
  scientific revolutions}}}\ (\bibinfo  {publisher} {University of Chicago
  Press Ltd},\ \bibinfo {year} {1970})\BibitemShut {NoStop}%
\bibitem [{\citenamefont {Ishii}\ \emph {et~al.}(2012)\citenamefont {Ishii},
  \citenamefont {Arakaki}, \citenamefont {Matsuda}, \citenamefont {Umemura},
  \citenamefont {Urushidani}, \citenamefont {Yamagata},\ and\ \citenamefont
  {Yoshida}}]{Ishii2012}%
  \BibitemOpen
  \bibfield  {author} {\bibinfo {author} {\bibfnamefont {A.}~\bibnamefont
  {Ishii}}, \bibinfo {author} {\bibfnamefont {H.}~\bibnamefont {Arakaki}},
  \bibinfo {author} {\bibfnamefont {N.}~\bibnamefont {Matsuda}}, \bibinfo
  {author} {\bibfnamefont {S.}~\bibnamefont {Umemura}}, \bibinfo {author}
  {\bibfnamefont {T.}~\bibnamefont {Urushidani}}, \bibinfo {author}
  {\bibfnamefont {N.}~\bibnamefont {Yamagata}}, \ and\ \bibinfo {author}
  {\bibfnamefont {N.}~\bibnamefont {Yoshida}},\ }\href {\doibase
  10.1088/1367-2630/14/6/063018} {\bibfield  {journal} {\bibinfo  {journal}
  {New Journal of Physics}\ }\textbf {\bibinfo {volume} {14}},\ \bibinfo
  {pages} {063018} (\bibinfo {year} {2012})}\BibitemShut {NoStop}%
\bibitem [{\citenamefont {Leskovec}\ \emph {et~al.}(2007)\citenamefont
  {Leskovec}, \citenamefont {Adamic},\ and\ \citenamefont
  {Huberman}}]{Leskovec2007}%
  \BibitemOpen
  \bibfield  {author} {\bibinfo {author} {\bibfnamefont {J.}~\bibnamefont
  {Leskovec}}, \bibinfo {author} {\bibfnamefont {L.~A.}\ \bibnamefont
  {Adamic}}, \ and\ \bibinfo {author} {\bibfnamefont {B.~A.}\ \bibnamefont
  {Huberman}},\ }\href {\doibase 10.1145/1232722.1232727} {\bibfield  {journal}
  {\bibinfo  {journal} {{ACM} Trans. Web}\ }\textbf {\bibinfo {volume} {1}},\
  \bibinfo {pages} {5} (\bibinfo {year} {2007})}\BibitemShut {NoStop}%
\bibitem [{\citenamefont {Pastor-Satorras}\ \emph {et~al.}(2015)\citenamefont
  {Pastor-Satorras}, \citenamefont {Castellano}, \citenamefont {Mieghem},\ and\
  \citenamefont {Vespignani}}]{Pastor2015}%
  \BibitemOpen
  \bibfield  {author} {\bibinfo {author} {\bibfnamefont {R.}~\bibnamefont
  {Pastor-Satorras}}, \bibinfo {author} {\bibfnamefont {C.}~\bibnamefont
  {Castellano}}, \bibinfo {author} {\bibfnamefont {P.~V.}\ \bibnamefont
  {Mieghem}}, \ and\ \bibinfo {author} {\bibfnamefont {A.}~\bibnamefont
  {Vespignani}},\ }\href {\doibase 10.1103/revmodphys.87.925} {\bibfield
  {journal} {\bibinfo  {journal} {Reviews of Modern Physics}\ }\textbf
  {\bibinfo {volume} {87}},\ \bibinfo {pages} {925} (\bibinfo {year}
  {2015})}\BibitemShut {NoStop}%
\bibitem [{\citenamefont {Centola}\ \emph {et~al.}(2007)\citenamefont
  {Centola}, \citenamefont {Egu{\'{\i}}luz},\ and\ \citenamefont
  {Macy}}]{Centola2007}%
  \BibitemOpen
  \bibfield  {author} {\bibinfo {author} {\bibfnamefont {D.}~\bibnamefont
  {Centola}}, \bibinfo {author} {\bibfnamefont {V.~M.}\ \bibnamefont
  {Egu{\'{\i}}luz}}, \ and\ \bibinfo {author} {\bibfnamefont {M.~W.}\
  \bibnamefont {Macy}},\ }\href {\doibase 10.1016/j.physa.2006.06.018}
  {\bibfield  {journal} {\bibinfo  {journal} {Physica A: Statistical Mechanics
  and its Applications}\ }\textbf {\bibinfo {volume} {374}},\ \bibinfo {pages}
  {449} (\bibinfo {year} {2007})}\BibitemShut {NoStop}%
\bibitem [{\citenamefont {P{\'{e}}rez-Reche}\ \emph {et~al.}(2011)\citenamefont
  {P{\'{e}}rez-Reche}, \citenamefont {Ludlam}, \citenamefont {Taraskin},\ and\
  \citenamefont {Gilligan}}]{Perez2011}%
  \BibitemOpen
  \bibfield  {author} {\bibinfo {author} {\bibfnamefont {F.~J.}\ \bibnamefont
  {P{\'{e}}rez-Reche}}, \bibinfo {author} {\bibfnamefont {J.~J.}\ \bibnamefont
  {Ludlam}}, \bibinfo {author} {\bibfnamefont {S.~N.}\ \bibnamefont
  {Taraskin}}, \ and\ \bibinfo {author} {\bibfnamefont {C.~A.}\ \bibnamefont
  {Gilligan}},\ }\href {\doibase 10.1103/physrevlett.106.218701} {\bibfield
  {journal} {\bibinfo  {journal} {Phys. Rev. Lett.}\ }\textbf {\bibinfo
  {volume} {106}},\ \bibinfo {pages} {218701} (\bibinfo {year}
  {2011})}\BibitemShut {NoStop}%
\bibitem [{\citenamefont {Broder-Rodgers}\ \emph {et~al.}(2015)\citenamefont
  {Broder-Rodgers}, \citenamefont {P{\'{e}}rez-Reche},\ and\ \citenamefont
  {Taraskin}}]{Broder2015}%
  \BibitemOpen
  \bibfield  {author} {\bibinfo {author} {\bibfnamefont {D.}~\bibnamefont
  {Broder-Rodgers}}, \bibinfo {author} {\bibfnamefont {F.~J.}\ \bibnamefont
  {P{\'{e}}rez-Reche}}, \ and\ \bibinfo {author} {\bibfnamefont {S.~N.}\
  \bibnamefont {Taraskin}},\ }\href {\doibase 10.1103/physreve.92.062814}
  {\bibfield  {journal} {\bibinfo  {journal} {Physical Review E}\ }\textbf
  {\bibinfo {volume} {92}},\ \bibinfo {pages} {062814} (\bibinfo {year}
  {2015})}\BibitemShut {NoStop}%
\bibitem [{\citenamefont {Ludlam}\ \emph {et~al.}(2011)\citenamefont {Ludlam},
  \citenamefont {Gibson}, \citenamefont {Otten},\ and\ \citenamefont
  {Gilligan}}]{Ludlam2011}%
  \BibitemOpen
  \bibfield  {author} {\bibinfo {author} {\bibfnamefont {J.~J.}\ \bibnamefont
  {Ludlam}}, \bibinfo {author} {\bibfnamefont {G.~J.}\ \bibnamefont {Gibson}},
  \bibinfo {author} {\bibfnamefont {W.}~\bibnamefont {Otten}}, \ and\ \bibinfo
  {author} {\bibfnamefont {C.~A.}\ \bibnamefont {Gilligan}},\ }\href {\doibase
  10.1098/rsif.2011.0506} {\bibfield  {journal} {\bibinfo  {journal} {Journal
  of The Royal Society Interface}\ }\textbf {\bibinfo {volume} {9}},\ \bibinfo
  {pages} {949} (\bibinfo {year} {2011})}\BibitemShut {NoStop}%
\bibitem [{\citenamefont {Cs{\'{a}}rdi}\ \emph {et~al.}(2007)\citenamefont
  {Cs{\'{a}}rdi}, \citenamefont {Strandburg}, \citenamefont {Zal{\'{a}}nyi},
  \citenamefont {Tobochnik},\ and\ \citenamefont {{\'{E}}rdi}}]{Csardi2007}%
  \BibitemOpen
  \bibfield  {author} {\bibinfo {author} {\bibfnamefont {G.}~\bibnamefont
  {Cs{\'{a}}rdi}}, \bibinfo {author} {\bibfnamefont {K.~J.}\ \bibnamefont
  {Strandburg}}, \bibinfo {author} {\bibfnamefont {L.}~\bibnamefont
  {Zal{\'{a}}nyi}}, \bibinfo {author} {\bibfnamefont {J.}~\bibnamefont
  {Tobochnik}}, \ and\ \bibinfo {author} {\bibfnamefont {P.}~\bibnamefont
  {{\'{E}}rdi}},\ }\href {\doibase 10.1016/j.physa.2006.08.022} {\bibfield
  {journal} {\bibinfo  {journal} {Physica A: Statistical Mechanics and its
  Applications}\ }\textbf {\bibinfo {volume} {374}},\ \bibinfo {pages} {783}
  (\bibinfo {year} {2007})}\BibitemShut {NoStop}%
\bibitem [{\citenamefont {Valverde}\ \emph {et~al.}(2007)\citenamefont
  {Valverde}, \citenamefont {Sol{\'{e}}}, \citenamefont {Bedau},\ and\
  \citenamefont {Packard}}]{Valverde2007}%
  \BibitemOpen
  \bibfield  {author} {\bibinfo {author} {\bibfnamefont {S.}~\bibnamefont
  {Valverde}}, \bibinfo {author} {\bibfnamefont {R.~V.}\ \bibnamefont
  {Sol{\'{e}}}}, \bibinfo {author} {\bibfnamefont {M.~A.}\ \bibnamefont
  {Bedau}}, \ and\ \bibinfo {author} {\bibfnamefont {N.}~\bibnamefont
  {Packard}},\ }\href {\doibase 10.1103/physreve.76.056118} {\bibfield
  {journal} {\bibinfo  {journal} {Physical Review E}\ }\textbf {\bibinfo
  {volume} {76}},\ \bibinfo {pages} {056118} (\bibinfo {year}
  {2007})}\BibitemShut {NoStop}%
\bibitem [{\citenamefont {Sheridan}\ \emph {et~al.}(2012)\citenamefont
  {Sheridan}, \citenamefont {Yagahara},\ and\ \citenamefont
  {Shimodaira}}]{Sheridan2012}%
  \BibitemOpen
  \bibfield  {author} {\bibinfo {author} {\bibfnamefont {P.}~\bibnamefont
  {Sheridan}}, \bibinfo {author} {\bibfnamefont {Y.}~\bibnamefont {Yagahara}},
  \ and\ \bibinfo {author} {\bibfnamefont {H.}~\bibnamefont {Shimodaira}},\
  }\href {\doibase 10.1016/j.physa.2012.05.041} {\bibfield  {journal} {\bibinfo
   {journal} {Physica A: Statistical Mechanics and its Applications}\ }\textbf
  {\bibinfo {volume} {391}},\ \bibinfo {pages} {5031} (\bibinfo {year}
  {2012})}\BibitemShut {NoStop}%
\bibitem [{\citenamefont {Zhou}\ and\ \citenamefont
  {Mondragon}(2004)}]{Zhou2004}%
  \BibitemOpen
  \bibfield  {author} {\bibinfo {author} {\bibfnamefont {S.}~\bibnamefont
  {Zhou}}\ and\ \bibinfo {author} {\bibfnamefont {R.~J.}\ \bibnamefont
  {Mondragon}},\ }\href {\doibase 10.1103/PhysRevE.70.066108} {\bibfield
  {journal} {\bibinfo  {journal} {Phys. Rev. E}\ }\textbf {\bibinfo {volume}
  {70}},\ \bibinfo {pages} {066108} (\bibinfo {year} {2004})}\BibitemShut
  {NoStop}%
\bibitem [{\citenamefont {Larsen}\ and\ \citenamefont {von
  Ins}(2010)}]{Larsen2010}%
  \BibitemOpen
  \bibfield  {author} {\bibinfo {author} {\bibfnamefont {P.~O.}\ \bibnamefont
  {Larsen}}\ and\ \bibinfo {author} {\bibfnamefont {M.}~\bibnamefont {von
  Ins}},\ }\href {\doibase 10.1007/s11192-010-0202-z} {\bibfield  {journal}
  {\bibinfo  {journal} {Scientometrics}\ }\textbf {\bibinfo {volume} {84}},\
  \bibinfo {pages} {575} (\bibinfo {year} {2010})}\BibitemShut {NoStop}%
\bibitem [{\citenamefont {Evans}\ \emph {et~al.}(2012)\citenamefont {Evans},
  \citenamefont {Hopkins},\ and\ \citenamefont {Kaube}}]{Evans2012}%
  \BibitemOpen
  \bibfield  {author} {\bibinfo {author} {\bibfnamefont {T.~S.}\ \bibnamefont
  {Evans}}, \bibinfo {author} {\bibfnamefont {N.}~\bibnamefont {Hopkins}}, \
  and\ \bibinfo {author} {\bibfnamefont {B.~S.}\ \bibnamefont {Kaube}},\ }\href
  {\doibase 10.1007/s11192-012-0694-9} {\bibfield  {journal} {\bibinfo
  {journal} {Scientometrics}\ }\textbf {\bibinfo {volume} {93}},\ \bibinfo
  {pages} {473} (\bibinfo {year} {2012})}\BibitemShut {NoStop}%
\bibitem [{\citenamefont {Burrell}(2013)}]{Burrell2013}%
  \BibitemOpen
  \bibfield  {author} {\bibinfo {author} {\bibfnamefont {Q.~L.}\ \bibnamefont
  {Burrell}},\ }\href {\doibase 10.1016/j.joi.2013.03.001} {\bibfield
  {journal} {\bibinfo  {journal} {Journal of Informetrics}\ }\textbf {\bibinfo
  {volume} {7}},\ \bibinfo {pages} {676} (\bibinfo {year} {2013})}\BibitemShut
  {NoStop}%
\bibitem [{\citenamefont {Seglen}(1992)}]{Seglen1992}%
  \BibitemOpen
  \bibfield  {author} {\bibinfo {author} {\bibfnamefont {P.~O.}\ \bibnamefont
  {Seglen}},\ }\href {\doibase
  10.1002/(sici)1097-4571(199210)43:9<628::aid-asi5>3.0.co;2-0} {\bibfield
  {journal} {\bibinfo  {journal} {Journal of the American Society for
  Information Science}\ }\textbf {\bibinfo {volume} {43}},\ \bibinfo {pages}
  {628} (\bibinfo {year} {1992})}\BibitemShut {NoStop}%
\bibitem [{\citenamefont {Bagrow}\ and\ \citenamefont
  {Brockmann}(2013)}]{Bagrow2013}%
  \BibitemOpen
  \bibfield  {author} {\bibinfo {author} {\bibfnamefont {J.~P.}\ \bibnamefont
  {Bagrow}}\ and\ \bibinfo {author} {\bibfnamefont {D.}~\bibnamefont
  {Brockmann}},\ }\href {\doibase 10.1103/PhysRevX.3.021016} {\bibfield
  {journal} {\bibinfo  {journal} {Phys. Rev. X}\ }\textbf {\bibinfo {volume}
  {3}},\ \bibinfo {pages} {021016} (\bibinfo {year} {2013})}\BibitemShut
  {NoStop}%
\end{thebibliography}%
\pagebreak
\section{Supplementary Material}
\subsection{Number of publications}

\begin{figure}[!ht]
\includegraphics*[width=0.4\textwidth]{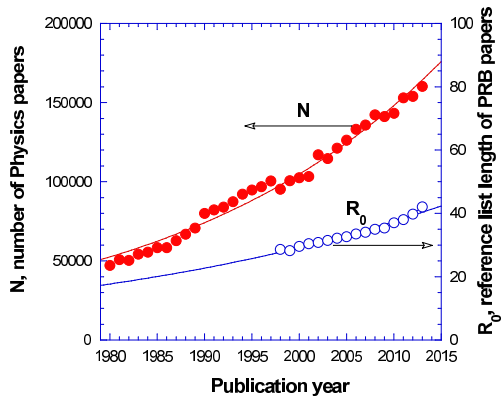}
\caption{Time dependence of the number of Physics papers published in 1980-1989 (full red circles). The red continuous line shows exponential approximation $N_{0}\propto e^{\alpha t}$ where $\alpha=0.046$. Open circles show time dependence of the average length of the reference list of a Physical Review paper. The blue continuous line shows exponential approximation $R_{0}\propto e^{\beta t}$ where $\beta=0.02$. 
}
\label{fig:N-R}
\end{figure}

To find how the number of publications depends on time we used Thomson-Reuters Web of Science and measured $N_{0}(t_{0})$, the total number of Physics papers  published annually during the period  1980-2013. Figure \ref{fig:N-R} shows exponential growth $e^{\alpha t_{0}}$ with $\alpha=0.046$ ($2\%$ annual growth) consistent with previous estimates of the growth of Physics publications in the period 1980-2010 \cite{Larsen2010,Martin2013}.

To find how the number of references depends on time we used Scopus database  and measured  $R_{0}(t_{0})$, the average length of the reference list of the Physical Review B papers published in  1996-2013. Figure \ref{fig:N-R} shows a very weak dependence  that agrees well with the measurements of Ref. \cite{Krapivsky2005,Martin2013} for the Physical Review papers. Ref. \cite{Krapivsky2005} claimed logarithmic time dependence while  Ref. \cite{Evans2012} claimed that there is a very slow growth of the reference list length before 2000 and subsequent acceleration following the advent of open access and electronic format journals that have no page limit. 
\subsection{Citation distribution of the Physical Review B papers represents the whole Physics field}
\begin{figure}[!ht]
\begin{center}
\includegraphics*[width=0.35\textwidth]{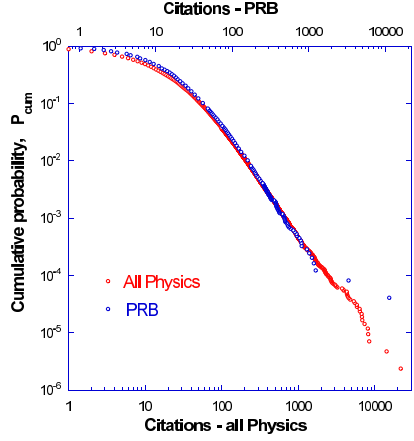}
\caption{Cumulative citation distribution  for  418,438 Physics papers published in 1980-1989. Citations were counted in July 2008. Blue points show corresponding distribution only for Physical Review B papers published in 1984. Both distributions are very similar and differ only in scale.
}
\label{fig:PRB-physics}
\end{center}
\end{figure}

We performed many measurements using the papers published in the Physical Review B. To what extent these measurements are generic, namely, do citation patterns of the PRB papers  represent Physics? Figure \ref{fig:PRB-physics} compares cumulative citation distributions of the  PRB papers and of all Physics papers published in 1980-1989. Although PRB papers garner $\sim$ 40 $\%$ more citations than an average Physics paper, citation distributions for the PRB papers and the whole Physics are very similar.

\begin{figure}[ht]
\includegraphics*[width=0.35\textwidth]{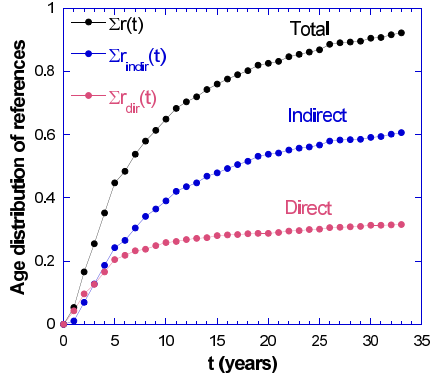}
\caption{Cumulative number of citations of a representative Physics paper, F. Himpsel et al., Phys. Rev. B \textbf{30}, 2257 (1984). The direct citations shoot up immediately  after publication of the parent paper and then their rate slowly decays. The indirect citations shoot up with 1-2  year delay, their rate achieves maximum in another couple of years and then  decays.}
\label{fig:Himpsel1}
\end{figure}

Dynamics of direct and indirect citations for the PRB papers is also generic. Figure \ref{fig:Himpsel1} shows  citation dynamics of one of such papers. The direct citations shoot up soon after publication  and then their growth slows down (surprisingly, it  does not  come to saturation even after 28 years), while indirect citations appear  after 1-2 year delay.
\subsection{Verification of the stochastic model of citation dynamics}
The agreement between the measured and simulated citation distributions is not enough to prove our model. Indeed, while early models of  complex networks growth were validated  by comparing measured and simulated aggregate characteristics, such as  degree distribution, our model belongs to next-generation, it is much more detailed and the comparison to real data is more demanding.  To the best of our knowledge, the methodology of comparing stochastic model/simulation to stochastic data is not well-established. Following Ref. \cite{Medo2014} we believe that the proper validation of a stochastic model shall include multidimensional analysis where several predictions of the model are compared to  measurements.  In the  paper we demonstrate that our model predicts the cumulative citation distribution fairly well. In what follows we verify our model in several other dimensions.

\emph{Stochastic component of the citation dynamics}.
Microscopic citation dynamics is usually considered in relation to the preferential attachment mechanism captured by the following equation
\begin{equation}
\lambda_{i}(t)\propto(K_{i}(t)+K_{0})^{\delta}
\label{Barabasi}
\end{equation}
where $K_{i}$ is the number of citations of the target paper $i$, $K_{0}$ is  the initial attractivity,, and $\delta$ is the growth exponent. Therefore, we measured citation dynamics of papers using the set of dependent and independent variables suggested Eq. \ref{Barabasi}. In particular, for each $t$ we sorted the papers into bins containing the papers with the same $K(t)$, the number of citations garnered by the time $t$. For each bin we  considered distribution of additional citations $k$ garnered in year $t+1$ and calculated the mean and the variance of  this distribution.  This was done both for measured and for simulated data.
\begin{figure}[!ht]
\includegraphics*[width=0.35\textwidth]{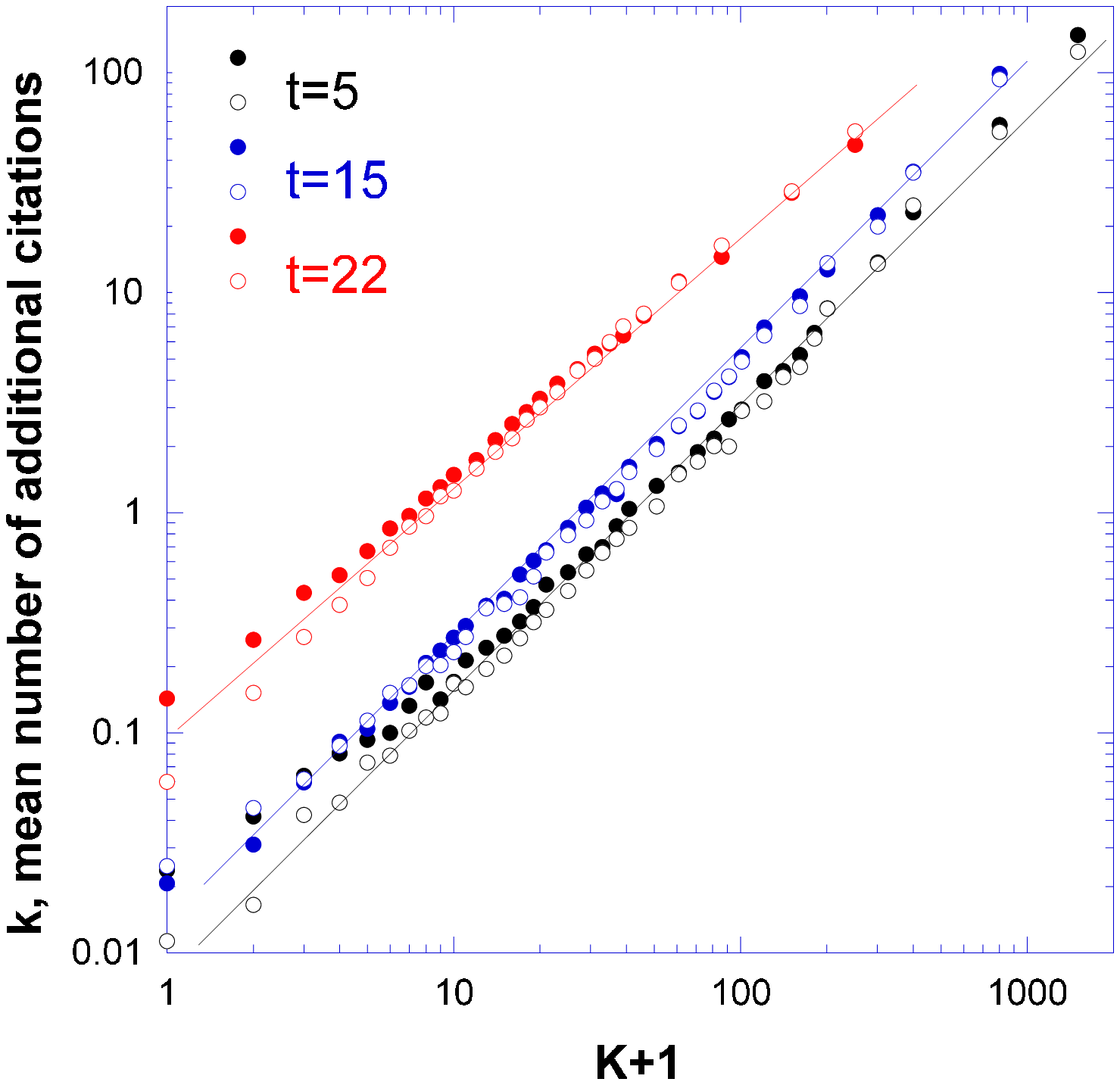}
\includegraphics*[width=0.35\textwidth]{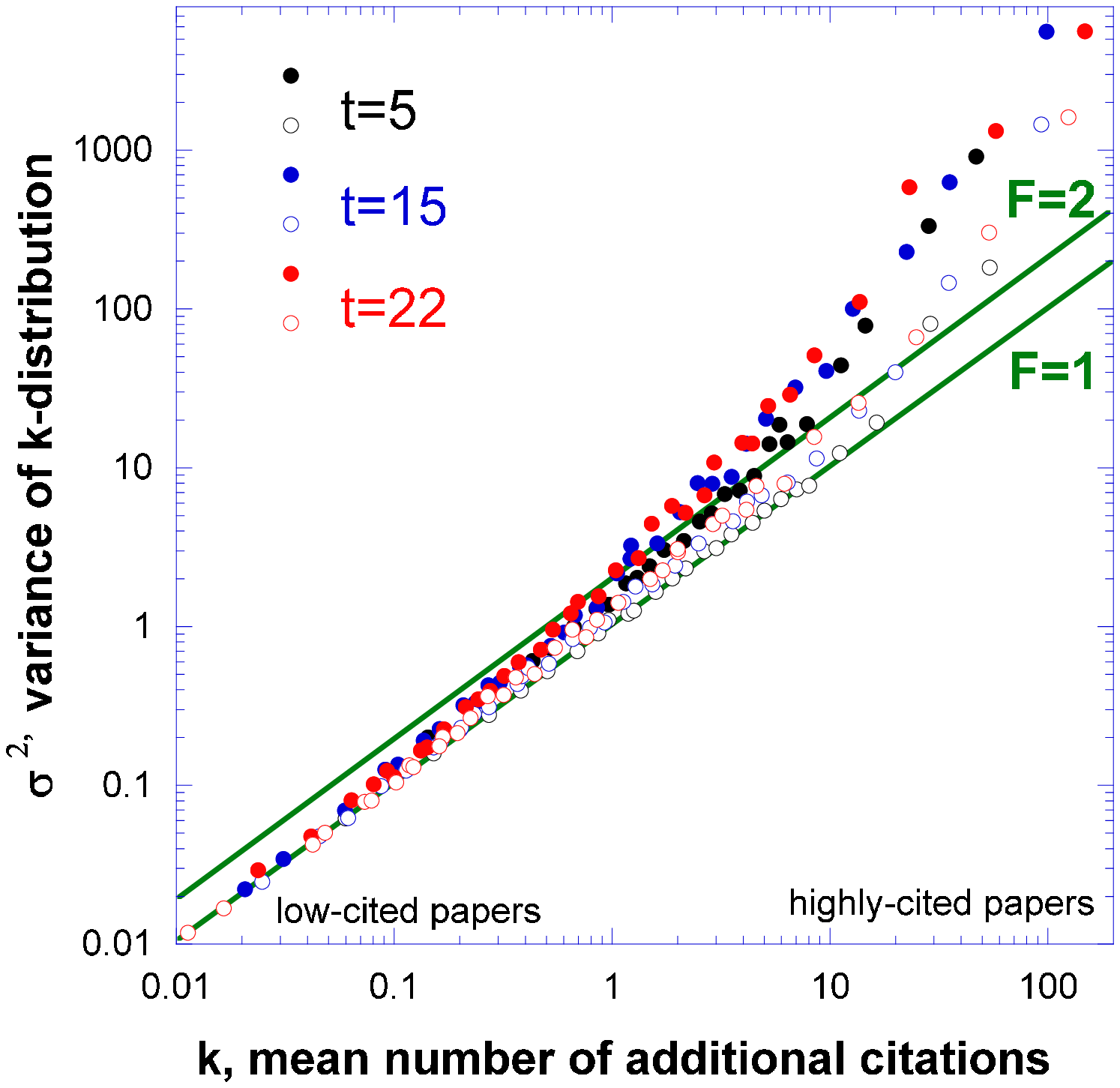}
\caption{(a) Mean number of  additional citations  $\overline{k_{i}(t)}$, in dependence of the number of previous citations $K(t)$, where $t$ is the time after publication. The straight lines show fit to Eq. \ref{Barabasi} with $K_{0}=1$. (b) Variance of the additional citation  distribution, $\sigma^{2}(t)=\overline{\left(k_{i}(t)-\overline{k_{i}(t)}\right)}$
versus mean, $\overline{k_{i}(t)}$. Full circles show measured values, open circles show results of numerical simulation, straight lines indicate constant variance-to-mean ratio (Fano number) where $F = 1$ corresponds to the Poisson distribution. The data are for  40,195 Physics papers published in 1984,  $t$ is the number of years after publication.
}
\label{fig:Fano}
\end{figure}

Figure \ref{fig:Fano}a shows that the mean number of additional citations for the measured and simulated data are very close: both follow Eq. \ref{Barabasi} with $\delta\sim 1.25$ and $K_{0}=1$. Figure \ref{fig:Fano}b  plots the variance versus mean for these distributions. The rationale for such plot is the fact that for the Poisson distribution, the variance-to-mean ratio (Fano number) is $F=1$, while for many other distributions $F>1$. Hence, any deviation from the Poisson distribution can be easily noticed.

For small $\overline{k_{i}(t)}$ the measured and simulated data are much more the same  and both are close to $F=1$ line.  This demonstrates a good agreement between the measurement and the model. It also means that the stochastic component of citation dynamics is Poissonian, namely random. For large $\overline{k_{i}(t)}$ the measured data deviate upwards from the $F=1$ line. This means that the variability of citation dynamics arises more from the differences in the citation history of the papers than from the chance. Although the simulated data for highly-cited papers also deviate upwards from the $F=1$ dependence, this deviation is smaller that that for measured data. Hence, our model captures well the mean citation dynamics of all papers, correctly predicts the variability of  citation dynamics of the low- and moderately-cited papers and underestimates it for highly-cited papers.

\emph{Autocorrelation.} Another point of comparison is the autocorrelation of additional citations acquired by a paper in subsequent years. We characterize it by the  Pearson autocorrelation coefficient, $c_{t,t-1}$ and considered it for annual citations measured for the sets of papers which have  the same number of previous citations $K(t)$. Specifically, we determined  the number of citations garnered by each paper in such set during two subsequent years, $t$ and $t-1$:   $k_{i}(t)$ and $k_{i}(t-1)$, correspondingly. Then we calculated the Pearson autocorrelation coefficient
\begin{figure}[!ht]
\includegraphics*[width=0.35\textwidth]{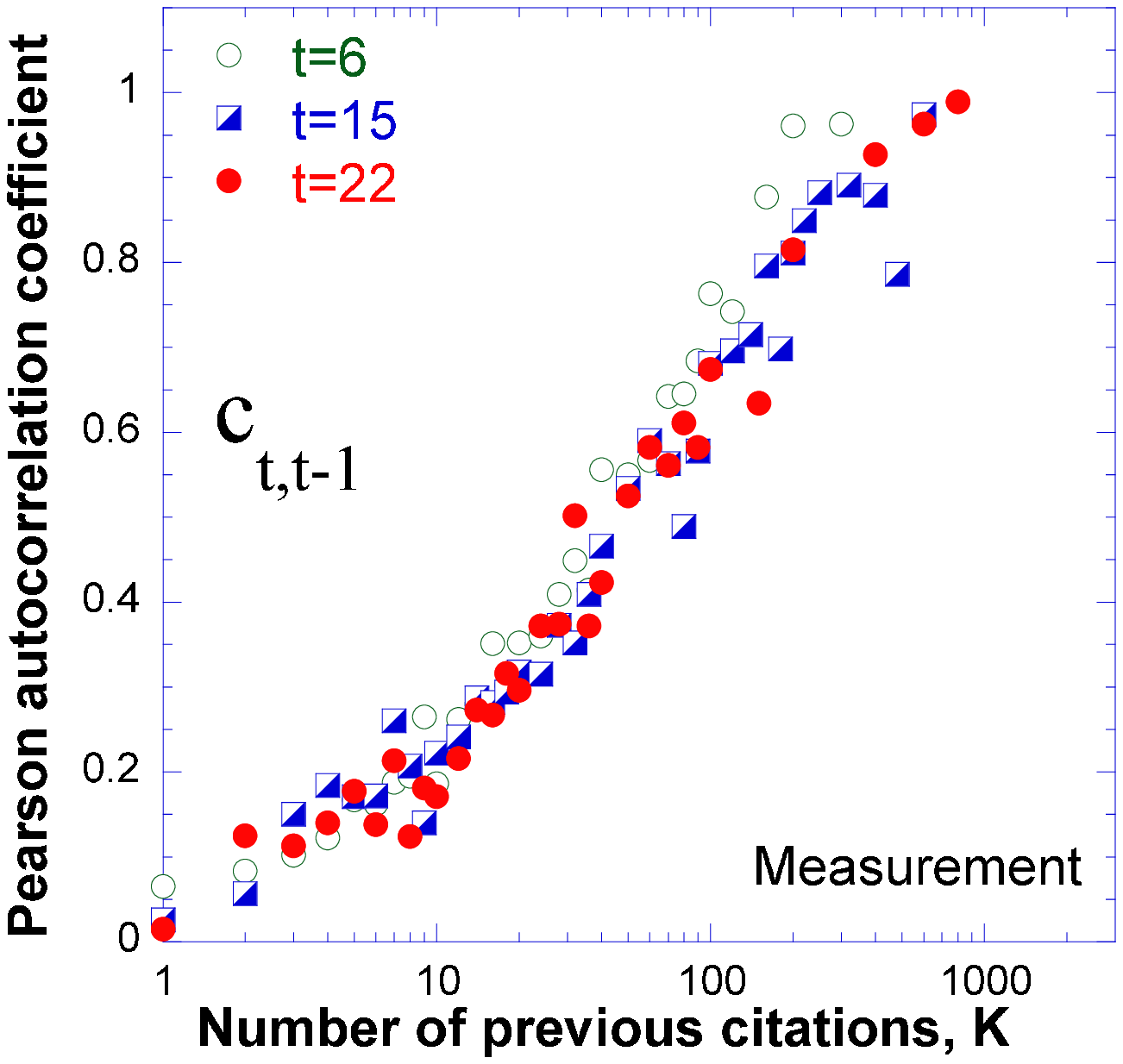}
\includegraphics*[width=0.35\textwidth]{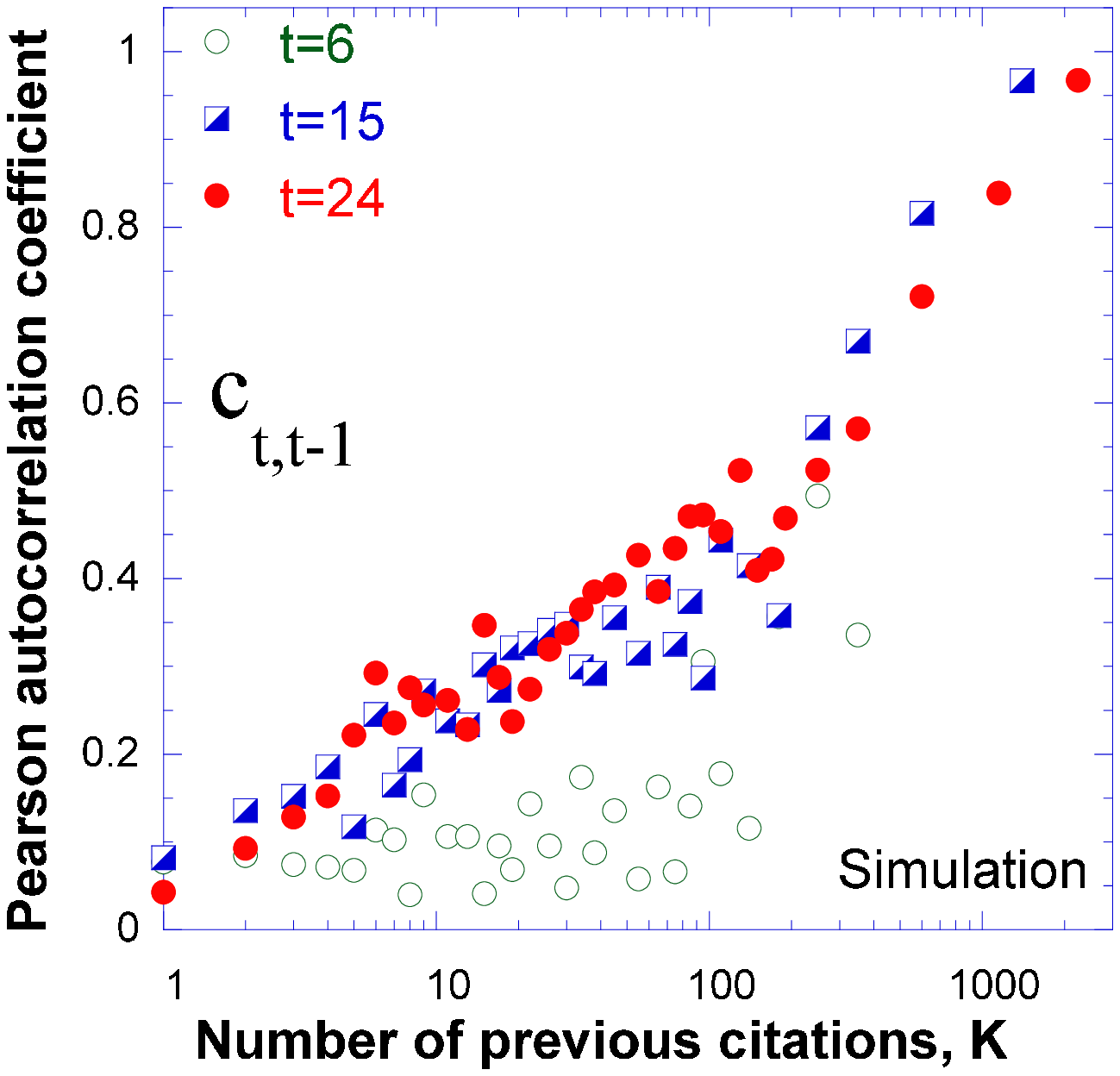}
\caption{The Pearson autocorrelation coefficient, $c_{t,t-1}$,  for additional citations $k_{i}(t),k_{i}(t-1)$. Each point corresponds to the set of papers with the same  number of previous citations  $K$ garnered during $t$ years after publication. (a) Measurements. (b) Numerical simulation. The simulation agrees with the measurements for $t>10$ yr and underestimates $c_{t,t-1}$ for $t<10$ yr.
}
\label{fig:correlation}
\end{figure}
\begin{equation}
c_{t,t-1}=\frac{\overline{\left(k_{i}(t)-\overline{k_{i}(t)}\right)\left( k_{i}(t-1)-\overline{k_{i}(t-1)}\right)}}{\sigma_{t}\sigma_{t-1}}
\label{c}
\end{equation}
where,  $\sigma_{t},\sigma_{t-1}$ are, correspondingly, the standard deviations of the $k_{i}(t)$ and $k_{i}(t-1)$ distributions and the averaging is performed over all papers in the set.

Figure \ref{fig:correlation} shows our results. We do not know why $c_{t,t-1}(K)$ dependences for different years $t$  collapse onto a single curve. A more important  fact is that  $c_{t,t-1}$ grows with $K$. This is a direct consequence of the nonlinear $\tilde{P_{0}}(K)$ dependence  (Eq. \ref{probability}) and our model reproduces this growth fairly well for $t>10$. For $t>10$ there is discrepancy between the measured and simulated $c_{t,t-1}(K)$ that can be lifted  by assuming that $\gamma$ in Eq. \ref{P0} depends on $K$. We reserve this topic for future studies.

What is the meaning of $c_{t,t-1}$? Low $c_{t,t-1}$ indicates that the stochastic component of  citation dynamics is random, high $c_{t,t-1}$ indicates that it is determined by previous history. Consequently,  small $c_{t,t-1}$ is associated with jerky,  and $c_{t,t-1}\sim 1$ is associated with smooth citation trajectories. In the main body of the paper we compare the measured and numerically simulated citation trajectories of the Physics papers that were published in 1984.  For moderately-cited papers  the measured and simulated  trajectories look  similar- the fluctuations are of the same size and the spread in trajectories is  the same. Trajectories are jerky, consistent with low $c_{t,t-1}$. For highly-cited papers  both sets are smooth, consistent with high $c_{t,t-1}$, although the spread of the measured trajectories exceeds the  spread of the simulated ones.

\begin{figure}[ht]
\includegraphics*[width=0.35\textwidth]{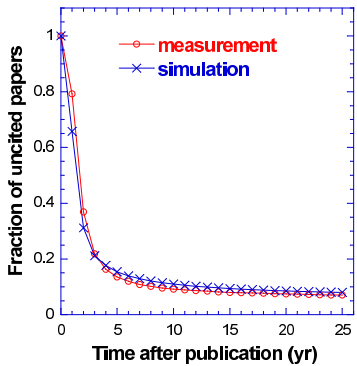}
\caption{Time dependence of the fraction of Physics papers that remained uncited 25 years after publication. The data is for the ensemble of 40, 195 Physics papers published in 1984. Note a good agreement between the measurements and simulations.}
\label{fig:uncited}
\end{figure}

\emph{Uncited papers}. Figure \ref{fig:uncited} shows that our model correctly predicts the number of uncited papers. We found that only a small fraction -$7.5\%$ of the Physics papers published in 1984- remained uncited after 25 years. The good correspondence between the measured number of uncited papers and the model prediction indicates that  uncited papers are a natural outcome of the stochastic Poisson process,\cite{Burrell2013} they are not unread and contribute to scientific progress being an integral part of scientific enterprise.\cite{Seglen1992}

\subsection{Our results in the context of network science}
We consider our measurements of the direct and indirect citations  in the context of network science. On the one hand, the number of second-generation citations $M^{II}$  is nothing else but the average nearest-neighbor connectivity, $k_{nn}$. Increasing $M^{II}(K)$ dependence indicates assortativity of citation network. On another hand, the number of indirect citations is directly related to the local clustering coefficient $C_{K}$, which is the ratio of the number of transitive triples to the total  number of triples connected to a certain parent node. Indeed, consider a parent paper $i$ that has $K$ citations. The number of all triples connected to it is $N^{II}N^{I}$ where $N^{I}=K$ is the average number of the first generation citing papers (node degree) and  $N^{II}$ is the average number of citing papers per one first-generation citing paper. Among these $N^{II}N^{I}$ papers, there are some associated with  indirect citation that make a part of $j$-multiplet (Fig. 17). The number of the later is $\pi_{j}jf_{j}N^{II}N^{I}$, where $f_{j}$ is the fraction of $j$-multiplets among second-generation citing papers, $\pi_{j}$ is the probability of indirect citation, and factor $j$ in the sum appears because each indirect citation in the $j$-multiplet is associated with $j$ triangles. The number of all triangles is $K(K-1)/2$. Since $K=N^{I}$, then
\begin{equation}
C_{K}=\frac{2N^{II}\sum_{j=1}j\pi_{j}f_{j}}{K-1}
\label{clustering}
\end{equation}
\begin{figure}[!ht]
\includegraphics*[width=0.35\textwidth]{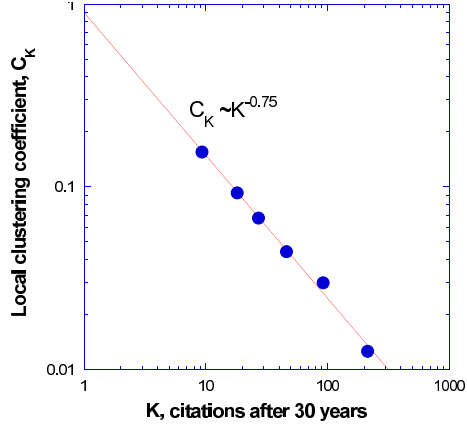}
\caption{$C_{k}$, local clustering coefficient, as yielded by Eq. \ref{clustering1}.
}
\label{fig:clustering}
\end{figure}

If we limit ourselves only to singlets and doublets and neglect higher-order multiplets, then $\pi_{2}=4\pi_{1}$, $f_{1}+f_{2}\approx1$, $s\approx1+f_{2}$ where $s=\frac{M^{II}}{N^{II}}$ is the ratio of the second generation citations to the second-generation citing papers (Sec. VIII). Finally,
\begin{equation}
C_{K}\approx\frac{2N^{II}\pi_{1}[1+7(s-1)]}{K-1}
\label{clustering1}
\end{equation}

Our measurements  indicate that $N^{II}$ is almost independent on $K$ while $s$ increases logarithmically with $K$. Figure \ref{fig:clustering} shows that $C_{K}$, which was calculated according to Eq. \ref{clustering1} using data of  Sec. VIII,  follows $K^{-0.75}$ dependence. These power-law dependence  agrees with the findings of Ref. \cite{Barabasi2015} for PR to PR citation network.

The direct relation between $C_{K}$ and $s$ suggests alternative interpretation of the probability of indirect citation $P_{0}$. Indeed, in Sec. VIIIb we showed that $P_{0}$ is directly related to $s$:
\begin{equation}
\tilde{P_{0}}\propto\pi_{1}[1+3(s-1)]
\label{3s}
\end{equation}
By excluding $s$ from Eqs. \ref{clustering1},\ref{3s} we achieve direct relation between $P_{0}$ and clustering coefficient  $C_{K}$. In particular, $P_{0}\propto C_{K}(K-1)$ for highly-cited papers with high $s$.  This relation indicates that among the papers with the same number of previous citations, those with high clustering coefficient are cited more intensively- the possibility already prevised by Bagrow and Brockmann \cite{Bagrow2013}.

\subsection{Hand-waving explanation of the nonlinearity}
\begin{figure}[!ht]
\begin{center}
\includegraphics*[width=0.3\textwidth]{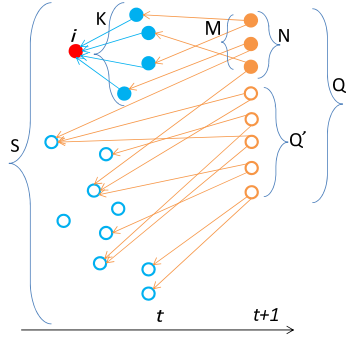}
\caption{A fragment of citation network showing a parent paper $i$ and its first- and second-generation citations. There are $K$ first-generation citing papers  and they are a part of all $S$ papers published in this field by the time $t$. There are $Q$ papers in this field that were published in year $t+1$ and that cite $S$ papers published earlier.  Among those $Q$ papers there are $N$ second-generation citing papers that cite one of the $K$ first-generation citing papers and there are $Q'$ papers that do not cite them.  $M$ is the number of the second-generation citations published in year $t+1$.
}
\label{fig:saturation}
\end{center}
\end{figure}
We present here a toy model explaining nonlinear dynamics of indirect citations. The model serves for purely illustrative purposes. Consider a parent paper \emph{i} that has  $K$  citing papers published by year $t$. These $K$ first-generation citing papers are a part of a large set of  all $S$  papers that were published in this research field by year $t$. Consider all $Q$ papers in this field that were published in the year $t+1$. We neglect obsolescence and assume that each of these papers issues $\sim m$ citations to the papers published previously.  The number of \emph{citations} of the $K$ papers (second-generation citations with respect to the parent paper \emph{i}) is $M\approx mQ\frac{K}{S}$. The number of second-generation \emph{citing papers} of the paper $i$ is  $N=Q-Q'$ where $Q'$ is the number of papers published in $t+1$ that do not cite our $K$ papers. Assuming Poissonian distribution  of  citations issued by each paper from the $Q$-set, we find $Q'=Q\sum_{n=0}^{\infty}(1-\frac{K}{S})^{n}\frac{m^{n}}{n!}e^{-m}= Qe^{-\frac{mK}{S}}\sum_{n=0}^{\infty}\frac{[m(1-\frac{K}{S})]^{n}}{n!}e^{-m(1-\frac{K}{S})}$.
According to the properties of the Poisson distribution, $\sum_{n=0}^{\infty}\frac{[m(1-\frac{K}{S})]^{n}}{n!}e^{-m(1-\frac{K}{S})}=1$, hence $N=Q(1-e^{-\frac{mK}{S}})$.

We consider now the parameter $s$ which measures the number of paths leading from the second-generation citing papers to the parent paper $i$, namely, $s=\frac{M}{N}=\frac{m \frac{K}{S}}{1-e^{-m\frac{K}{S}}}$. The series expansion of this expression in small parameter $m\frac{K}{S}$  yields $s\approx 1+K\frac{m}{2S}+K^{2}\frac{m^2}{12S^2}+..$. Thus $s$ increases with $K$ meaning that the highly-cited papers have higher proportion of multiple paths than the low-cited papers. The source of nonlinear citation dynamics is this $s(K)$ dependence.

Of course, this hand-waving explanation of the $s(K)$ dependence does not account for all our results. It assumes that  the number of second-generation citations of a given paper grows linearly with $K$ while the number of its second-generation citing papers grows slowly than linear with $K$. Our measurements indicate exactly the opposite behavior - the number of second-generating citing papers grows linearly with $K$ and the number of second-generation citations grows faster than linear, indicating on assortativity of citation network.
\end{document}